\documentclass[fleqn,usenatbib]{mnras}

\usepackage{newtxtext,newtxmath}

\usepackage[T1]{fontenc}
\usepackage{ae,aecompl}

\usepackage{graphicx}	
\usepackage{amsmath}	
\usepackage{amssymb}	
\usepackage{color}
\usepackage[utf8]{inputenc}
\usepackage{natbib}
\usepackage[normalem]{ulem}
\usepackage{hyperref}
\usepackage[dvipsnames]{xcolor}

\usepackage{etoolbox}
\makeatletter
\patchcmd\@combinedblfloats{\box\@outputbox}{\unvbox\@outputbox}{}{%
   \errmessage{\noexpand\@combinedblfloats could not be patched}%
}%
 \makeatother

\newcommand{\NconfirmedDNS}{15}

\title[DNSs]{Modelling Double Neutron Stars: Radio and Gravitational Waves}

\author[Debatri Chattopadhyay et al.]{
\newauthor  Debatri Chattopadhyay,$^{1,2}$\thanks{E-mail: dchattopadhyay@swin.edu.au}
Simon Stevenson,$^{1,2}$
Jarrod R. Hurley,$^{1,2}$
Luca J. Rossi,$^{1}$
\newauthor and Chris Flynn$^{1}$
\\
$^{1}$ Centre for Astrophysics and Supercomputing, Swinburne University of Technology, John St, Hawthorn, Victoria- 3122, Australia \\
$^{2}$ The ARC Centre of Excellence for Gravitational Wave Discovery,  OzGrav
}

\date{Accepted XXX. Received YYY; in original form ZZZ}

\pubyear{2019}

\begin{document}
\label{firstpage}
\pagerange{\pageref{firstpage}--\pageref{lastpage}}
\maketitle

\begin{abstract}
We have implemented prescriptions for modelling pulsars in the rapid binary population synthesis code COMPAS. We perform a detailed analysis of the double neutron star (DNS) population, accounting for radio survey selection effects. The surface magnetic field decay timescale (${\sim}1000$\,Myr) and mass scale (${\sim}0.02$\,M$_\odot$) are the dominant uncertainties in our model. Mass accretion during common envelope evolution plays a non-trivial role in recycling pulsars. We find a best-fit model that is in broad agreement with the observed Galactic DNS population. Though the pulsar parameters (period and period derivative) are strongly biased by radio selection effects, the observed orbital parameters (orbital period and eccentricity) closely represent the intrinsic distributions. The number of radio observable DNSs in the Milky Way at present is about 2500 in our model, corresponding to approximately 10\% of the predicted total number of DNSs in the galaxy. Using our model calibrated to the Galactic DNS population, we make predictions for DNS mergers observed in gravitational waves. The DNS chirp mass distribution varies from 1.1M$_\odot$ to 2.1M$_\odot$ and the median is found to be 1.14\,M$_\mathrm{\odot}$. The expected effective spin $\chi_\mathrm{eff}$ for isolated DNSs is $\lesssim$0.03 from our model. We predict that 34\% of the current Galactic isolated DNSs will merge within a Hubble time, and have a median total mass of 2.7\,M$_\mathrm{\odot}$. Finally, we discuss implications for fast radio bursts and post-merger remnant gravitational-waves.
\end{abstract}





\section{Introduction}
\label{sec:introduction}

\begin{table*}
    \scriptsize
	\centering
	\caption{Observed pulsars in Milky Way DNSs.
	References: 
	a \citep{Breton:2008xy, Ferdman:2013xia}, 
	b \citep{Cameron:2017ody}, 
	c \citep{Champion:2004hc}, 
	d \citep{Faulkner:2004ha, Ferdman:2014rna}, 
	e \citep{Nice:1996qe, Janssen:2008mh}, 
	f \citep{Lazarus:2016hfu, Ferdman:2017}, 
	g \citep{Lyne:1999hu, Corongiu:2006rd}, 
	h \citep{Martinez:2015mya}, 
	i \citep{Martinez:2017jbp}, 
	j \citep{Stovall:2018ouw}, 
	k \citep{Swiggum:2015yra}, 
	l \citep{Fonseca:2014qla}, 
	m \citep{Lynch:2018zxo},
	n \citep{HulseTaylor:1975, Weisberg:2016jye},
	o \citep{Keith:2008ga}, 
	p \citep{Ng:2018lkc}, 
	q \citep{vanLeeuwen:2014sca}, 
	r \citep{LynchR:2012},
	s \citep{Anderson:1990}.
	Systems marked $^*$ are in globular clusters, whereas systems marked $^\dagger$ may contain a white dwarf companion.
	$P$: pulsar spin period, $\dot{P}$: pulsar spin down rate, $e$: orbital eccentricity, $L_\mathrm{1400}$: Luminosity in 1400 MHz, $B_\mathrm{surf}$: pulsar surface magnetic field, $P_\mathrm{orb}$: orbital period, $M_\mathrm{p}$: pulsar mass, $M_\mathrm{c}$: companion mass.}
	
	\begin{tabular}{llcccccccr} 
		\hline
		Index & Name & $P$(s) & $\dot{P}$($10^{-18}$s/s) & $e$ &  $L_\mathrm{1400}$ (mJy $\times$ kpc$^2$) & $B_\mathrm{surf}$($10^{9}$G) & $P_\mathrm{orb}$(days) & $M_\mathrm{p}$(M$_\mathrm{\odot}$) & $M_\mathrm{comp}$(M$_\mathrm{\odot}$)\\
		\hline
        1 & J0737$-$3039A$^a$  &  0.022  &  1.75993  &  0.087  &    1.94  &    6.4  &  0.102 & 1.338 & 1.248\\
        2 & J0737$-$3039B$^a$ &  2.773  &  892.0  &  0.087  &    1.57  &    1590  &  0.102 & 1.248 & 1.338\\
        3 & J1757$-$1854$^b$  &  0.021  &  2.6303  &  0.605  &    95.84  &    7.61  &  0.183 & 1.338 & 1.394\\
        4 & J1829+2456$^c$  &  0.041  &  0.0525  &  0.139  &    *  &  1.48  &  1.176 & $<$1.34 & $>$1.26\\
        5 & J1756$-$2251$^d$  &  0.028  &  1.017502  &  0.180  &    0.32  &    5.45  &  0.319 & 1.341 & 1.230 \\
        6 & J1518+4904$^e$  &  0.040  &  0.027190  &  0.249  &   3.69  &    1.07  &  8.634 & $<$1.42 &$>$1.29 \\
        7 & J1913+1102$^f$  &  0.027  &  0.161  &  0.089  &   1.02  &    2.12  &  0.206 & $<$1.84 &$>$1.04 \\
        8 & J1811$-$1736$^g$  &  0.104  &  0.901  &  0.828  &    25.51  &    9.8  &  18.779 & $<$1.74 &$>$0.93 \\
        9 & J0453+1559$^h$  &  0.045  &  0.18612  &  0.112  &    *  &    2.95  &  4.072 & 1.559 & 1.174\\
        10 & J1411+2551$^i$  &  0.062  &  0.0956  &  0.169  &   *  &    2.47  &  2.615 & $<$1.62 &$>$0.92\\
        11 & J1946+2052$^j$  &  0.016  &  0.92  &  0.063  &   0.76  &    4.0  &  0.078 & $<$1.31 &$>$1.18 \\
        12 & J1930$-$1852$^k$  &  0.185  &  18.001  &  0.398  &   *  &    58.5  &  45.060 & $<$1.25 & $>$1.30 \\
        13 & B1534+12$^l$  &  0.037  &  2.422494  &  0.273  &    0.66  &  9.7  &  0.420 & 1.333 & 1.345\\
        14 & J0509+3801$^m$  &  0.076  &  7.931  &  0.586  &    *  &  24.9  &  0.379 & 1.36 & 1.46\\
        15 & B1913+16$^n$ &  0.059  &  8.6183  &  0.617  &   24.81  &    22.8  &  0.322 & 1.438 & 1.390 \\
        16 & J1753$-$2240$^o$ $^\dagger$  &  0.095  &  0.97  &  0.303  &  1.56  &    9.72  &  13.637 & * & *\\
        17 & J1755$-$2550 $^p$ $^\dagger$  &  0.315  &  2433.7  &  0.089  & 4.78  &    886  &  9.696 & * & $>$ 0.40\\
        18 & J1906+0746 $^q$ $^\dagger$  &  0.144  &  20267.8  &  0.085  &   30.12  &    1730  &  0.165 & 1.291 & 1.322 \\
        19 & J1807$-$2500B $^r$ $^*$ $^\dagger$  &  0.004  &  0.0823  &  0.747  &   *  &    0.594  &  9.956 & 1.366 & 1.206\\
        20 & B2127+11C $^s$ $^*$  &  0.030  &  4.98789  &  0.681  &   *  &    12.5  &  0.335 & 1.358 & 1.354\\
		\hline
	\end{tabular}

    \label{tab:observed_DNSs}
\end{table*}

Much of what we know about neutron stars (NSs) has come from radio telescope observations of pulsars---rapidly rotating, highly magnetised NSs \citep{Hewish:1968bj}. Pulsars are extraordinarily regular in their spin. Their period and period derivatives can be measured with phenomenal precision. For some binary pulsars, their orbital properties (orbital period, eccentricity and masses) are also well-measured quantities. Pulsar timing is comparable in precision to terrestrial atomic clocks \citep{Hobbs:2012, Hobbs:2019pulsarbased}. An aggregation of pulsars, with spin periods of the order of milliseconds (called millisecond pulsars, MSPs), scattered across the Milky Way can be analysed to detect low frequency gravitational waves from merging supermassive black-holes at the centre of galaxies \citep[e.g.][]{Mingarelli:2019pyd}. This collective ensemble of millisecond pulsars is called a pulsar timing array \citep{Arzoumanian:2018, Desvignes:2016, Hobbs:2013aka}. 

Within the observed pulsar population of the Milky Way there are \NconfirmedDNS{} confirmed double neutron star (DNS) binaries  (see Table~\ref{tab:observed_DNSs}), including one special system where both binary members are pulsars; the double pulsar PSR J0737$-$3039 \citep{Burgay:2003jj}. Out of the 15 confirmed DNSs, 14 are in the Galactic field \citep[e.g.][]{Martinez:2015mya, HulseTaylor:1975} and one (B2127+11C) is in the Milky Way globular cluster M15 \citep{Anderson:1990,Jacoby:2006dy}. There are four additional NS binaries in which the companion may be either a NS or a white dwarf, three of which are in the field.

DNSs are one of the most interesting classes of astrophysical systems known. These systems provide information in a number of areas of fundamental physics and astrophysics. The extreme gravity in the proximity of these binary systems is unlike any terrestrial laboratory system and allows tests of General Relativity \citep[GR,][]{Kramer97}. In particular, the double pulsar system has repeatedly shown excellent agreement to predictions by GR \citep[e.g.][]{KramerStairs:2008, Kramer:2005ez}. 

DNSs emit high frequency gravitational waves when they merge, and the signals can be identified by present-day ground-based gravitational wave detectors such as the Advanced Laser Interferometer Gravitational-Wave Observatory \citep[aLIGO,][]{TheLIGOScientificDetector:2014jea} and Advanced Virgo \citep[aVirgo,][]{TheVirgoDetector:2014hva}. In addition to emitting detectable gravitational waves, DNS mergers may also produce counterparts in electromagnetic radiation. GW170817 \citep{Monitor:2017mdv} became the first astronomical event to be detected both in electromagnetic radiation and gravitational waves \citep{TheLIGOScientific:2017qsa}, opening a new era of multi-messenger astronomy \citep{GW170817MultiMessenger:GBM:2017lvd}. The event allowed measurements of the Hubble constant \citep{Abbott:2017xzu, Hotokezaka:2018dfi} and confirmed the long standing hypothesis that DNS mergers are progenitors of at least some short gamma ray bursts \citep[e.g.][]{Murase:2017snw}. Another DNS merger candidate GW190425 was detected by the third observing run of aLIGO and aVIRGO \citep{Abbott:2020uma}, however, no associated electromagnetic counterpart was identified.

The formation of DNSs occurs predominantly through isolated binary evolution. Two massive stars need to remain gravitationally bound to one another as they evolve through two separate supernovae (see \citealp{Tauris:2017omb} for a recent review). Velocity kicks imparted to the NSs at birth may lead to highly eccentric orbits. The orbits of DNSs shrink by losing energy through gravitational wave emission, decreasing both the orbital period and eccentricity \citep{Peters:1964}. Gravitational wave emission is enhanced in highly eccentric binaries, leading them to merge more quickly \citep{2005ApJ...632.1054C}. Dynamical formation of DNSs in star cluster environments is expected to be inefficient \citep{Phinney:1991, Belczynski:2017mqx, 2019arXiv191010740Y}; although the parameter space still requires further investigation. In this work we focus on DNSs formed through isolated binary evolution. We therefore exclude the globular cluster DNSs J1807$-$2500B and B2127+11C from our sample.

In order to predict the properties and merging rates of such binaries, detailed modelling of sources from field binaries is required. We study the properties of DNSs by modelling ensembles of such systems, incorporating the physics of binary evolution and pulsar evolution. The ensembles are generated under different initial assumptions, and are studied to understand the properties and statistics of the population. This method is called population synthesis. Our population synthesis results can be used to predict the properties and detection rates for both ground and space based gravitational-wave \citep[see also][]{Lau:2019} observatories across the parameter space. These theoretical predictions can be compared against the observed radio and gravitational-wave populations of DNSs in order to help constrain uncertain physics such as the decay of NS magnetic fields.

We use the rapid population synthesis code Compact Object Mergers: Population Astrophysics and Statistics \citep[COMPAS,][]{Stevenson:2017tfq,Vigna-Gomez:2018dza,Neijssel2019MNRAS,Broekgaarden:2019MNRAS}, which implements Single Stellar Evolution \citep[SSE,][]{2000MNRAS.315..543H} and Binary Stellar Evolution \citep[BSE,][]{2002MNRAS.329..897H}. COMPAS has previously been used by \citet{Vigna-Gomez:2018dza} to study the formation history of Galactic DNSs. 

The paper is organised as follows: in Section~\ref{sec:pulsar_modelling} we describe the model for pulsar evolution we have implemented in COMPAS. We use this to study the properties of pulsar-NS binaries. We follow the orbits of the simulated DNSs in the Galaxy using the Numerical Integrator of Galactic Orbits  \citep[NIGO,][see Section~\ref{subsec:GalacticPotential}]{2015A&C....12...11R}, and account for radio selection effects using PSREvolve \citep[][see Section~\ref{subsec:radio_selection_effects}]{2011MNRAS.413..461O}. We generate a suite of models, varying our assumptions about pulsar evolution, and compare our models to the observed sample of Galactic DNSs in Section~\ref{sec:RadioPopulation}. In Section~\ref{sec:grav_waves} we use our model calibrated to the Galactic DNS population to make predictions for DNSs observable in gravitational-waves \citep[see also][]{Lau:2019}. We summarize our findings in Section~\ref{sec:discussion_conclusion}.

\begin{figure*}
\includegraphics[width=0.99\textwidth]{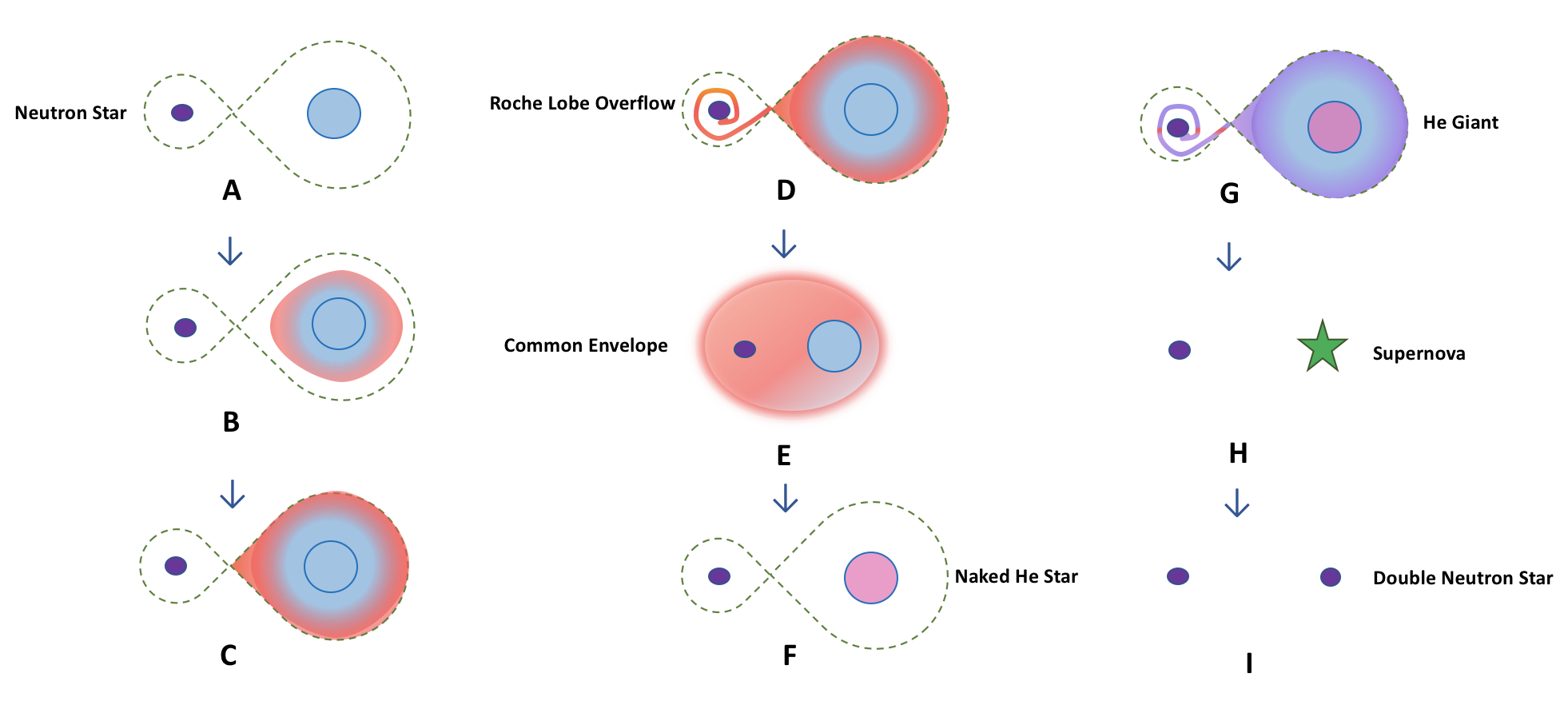}
\caption{
Cartoon of the evolutionary phases in the formation of a DNS (also see \citealp{Vigna-Gomez:2018dza}). Here we focus on the evolution of the binary after the formation of the first neutron star (A). The envelope of the main-sequence companion of the neutron star expands (B) and fills its Roche Lobe (C). Matter from the companion then falls onto the neutron star by the formation of an accretion disk (D). This mass transfer then becomes a run-away process forming a common envelope, engulfing both the stars (E). Further, the common envelope ejection leads to the formation of a naked Helium (He) star (F), which evolves to a He-giant. Mass transfer onto the neutron star also occurs from the He-giant companion star (G). The episode of case BB mass transfer recycles the first formed neutron star. After the second supernova event (H), a DNS is formed (I).
}
\label{DNSFormation}
\end{figure*} 


\section{Modelling Pulsar Evolution}
\label{sec:pulsar_modelling}

To date, COMPAS allowed a remnant to be predicted as a black hole \citep[e.g.][]{Stevenson:2017tfq, Barrett:2018MNRAS,  Stevenson:2019rcw, Bavera:2019} or a NS \citep[e.g.][]{Vigna-Gomez:2018dza,Neijssel2019MNRAS,Broekgaarden:2019MNRAS}, but did not follow any subsequent evolution of these objects. To analyse whether a NS is also a pulsar, other properties such as the magnetic field and spin period need to assigned to the NS and calculated over time, e.g. to determine the spin down/up rate.

In COMPAS, a pulsar-NS binary can be formed through several channels. In the most dominant channel \citep[c.f.][]{Vigna-Gomez:2018dza}, the initially more massive star in the binary loses some of its mass through mass transfer onto the initally less massive star, and then undergoes a supernova (SN) explosion. In some cases, this mass transfer may lead to a reversal of the mass ratio, with the initially less massive star becoming more massive and undergoing SN first to produce a NS. For further details on evolutionary channels of massive binaries forming DNSs prior to the first SN, we refer to \citet{Vigna-Gomez:2018dza}. 

In this paper, we focus on the evolution of the binary after the first SN event that produces the first neutron star in the binary (Fig.~\ref{DNSFormation}-A). The first born neutron star may be spun-up through mass transfer from its companion via Roche lobe overflow (Fig.~\ref{DNSFormation}-D). In many cases, the extreme mass ratio leads to common envelope evolution (Fig.~\ref{DNSFormation}-E: for more details on assumptions of mass transfer and common envelope evolution see section~\ref{subsec:mass_transfer_case}). After the ejection of the common envelope, the companion becomes a naked Helium (He) star (Fig.~\ref{DNSFormation}-F). It subsequently evolves to become a He-giant star, that again overflows its Roche lobe in an episode of case BB mass transfer \footnote{Stable mass transfer from a He-star \citep{Dewi:2002rx}.}. The first born neutron star may be spun up through mass accretion during this phase (Fig.~\ref{DNSFormation}-G). Finally, the companion undergoes SN as well (Fig.~\ref{DNSFormation}-H) and a DNS system is formed (Fig.~\ref{DNSFormation}-I).

All NSs are born in a SN as a radio-observable pulsar in our simulations. We assign them a spin period and magnetic field at formation. We assume the canonical magnetic dipole model for pulsars \citep{OstrikerGunn:1969}. Their rotational deceleration (spin down) is computed as a function of time, and at the time of observation the pulsar may have become a non-radio NS. 

Accretion onto a pulsar during mass transfer or common envelope evolution causes an exchange of angular momentum between the accreting pulsar and infalling matter \citep{1994MNRAS.269..455J} that can modify the pulsar's spin and magnetic field \citep{ZhangKojima:2006}. Pulsars that get spun-up by mass transfer are called `recycled' pulsars, and those that are not are `non-recycled' pulsars. During mass accretion onto a pulsar, the binary emits in X-ray \citep{Nagase:1989}, contrary to the usual radio emission. Since the accretion phase is short lived compared to the entire evolution timescale, we do not model the X-ray emission phase. In a binary system, only the first born neutron star has the possibility of becoming a recycled pulsar. The second born NS already has a NS companion and thus has no possibility of becoming a recycled pulsar. We use the term `primary' to indicate the first born NS which may or may not be a recycled pulsar. Likewise, the term `secondary' is used in this paper to indicate the star that becomes a NS second. 

In the following, we describe the two separate cases of pulsar evolution we have modelled: 
\begin{enumerate}
    \item \textbf{Isolated pulsar evolution} when the pulsar and its companion are evolving independently (see Section~\ref{subsec:isolated})
    \item \textbf{Pulsar recycling through mass transfer} when there is mass transfer from the companion onto the pulsar (see Section~\ref{subsec:mass_transfer_case})
\end{enumerate}

\subsection{Isolated pulsar evolution}
\label{subsec:isolated}

For a pulsar evolving without any interaction involving mass transfer from its companion we consider the evolution to be ``Isolated". Although the binary system is bound together gravitationally, the pulsar parameters---spin, spin down rate and magnetic field---remain unaffected by the presence of the companion. The pulsar can thus be assumed to be a magnetized, rotating, spherical body spinning down solely due to magnetic dipole radiation. This is called the ``spin down" phase of the pulsar. The rate of change of angular velocity, i.e. angular acceleration ($\Dot{\Omega}$) and the angular velocity (we use the terms angular velocity and angular frequency interchangeably in this paper) $\Omega$ are related by
\begin{equation}
    \Dot{\Omega} \propto \Omega^n ,
    \label{BrakingIndex}
\end{equation}
where $n$ is the magnetic braking index. For magnetic dipole emission assumed in our model, $n=3$. Observationally, $n$ shows a range of values from 2.5--3.5 \citep{Manchester:2005yCat}. Angular momentum loss through gravitational waves results in a braking index $n = 5$ and is negligible for most pulsars \citep{Woan:2018tey}. If the spinning down of the pulsar occurs due to stellar winds, the braking index $n = 1$ \citep{1969ApJ...157..869G}. 
While young pulsars are more likely to show higher braking indices \citep[e.g.][]{Archibald:2016hxz},the magnetic braking index model with $n=3$ is a good description for pulsars towards the middle of their life, as well as for observed DNSs. 

The rate of change of angular frequency for our pulsar model is given by
\begin{equation}
     \dot{\Omega} = -\frac{8\pi B^2R^6\sin^2 \alpha \Omega^3}{3 \mu_0 c^3 I } ,
     \label{SpinDownIsolated}
 \end{equation}
where $\Omega$ is the angular frequency, $\Dot{\Omega}$ is the rate of change of $\Omega$, $B$ is the surface magnetic field of the pulsar, $R$ is the radius of the pulsar, $\alpha$ is the angle between the axis of rotation and magnetic axis, $c$ is the speed of light, $\mu_0$ is the permeability of free space and $I$ is the moment of inertia of the pulsar. The equation is in SI units. We calculate the spin $P$ and spin down rate $\Dot{P}$ from $\Omega$ and $\Dot{\Omega}$ using
\begin{equation}
    P = \frac{2\pi}{\Omega} ,
    \label{eq:PfromOmega}
\end{equation}
and
\begin{equation}
    \dot{P}= -\frac{\dot{\Omega}P}{\Omega} .
    \label{eq:PdotFromOmegaDot}
\end{equation}

The surface magnetic field of the pulsar decreases due to ohmic dissipation \citep{Urpin:1997fg, KonarBhattacharya:1997a, KonarBhattacharya:1999a, KonarBhattacharya:1999b}, where the presence of an electric field creates resistance and results in the decay of the surface magnetic field. The radio observations of both single and binary pulsars show the older population having a surface magnetic field lower than younger pulsars. This has motivated the magnetic decay models of pulsars in older studies \citep{GunnOstriker:1970, Stollman:1987} and comparatively more recent studies as well \citep{Kiel:2008xw, 2011MNRAS.413..461O}. We assume that the surface magnetic field decays with time according to
\begin{equation}
    B = (B_0 - B_\mathrm{min})\times\exp(-t/\tau_d)+B_\mathrm{min} ,
    \label{MagneticFieldIsolated} 
\end{equation}
where $\tau_d$ is the magnetic field decay timescale and is a free parameter in our model, $B_0$ is the initial surface magnetic field and $B_\mathrm{min}$ is the minimum surface magnetic field strength at which we assume the magnetic field decay ceases and is also a free parameter.  \citet{ZhangKojima:2006} showed that there is a lower limit to the surface magnetic field strength of a pulsar, with a typical value of $B_\mathrm{min}= 10^8$\,G \citep{2011MNRAS.413..461O}.

Substituting $B$ from Equation~\ref{MagneticFieldIsolated} into Equation~\ref{SpinDownIsolated} and integrating gives
\begin{equation}
   \frac{1}{\Omega_f ^2} = \frac{8\pi R^6\sin^2\alpha}{3\mu_0c^3I}[B_\mathrm{min}^2\Delta t - \tau_d B_\mathrm{min} (B_f - B_i) - \frac{\tau_d}{2}(B_f ^2 - B_i ^2)] + \frac{1}{\Omega_i ^2}  .
   \label{SpinIsolatedIntegrated}
\end{equation}
Equation~\ref{SpinIsolatedIntegrated} gives an analytic solution and direct equation for spin\footnote{The exact solution liberates us from using any numerical integrator, and thus our model is both computationally efficient, and the results are free from the build up of numerical errors.}. Here, $\Omega_i$ and $\Omega_f$ are the initial and final spins, $B_i$ and $B_f$ are the initial and final magnetic fields, and $\Delta t$ is time difference between the two states. We calculate the angular momentum $J = I \Omega$ from Equation~\ref{SpinIsolatedIntegrated} directly, using the equation of state insensitive relation from \citet{Lattimer:2004nj} for the moment of inertia $I$. 

The value of the magnetic field decay timescale $\tau_d$ is one of our key uncertainties. Previous works by \cite{Kiel:2008xw} and \cite{2011MNRAS.413..461O} have assumed different values of this parameter. We have varied our models with $\tau_d$ = 10, 100, 500, 1000 and 2000 Myr (see Table~\ref{tab:TableModelSpecifications}). We show that our best fit model has a magnetic field decay timescale of $\tau_d = 1000$\,Myr, but is highly dependent on other parameters (see Section~\ref{subsubsec:pulsar_params}).

The birth magnetic field $B_0$ and initial angular velocity $\Omega_0$ for pulsars are based on radio observations (see Section~\ref{subsec:initial_model}). We vary our choices of these parameters and discuss their impact on our results in Section~\ref{subsubsec:pulsar_params}.

\subsection{Pulsar recycling through mass transfer}
\label{subsec:mass_transfer_case}

If the companion of the pulsar in the binary system is still evolving, for systems of sufficiently short orbital period there can be mass transfer onto the pulsar. Mass accretion changes the spin of the pulsar primarily due to exchange of angular momentum. The accumulation of mass also buries the magnetic field. Mass transfer may happen through two main channels:
\begin{enumerate}
    \item \textbf{ Roche Lobe Overflow (RLOF)} see Section~\ref{subsubsec:roche_lobe_overflow}
    \item \textbf{Common Envelope (CE) evolution} see Section~\ref{subsubsec:common_envelope}
\end{enumerate}
Pulsars can also undergo wind accretion through stellar winds \citep[e.g.][]{Stella:1985, Li:1995}, but we have not modelled this. 

\subsubsection{Roche Lobe Overflow}
\label{subsubsec:roche_lobe_overflow}

During stellar evolution in a binary system, a star can expand and fill its Roche Lobe. Any further expansion results in matter overflowing through the inner Lagrangian point (where the gravitational potential of the two members of the binary star system balance each other out) to the other star of the binary system. The infalling matter forms an accretion disk \citep{2015ebss.book..179I} before reaching the surface of the companion. In the case of accretion onto a NS, the NS is always the accretor and the companion star is the donor. We use the approximation to the Roche Lobe radius from \citet{Eggleton:1983}
\begin{equation}
   \frac{R_\mathrm{L}}{a} = \frac{0.49 q^\frac{2}{3}}{0.6q^\frac{2}{3} + \ln(1+q^\frac{1}{3})} , 
   \label{RocheLobeRadiusEgg}
\end{equation}
where $R_\mathrm{L}$ is the radius of a representative sphere of volume equal to that of the Roche Lobe of the companion to the NS (with mass $M_\mathrm{comp}$), $a$ is the orbital separation of the system, and $q = M_\mathrm{comp} / M_\mathrm{NS}$ is the mass ratio of the two stars. 

RLOF onto a pulsar results in an exchange of angular momentum. This in turn will affect the spin of the pulsar. As expected from the dynamics, the change in the pulsar's rotational velocity is dependent on the initial direction and magnitude of angular momentum of the pulsar and the accretion disk surrounding it. Thus, mass transfer may either spin up or spin down a pulsar.

We follow the modelling and prescription given by \citet{1994MNRAS.269..455J}, also used by \citet{Kiel:2008xw}, for calculating the change in angular momentum of the pulsar due to infalling matter from the companion star. The rate of change in angular momentum ($\dot{J}_\mathrm{acc}$) is given by
\begin{equation}
   \dot{J}_\mathrm{acc} = \epsilon V_\mathrm{diff} R_A^2 \dot{M}_\mathrm{NS} ,
   \label{AngularMomentumAccretion}
\end{equation}
and
\begin{equation} 
    V_\mathrm{diff} = \Omega_K|_{R_A}  -  \Omega_\mathrm{co} ,
    \label{VdiffAccretion}
\end{equation}
where $\epsilon$ is the efficiency factor (we consider $\epsilon$=1.0 for all models), $\dot{M}_\mathrm{NS}$ is the mass accretion rate onto the pulsar and $V_\mathrm{diff}$ is the difference between Keplerian angular velocity at the magnetic radius $\Omega_K|_{R_A}$ and the co-rotation angular velocity $\Omega_\mathrm{co}$. We assume that the magnetic radius $R_\mathrm{A}=R_\mathrm{Alfven}/2$ as in \citet{Kiel:2008xw}, where the Alfven radius---the radius at which the ram pressure of the fluid is balanced by magnetic pressure \citep{AlfvenRadius:Belenkaya2015}---is given by
\begin{equation}
   R_\mathrm{Alfven} = \left(\frac{2\pi^2}{G\mu_0^2}\right)^\frac{1}{7}\times\left(\frac{R^6}{\dot{M}_\mathrm{NS}M_\mathrm{NS}^\frac{1}{2}}\right)^\frac{2}{7}\times B^\frac{4}{7} .
   \label{AlfvenRadius}
\end{equation}
The components $\Omega_K|_{R_A}$ and $\Omega_\mathrm{co}$ are calculated by COMPAS and are system-specific. The resultant final spin after each timestep for the mass accretion case is given by 
\begin{equation}
   \Omega_\mathrm{i+1} = \Omega_\mathrm{i} + \frac{\Delta J_\mathrm{acc}}{I} ,
   \label{FinalSpinAccretion}
\end{equation}
where $\Delta J_\mathrm{acc}$ is the change in angular momentum owing to accretion within that timestep.

Millisecond pulsars are observed to have period derivatives (and therefore surface magnetic fields) much lower than other pulsars. The evolutionary path of a millisecond pulsar involves a pulsar being spun up by accreting matter from its companion. The infalling material from the companion star onto the pulsar buries the pulsar magnetic field. There have been several proposed explanations of this quenching of the pulsar's surface magnetic field by mass accretion. As discussed by \cite{ZhangKojima:2006} in the concept of a bottom field, the accreted matter might create a bulge at the equatorial region, thus disrupting the spherical symmetry of the idealized case. In turn, the magnetic lines of force change and are buried in the equatorial region due to the magnetic conductivity of the pulsar. Since radio observations only observe the magnetic lines of force from the magnetic poles, which decrease in magnitude at the cost of the equatorial burial, the older pulsars in a binary usually have a lower surface magnetic field. 

We assume that the surface magnetic field magnitude $B$ decays exponentially with accreted mass $\Delta M_\mathrm{NS}$ \citep{2011MNRAS.413..461O} as
\begin{equation}
   B = (B_0 - B_\mathrm{min})\times\exp(-\Delta M_\mathrm{NS}/\Delta M_d)+B_\mathrm{min} , 
   \label{MagneticFieldAccretion}
\end{equation}
where $\Delta M_d$ is the magnetic field mass decay scale, a free parameter in our model and $\Delta M_\mathrm{NS}$ is the total amount of accreted mass by the NS. In COMPAS, the mass transfer rate $\dot{M}$ for case BB mass transfer is calculated as
\begin{equation}
   \dot{M} = \frac{M_\mathrm{env}}{\tau_\mathrm{KH}} ,
   \label{eq:mdot}
\end{equation}
where $M_\mathrm{env}$ is the mass of the envelope and $\tau_\mathrm{KH}$ is the Kelvin-Helmholtz timescale of the donor star. For the systems of our interest, $M_\mathrm{env}\approx$few M$_{\odot}$ and $\tau_\mathrm{KH}\approx10^4$\,yr, thus giving $\dot{M}\approx10^{-4}$\,M$_\mathrm{\odot}$\,yr$^{-1}$.

Mass transfer onto a compact object is limited to the Eddington rate. The Eddington luminosity is
\begin{equation}
    L_\mathrm{E} = \frac{4\pi GcM_\mathrm{NS}m_\mathrm{p}}{\sigma_\mathrm{T}} ,
    \label{eq:eddington_luminosity}
\end{equation}
where $M_\mathrm{NS}$ is the mass of the accreting star (in this case the NS), $m_\mathrm{p}$ is the proton mass and $\sigma_\mathrm{T}$ is the Thomson Scattering cross-section of an electron. If the entire accretion energy is converted to luminosity, the luminosity can be expressed as 
\begin{equation}
    L_\mathrm{acc} = \frac{GM_\mathrm{NS}\dot{M}_\mathrm{NS}}{R} ,
    \label{eq:}
\end{equation}
where $\dot{M}_\mathrm{NS}$ is the mass accretion rate of the NS and $R$ is the radius of the NS. Equating $L_\mathrm{acc}$ to $L_\mathrm{E}$ for the fully efficient energy conversion, we obtain the mass accretion rate $\dot{M}$ for the Eddington mass accretion case
\begin{equation}
   \dot{M_\mathrm{E}} = \frac{4\pi cm_\mathrm{p}R}{\sigma_\mathrm{T}} \approx 1.4 \times 10^{-8} \, \mathrm{M}_\odot \, \mathrm{yr}^{-1} \, .
   \label{eq:eddington_accretion_rate}
\end{equation}
Only a small fraction of mass lost by the donor is actually accreted by the pulsar, and a significant portion of the matter is lost from the system. We consider the accretion rate onto a pulsar to be limited to $\dot{M_\mathrm{E}}$. It is also assumed that the mass lost from the system carries away the specific orbital angular momentum of the accretor \citep{Vigna-Gomez:2018dza}. $\dot{M}_\mathrm{NS}$ is multiplied by the time duration $\Delta t\approx\tau_\mathrm{KH}$ of the mass transfer to obtain $\Delta M_\mathrm{NS}$ in COMPAS. 

Almost all cases of the recycling of pulsars result in $\dot{M} = \dot{M_\mathrm{E}}$. The Eddington limit is calculated assuming spherically symmetric accretion onto a star. However, accretion through RLOF occurs via the formation of an accretion disk. Thus for the latter case, there remains the possibility of $\dot{M} > \dot{M_\mathrm{E}}$ \citep[see e.g.][for discussion]{Tauris:2017omb}.

\subsubsection{Common Envelope Evolution}
\label{subsubsec:common_envelope}

RLOF can become a runaway process if the mass-transfer rate is high enough and/or if after the initial mass transfer and subsequent orbital shrinkage the donor star continues to expand. The expansion results in further mass transfer and further reduction of the orbital separation of the binary. This leads to  unstable mass transfer on a short dynamical timescale over which the companion star is unable to accrete all the matter. The process results in the engulfing of the binary companion in an envelope of gaseous matter, such that both the stars (the companion star and the core of the donor) are inside a common envelope \citep[CE,][]{1988ApJ...329..764L, Ivanova2013}. 

The common envelope phase is important for massive compact binaries that are progenitors of gravitational waves \citep[see e.g.][for a recent review]{2018arXiv180605820M}. During the CE phase, the stars in the binary experience a drag force (fluid resistance due to motion of a body through it) from the surrounding gas that makes up the envelope. As part of the process orbital energy is transferred to the envelope \citep{1976IAUS...73...75P}. This decreases the orbital separation between the objects. The phase ends with either the ejection of the common envelope or by the merger of the two systems still inside the envelope. In case of the ejection of CE, the orbital separation of the binary maybe sufficiently reduced to allow for a subsequent gravitational-wave driven merger \citep{Ivanova2013} to occur in a Hubble time \citep{1993PASP..105.1373I}. The duration of the common envelope phase is uncertain, but it is much shorter than the total stellar evolution timescale; it is therefore assumed to be instantaneous in COMPAS. 

Using a parametrized formalism, the binding energy of the envelope $E_\mathrm{bind}$ can be expressed as \citep{Webbink:1984, deKool:1990}
\begin{equation}
   E_\mathrm{bind} = \alpha_\mathrm{CE}(\frac{GM_\mathrm{comp,c}M_\mathrm{NS}}{2a_\mathrm{f}} - \frac{GM_\mathrm{comp}M_\mathrm{NS}}{2a_\mathrm{i}}) \,
   = \frac{GM_\mathrm{comp}M_\mathrm{env}}{\lambda a_\mathrm{i}r_\mathrm{L}} \, ,
\end{equation}
where $\alpha_\mathrm{CE}$ is the efficiency denoting the fraction of the orbital energy of the companion star that may be used to eject the CE, the envelope binding energy is parameterised by $\lambda$, $M_\mathrm{comp}$ is the mass of the companion (donor), $M_\mathrm{comp,c}$ is the donor's core mass and $M_\mathrm{env}$ is its envelope mass, $M_\mathrm{NS}$ is the mass of the NS (pulsar), $a_\mathrm{i}$ is the initial orbital separation, $a_\mathrm{f}$ is the final orbital separation of the two stars and $r_\mathrm{L} = R_\mathrm{L}/a_\mathrm{i},$ $R_\mathrm{L}$ is the Roche Lobe radius.  

Due to the presence of the parameters $\alpha$ and $\lambda$ the formulation is often referred to as the `$\alpha-\lambda$' parametrization. We use fitting formulae from \cite{XuLi:2010}\footnote{we use their $\lambda_\mathrm{b}$ values} to determine the value of $\lambda$ as in \citet{Vigna-Gomez:2018dza} and \citet{HowittLRN}.  We assume $\alpha=1$ for all our models (the impact of varying $\alpha$ has been examined in \citealp{Vigna-Gomez:2018dza}).

There is a limit to how much mass a NS can accrete during a common envelope event. It has been previously suggested that a NS can accumulate enough mass during the CE phase for it to collapse to a black hole \citep{Chevalier1993, Armitage:1999qp, Bethe:2005ju}. \cite{MacLeod:2014yda} have shown that---when considering the density gradient of the accretion disk---the amount of mass a neutron star can accrete is constrained to $\lesssim 0.1$\,M$_\odot$. Thus \cite{MacLeod:2014yda} argue that most NSs would survive the CE phase. We assume that during the CE mass transfer, the spin of the NS and the surface magnetic field both get affected by infalling matter, as given by Equations~\ref{AngularMomentumAccretion} and \ref{MagneticFieldAccretion}. 

We have incorporated the effect of mass accretion onto a NS during the common envelope phase in COMPAS. Since the amount of mass accreted during a common envelope is uncertain, we test the following variations:
\begin{enumerate}
    \item \textbf{Zero} No mass accretion during common envelope evolution. This was the previous default model, as in \citet{Vigna-Gomez:2018dza}.
    \item \textbf{Uniform} The amount of mass accreted during common envelope is drawn from a uniform distribution between $M_\mathrm{acc}^\mathrm{min}$ and $M_\mathrm{acc}^\mathrm{max}$ \citep[similar to][]{2011MNRAS.413..461O}.
    \item \textbf{MacLeod} We use a prescription based on \cite{MacLeod:2014yda} which gives the mass accreted as a function of donor mass and radius (see Equation~\ref{eq:macleod_ce_deltaMd} below).
\end{enumerate}

For all models we assume $M_\mathrm{acc}^\mathrm{min} = 0.04$\,M$_\odot$ and $M_\mathrm{acc}^\mathrm{max} = 0.1$\,M$_\odot$ \citep{MacLeod:2014yda}.

We approximate the amount of mass accreted by a NS $\Delta M_\mathrm{NS}$ during a CE as a function of the companion mass $M_\mathrm{comp}$ and radius $R_\mathrm{comp}$ using a fit to Figure~4 in \cite{MacLeod:2014yda} as
\begin{equation}
    \Delta M_\mathrm{NS} / M_\odot = a (R_\mathrm{comp} / R_\odot) + b ,  \quad M_\mathrm{acc}^\mathrm{min} < \Delta M_\mathrm{NS} / \mathrm{M}_\odot < M_\mathrm{acc}^\mathrm{max}
    \label{eq:macleod_ce_deltaMd}
\end{equation}
where
\begin{equation}
    a = a_a (M_\mathrm{comp} / M_\odot) + b_a ,
    \label{eq:macleod_ce_m}
\end{equation}
and
\begin{equation}
    b = a_b (M_\mathrm{comp} / M_\odot) + b_b ,
    \label{eq:macleod_ce_c}
\end{equation}
with $a_a = -1.1 \times 10^{-5}$, $a_b = 1.5 \times 10^{-2}$, $b_a = 1.2 \times 10^{-4}$ and $b_b = -1.5 \times 10^{-1}$. 

\subsection{Pulsar Death}
\label{subsec:pulsar_death}

Rotation powered pulsars stop emitting in the radio band once they cross a `death line' in the $P\dot{P}$ diagram (the plot consists of logarithmic axes of spin $P$ and spin down rate $\dot{P}$). This is because the spin of the old pulsar is decelerated to a point where the magnetic field is not sufficiently strong to produce electron-positron pairs required for radio emission \citep{Chen:1993, RudakRitter1994, Medin:2010}.

For the $P \dot{P}$ diagram relevant for DNSs, we use the death lines given by \citet{RudakRitter1994}
\begin{equation}
    1) \log_{10}\Dot{P} = 3.29\times\log_{10}P - 16.55 ,
    \label{eq:deathline1}
\end{equation}
and
\begin{equation}
    2) \log_{10}\Dot{P} = 0.92\times\log_{10}P - 18.65 .
    \label{eq:deathline2}
\end{equation}
However, there are radio pulsars observed beyond the empirical death lines \citep{Young1999}. In addition, rejecting pulsars in our simulations once they cross the death lines leads to a pile up at the death line boundary, which is not observed \citep{Szary:2014dia}. We hence use a `hybrid' approach in determining radio pulsar death, described as follows. 

The model of \citet{Szary:2014dia} describes pulsars ceasing to emit in the radio once their radio efficiency $\xi$ crosses some threshold. The radio efficiency $\xi$ is the ratio of the pulsar radio luminosity $L$ (see Section~\ref{subsec:radio_selection_effects}) and the pulsar spin down power $\dot{E}$
\begin{equation}
    \xi \equiv \frac{L}{\dot{E}} ,
    \label{eq:pulsar_efficiency}
\end{equation}
where
\begin{equation}
    \dot{E} = 4\pi^2I\dot{P}P^{-3}  ,
    \label{eq:pulsar_spin_down_power}
\end{equation}  
and $I$ is the moment of inertia of the pulsar. The threshold radio efficiency is a free parameter in our model, and is assumed to be $\xi_\mathrm{max} = 0.01$ \citep{Szary:2014dia}. Pulsars with $\xi > \xi_\mathrm{max}$ are assumed to cease emitting in the radio. In our simulations we assume that DNSs that either cross the second death-line given by Equation~\ref{eq:deathline2}, or have $\xi \geq \xi_\mathrm{max}$, have stopped emitting in radio. However, for gravitational wave analysis (Section~\ref{sec:grav_waves}), we include all the NSs, including those that cross the death line and exceed the radio efficiency limit. 

\subsection{Life of a DNS}
\label{subsec:life_of_DNS}

\begin{figure*}
\includegraphics[width=0.47\textwidth]{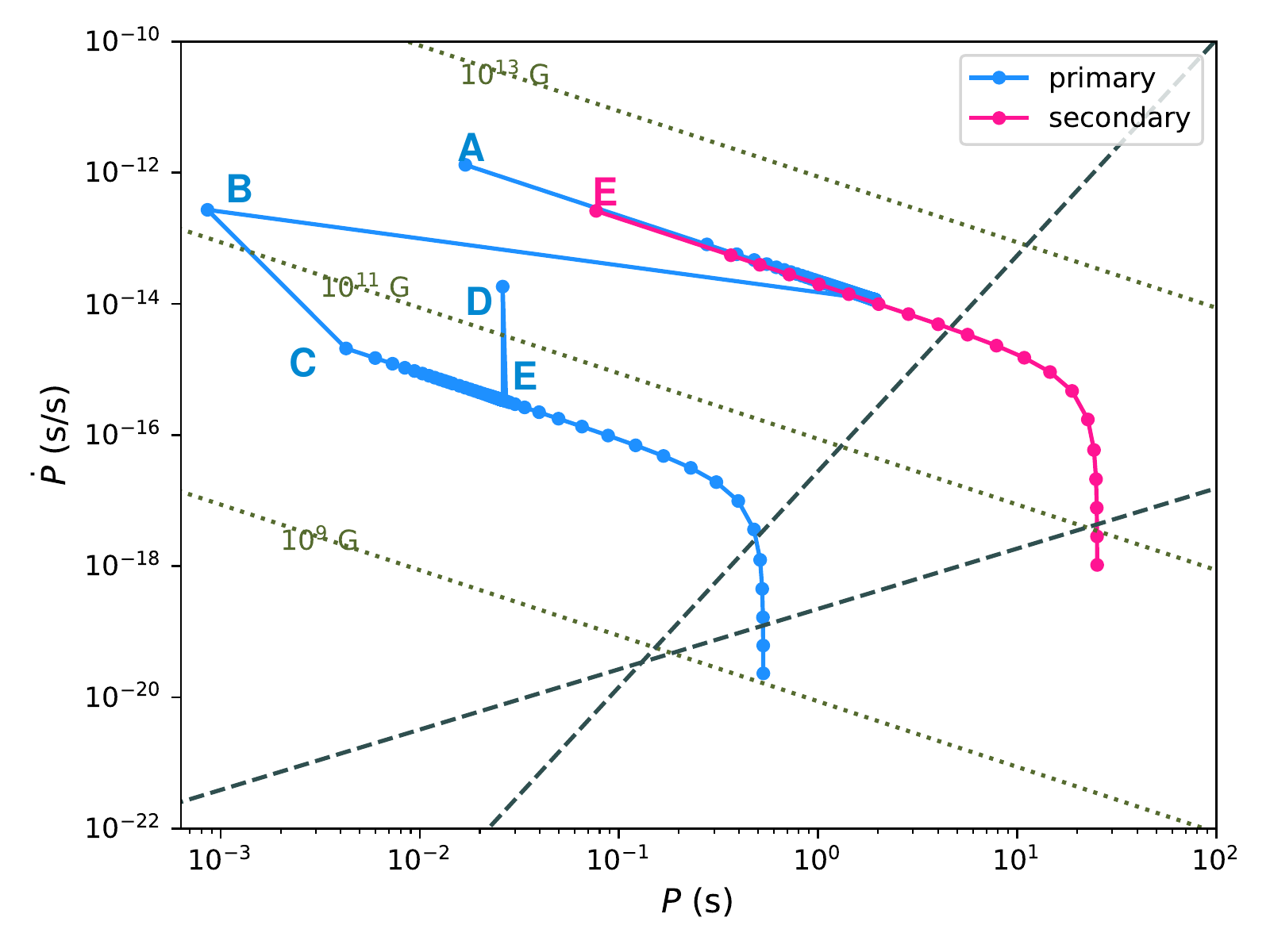}
\includegraphics[width=0.5\textwidth]{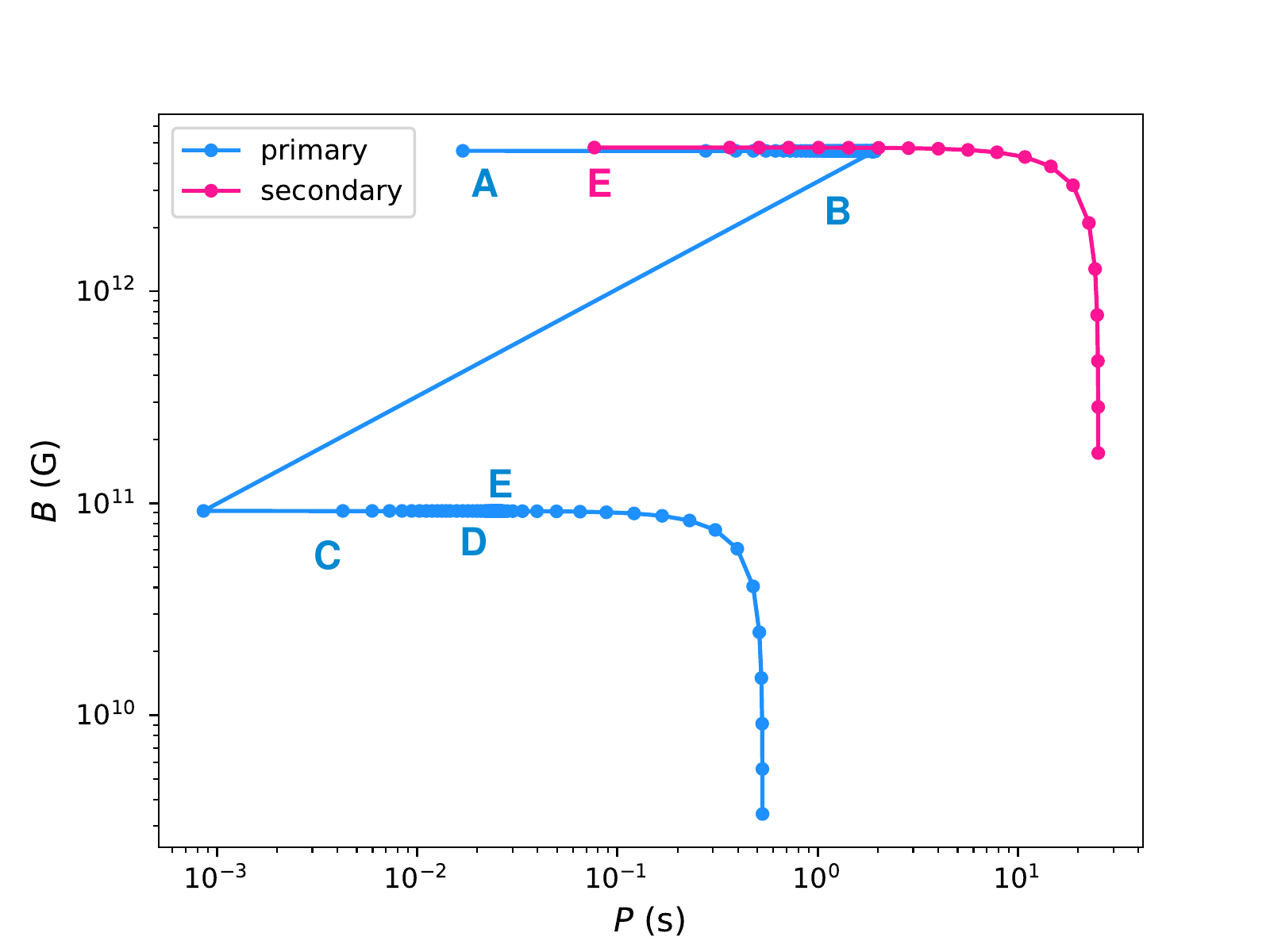}
\caption{The left figure shows the $P\dot{P}$ diagram for a DNS system. The blue line traces the time evolution of the first born neutron star (primary). The primary is born with a large magnetic field (A) and spins down quickly along a line of roughly constant magnetic field strength (the diagonal, olive green dotted lines are lines of constant magnetic field strength, calculated from Equation~\ref{MagneticFieldIsolated}, hence only valid during non-mass-transfer isolated pulsar evolution). The companion star fills its Roche Lobe and the binary undergoes common envelope evolution, leaving behind a pulsar-helium star binary. Accretion onto the primary pulsar during common envelope evolution buries its magnetic field and spins it up to a short spin period (B), after which the primary pulsar continues to spin down as an isolated pulsar from C. A second episode of mass transfer occurs (case BB) when the helium star fills its Roche Lobe, further spinning up the primary (D). Finally, the helium star explodes in a supernova, leaving behind a non-recycled pulsar (the secondary) at E. The secondary follows the pink line in the $P\dot{P}$ parameter space. The black dashed lines are the two death lines discussed in Section~\ref{subsec:pulsar_death}. The right figure shows the same binary in a $PB$ plot.
}
\label{PPdotEvolution}
\end{figure*} 

The typical evolution of a pulsar can be understood by tracking its movement in a $P\dot{P}$ diagram. Radio observations can only detect the pulsars at a particular snapshot in time (i.e. the current observation time) and cannot trace the entire history or future evolution through the parameter space (since the evolutionary timescale is $\mathcal{O}(\mathrm{Myrs})$). It is informative to trace the movement of a modelled pulsar through the $P\dot{P}$ diagram to illustrate our model across the stages of a pulsar's life. We show the $P\dot{P}$ diagram for two pulsars in a DNS binary in Fig.~\ref{PPdotEvolution}. It is a time integrated plot, capturing the complete time evolution of the individual pulsars. The angular frequency $\Omega$ and magnetic field $B$ of the primary (the first born NS) decay exponentially following Equations~\ref{SpinDownIsolated} and \ref{MagneticFieldIsolated}. If there is mass transfer from the companion (see Section~\ref{subsec:mass_transfer_case}), the movement of the primary in the phase space becomes discontinuous, as it now decays according to Equation~\ref{MagneticFieldAccretion}. In reference to Fig.~\ref{PPdotEvolution}, the primary pulsar is born at point `A'. As it spins down, there is CE mass transfer from its core-He-burning companion onto the pulsar at `B' which is notable by the discontinuity. The mass accretion is significant enough to create a considerable change in the surface magnetic field of the pulsar. The CE is ejected and the companion becomes a naked He-star. The primary pulsar then continues to spin down (from `C'). At `D', there is a second RLOF mass transfer phase from the companion which has become a He-giant. However, the mass accreted by the pulsar is less compared to the previous case, and hence the change in the magnetic field is smaller. At `E', the companion undergoes a SN and creates a newly formed pulsar, the `secondary', while the primary pulsar continues to spin down. 

The same evolutionary progress of these example pulsars can be visualized in $P-B$ (spin-magnetic field) space as well. The recycling of the primary causes an abrupt change in the magnetic field and spin due to mass transfer. The secondary pulsar, however has its surface magnetic field decay exponentially. 

Mass transfer may even recycle a pulsar that has already crossed the death line(s), reducing its surface magnetic field strength and spin period below that of non-recycled pulsars. The secondary (the second born NS) does not have any chance of mass transfer (the other star is already a NS) and hence evolves only following Equations~\ref{SpinDownIsolated} and \ref{MagneticFieldIsolated} and evolves through the $P\dot{P}$ space without any discontinuity. 

\subsection{Method of selecting a particular snapshot in the lifetime of a pulsar} 
\label{subsec:snapshot}

\begin{figure} 
\includegraphics[width=0.84\columnwidth]{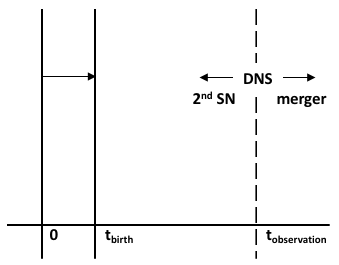}
\caption{Timeline of Pulsar-Neutron Star Binaries. The `0' signifies the modelling initiation time in COMPAS, when the Zero Age Main Sequence (ZAMS) star starts its life. $t_\mathrm{birth}$ is the birth time drawn from an uniform distribution for each system, to which we displace the origin of the evolution. $t_\mathrm{observation}$ denotes the current age of the Milky Way (taken to be $13\,$Gyr in our model) when we observe the systems. Only the systems existing as double neutron stars at $t_\mathrm{observation}$ are selected from the entire modelled population and analysed further.}
\label{fig:Double Neutron Star TimeLine}
\end{figure}

While computing the evolution of a particular binary in COMPAS, its birth time is set to be the origin of the time axis. Thus each and every binary is born at time zero, and evolves accordingly. However, when we actually compare our model to observations, we need to adjust the birth times, and select a point in the binary's lifetime in accordance to the current age of Milky Way. We assume a uniform star formation history for the Milky Way \citep{Vigna-Gomez:2018dza}, and hence randomly generate a birth time for each binary. We then select an observation time of ~13\,Gyr, approximate to the current age of the Milky Way. We only select the systems that exist as DNSs at that time. In order to obtain the exact values of the pulsar parameters at the selected time, we use linear interpolation in-between the two points of the relevant parameter that are closest and encompasses the observation time. By drawing many different birth times, we re-use different evolutionary phases of the same binary. This gives a statistically robust ensemble without being computationally expensive through running more systems. Figure~\ref{fig:Double Neutron Star TimeLine} shows the schematic diagram of the process.

\subsection{Galactic Potential}
\label{subsec:GalacticPotential}

After assigning each binary a random birth time, correcting for the new origin of the time frame, and calculating the relevant parameters at the observation time using linear interpolation, we put the binaries in a Galactic potential. We compute the orbits of the binaries in a Milky Way-like potential, accounting for the re-distribution (in position and velocity) due to the second SN until the selected observation time.

\subsubsection{Velocity Kicks}
\label{subsubsec:velocity_kicks}

NSs receive a `natal kick' when they form through a SN event \citep{Helfand:1977, Lyne:1994}. Although there are uncertainties associated with the kick velocity, and its correlation to the SN mechanism as well as the resultant birth properties of the pulsar \citep{Bailes:1989}, it has generally been accepted that the magnitude ranges from 90--500\,km s$^{-1}$, though can be as high as $\approx$~$ 1000$\,km s$^{-1}$ \citep{Arzoumanian:2002}. The velocity kicks are also expected to be dependant on the type of SN process that creates the NS. Electron Capture (EC) SN \citep{Nomoto:1984, Nomoto:1987} and Ultra-Stripped (US) SN \citep{Tauris:2013, Tauris:2015} are thought to generate a lower velocity kick than Core Collapse (CC) SN \citep{Fryer:2012}. For more details on how COMPAS models the individual SN events we refer to \cite{Vigna-Gomez:2018dza}. For the NSs at birth, we assume a 1-dimensional Maxwellian SN kick velocity distribution \citep{Hansen:1997zw} with root mean square of the velocity $\sigma_\mathrm{high} = 265$\,km s$^{-1}$ for CCSN \citep{Hobbs:2005} and $\sigma_\mathrm{low} = 30 $\,km s$^{-1}$ for both ECSN and USSN respectively \citep{Pfahl:2002, Podsiadlowski:2004}. The birth kicks are assumed to be isotropic in the reference frame of the star that is undergoing the SN event \citep{Vigna-Gomez:2018dza}. Depending on the magnitude and direction of the kick, the binary may be disrupted and the pulsar may be ejected. Since this study focuses on DNSs, we remove such cases from our data-set. Besides from disrupting the binary, a velocity kick of sufficient magnitude may also be greater than the escape velocity of the host galaxy, and hence eject the DNS, leading to potential hostless short gamma-ray bursts \citep[e.g.][]{2002ApJ...571..394B,2003MNRAS.342.1169V,2010ApJ...725L..91K,2013ApJ...776...18F,2019arXiv191003598Z}. 

\subsubsection{NIGO}
\label{subsubec:NIGO}

We account for these SN kicks and their effect on the final distribution of the systems in the Galactic potential of the Milky Way using NIGO \citep{2015A&C....12...11R, RossiHurley:2015}. With NIGO, we select a three dimensional gravitational potential for the Milky Way comprised of three components; a bulge at the Galactic center, a disc and a halo.

We model the bulge as a Plummer sphere \citep{Plummer:1911, MiyamotoNagai:1975}:
\begin{equation}
   \Phi_\mathrm{b} = -\frac{GM_\mathrm{b}}{\sqrt{R^2 + z^2 + b_\mathrm{b}^2 }} \, ,
   \label{bulge}
\end{equation}
where $G$ is the universal constant of gravitation, $M_\mathrm{b}$ is the mass of the bulge and $b_\mathrm{b}$ is the scale length and $R^2 = x^2 + y^2$.

We use an exponential disc, formed from the superposition of three individual Miyamoto-Nagai \citep{MiyamotoNagai:1975} potentials (as implemented by \citealp{Flynn:1996ej}):
\begin{equation}
   \Phi_\mathrm{d} = -\sum_{n=1}^{3} \frac{GM_\mathrm{d_\mathrm{n}}}{\sqrt{R^2 + [a_\mathrm{d_\mathrm{n}} + (\sqrt{b_\mathrm{d}^2 + z^2)}]^2 }} \, ,
   \label{disc} 
\end{equation}
where $M_\mathrm{d_\mathrm{n}}$ are the masses of each disk, $a_\mathrm{d_\mathrm{n}}$ are related the disk scale lengths of the three disc components and $b_\mathrm{d}$ is related to the disc scale height.

Finally, we use an NFW dark matter halo \citep{Navarro:1996gj}:
\begin{equation}
    \Phi_\mathrm{h} = -\frac{GM_\mathrm{h}}{r}\ln\left({1+\frac{r}{a_\mathrm{h}}}\right) \, ,
    \label{halo}
\end{equation}
where \(M_\mathrm{h}\) is the mass of the halo, \(a_\mathrm{h}\) is the length scale, and $r=\sqrt{R^2 + z^2}$.

The values of the parameters in Equations~\ref{bulge} and ~\ref{halo} are taken from \citet{Irrgang:2013} (Model III), and those in Equation~\ref{disc} are from the thin disc description of \citet{SmithFlynn:2015}.

The potentials \(\Phi_\mathrm{b}\), \(\Phi_\mathrm{d}\) and \(\Phi_\mathrm{h}\) are expressed here in right-handed, Galacto-centric, Cartesian coordinates, and the total gravitational potential, \(\Phi_\mathrm{total}\) of the galaxy is given by \(\Phi_\mathrm{total} = \Phi_\mathrm{b} + \Phi_\mathrm{d} + \Phi_\mathrm{h}\).

COMPAS generates the kick velocity while evolving the binaries, according to the type of SN each star undergoes (see Section~\ref{subsubsec:velocity_kicks}). Only if the binary is not disrupted to two isolated bodies after the SN event does it stay in our data-set. We first distribute the evolved DNS systems from COMPAS following the density distribution yielded by the exponential disc. We then assign the velocity components following the circular rotation curve of the galaxy and add $10$\,kms$^{-1}$ of velocity dispersion. NIGO then takes as input the velocity and time of the second SN event generated by COMPAS, accounts for the magnitude and direction of this kick and re-distributes the position and velocity of the systems. The program then continues to evolve the systems in the new orbits up to the present time. 

\begin{figure*}
\includegraphics[width=0.45\textwidth]{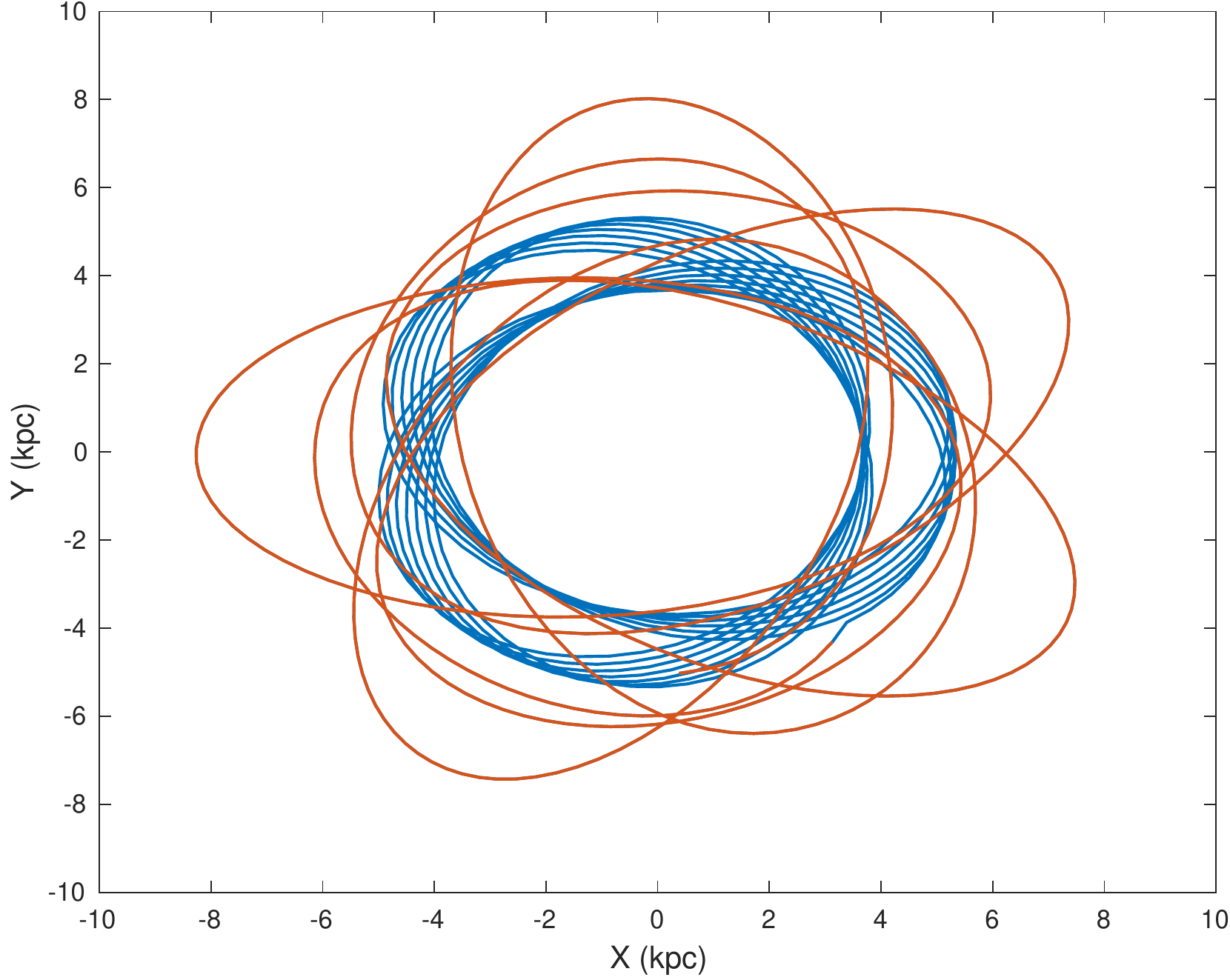}
\includegraphics[width=0.44\textwidth]{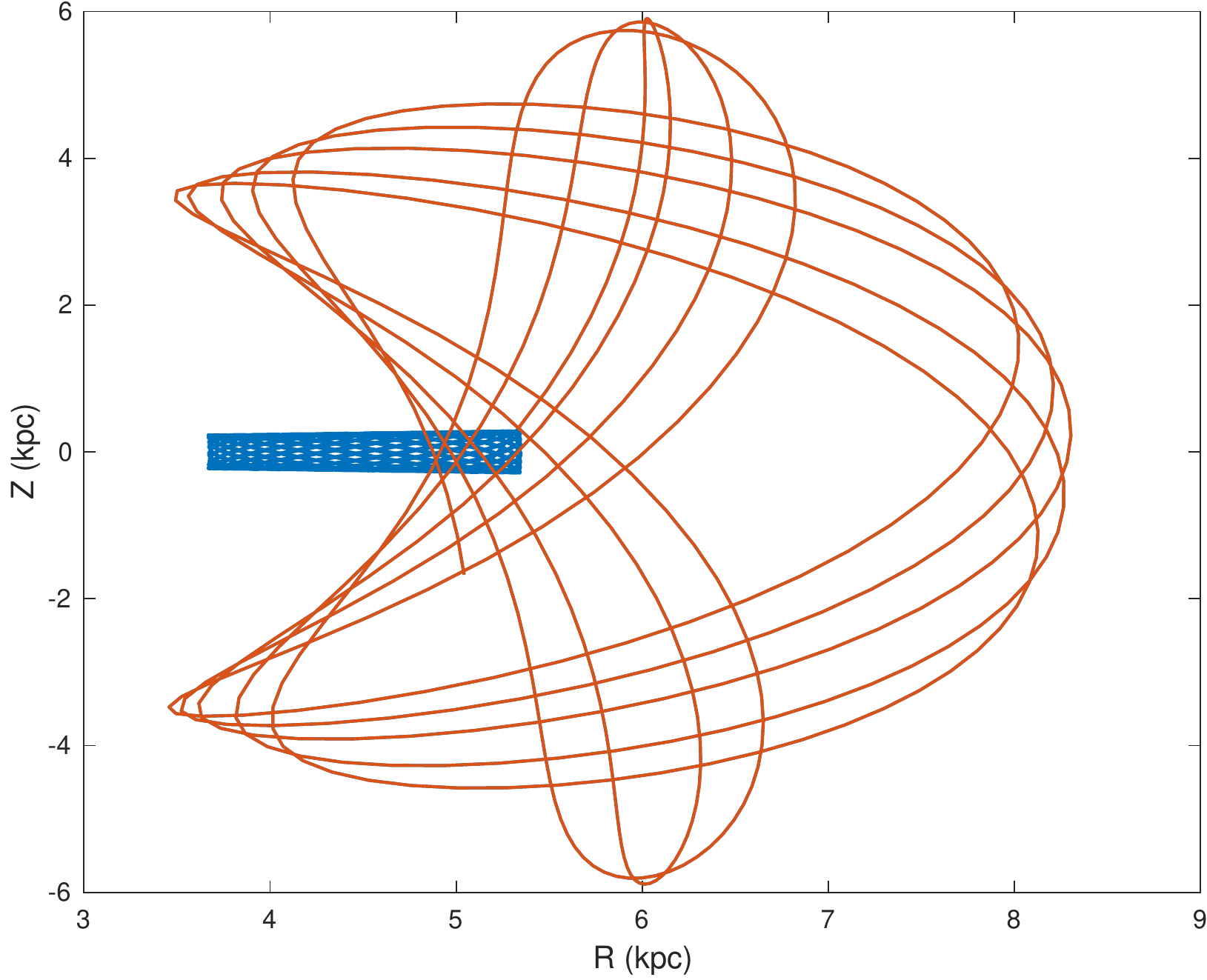}
\caption{The orbits of a binary system in the Galactic potential. The blue line traces the orbit in the pre-second-supernova stage, while the orange line traces the same after the formation of a double neutron star system, accounting for the velocity kicks. The left panel shows the $X$--$Y$ axes view in a Galactocentric Cartesian coordinate system. The right panel shows the same for the radial $R = \sqrt{x^2 + Y^2}$ vs. $Z$ components. We have selected a particular binary with the second supernova kick of $\approx$280 km/s for illustrative purposes, as this kick is high enough to enable the orbital change to be readily visualised yet smaller than the escape velocity of the Milky Way.
}
\label{fig:Orbits:NIGO}
\end{figure*}

There remains a very small probability of the first SN generating a strong enough velocity kick to eject the system out of the galaxy, yet oriented in such a way that the binary remains bound. Such kicks require very specific direction depending on individual system parameters and are not taken into account in our analysis. Figure~\ref{fig:Orbits:NIGO} shows the Galactic orbits for an example DNS system distinguishing  the orbit for the pre-second-SN stage and the orbit after the formation of the double compact object. Though the orbit changes, this particular binary remains bound and stays in the galaxy. 

\subsection{Radio Selection Effects: PSREvolve}
\label{subsec:radio_selection_effects}

To compare our modelled DNS systems with the catalogue of radio pulsars, observational selection bias by radio telescopes needs to be modelled. These selection effects include the dependence on sky location, interstellar radio scintillation, frequency dependence of scattering/ smearing of the pulses, orbital eccentricity and relativistic effect of pulsars in binaries, uncertainty in modelling luminosity and beam geometry. We use the code PSREvolve \citep{2011MNRAS.413..461O} to account for some of these selection effects namely, the dependence on sky location, the frequency dependence of scattering/smearing and the broadening of the beam.

The radiometer equation \citep{Dewey:1985, Lorimer:2004}
\begin{equation}
    S_\mathrm{min} = \beta\frac{(S/N_\mathrm{min})(T_\mathrm{rec} + T_\mathrm{sky})}{G\sqrt{n_\mathrm{p}t_\mathrm{int}\Delta f}}\sqrt{\frac{W_\mathrm{e}}{P - W_\mathrm{e}}} ,
    \label{Radiometer}
\end{equation}
gives the lower limit of flux $S_\mathrm{min}$ that a source must have in order to be detected for a given signal-to-noise ratio ($S/N_\mathrm{min}$). The parameter $\beta$ accounts for errors that increase the noise in the signal (digitisation errors, radio interference, band-pass distortion), $T_\mathrm{rec}$ and $T_\mathrm{sky}$ represents the receiver noise temperature and sky temperature in the direction of the particular pulsar respectively, $G$ is the gain of the telescope, $n_\mathrm{p}$ is the number of polarizations in the detector, $t_\mathrm{int}$ is the integration time, $\Delta f$ is the receiver bandwidth, $W_\mathrm{e}$ is pulse width and $P$ is the period of the pulsar. The sky temperature $T_\mathrm{sky}$ is determined by the location of the pulsar in the galaxy, and PSREvolve inputs the information calculated by NIGO, while the pulse period $P$ is computed by COMPAS. Assuming $\beta=1$ and $S/N_\mathrm{min} \geq 10$, we use the Parkes Multibeam Pulsar Survey \citep{Manchester:2001} specifications, to evaluate Equation~\ref{Radiometer}. Although not all DNSs were discovered by this survey, it remains one of the most successful pulsar surveys to date, and it is also meaningful to analyse our models using the specifications of one particular survey. The free electron distribution in the galaxy broadens the intrinsic pulse width ($W_\mathrm{i}$: \citet{Cordes:2002wz}), while the interstellar medium (ISM) scatters the pulsar beam. These effects, along with the sampling time of the survey produce an effective pulse width $W_\mathrm{e}$ expressed as \citep{Burgay:2003jj}
\begin{equation}
    W^2_\mathrm{e} = W^2_\mathrm{i} + \tau^2_\mathrm{samp} + \left(\tau_\mathrm{samp}\frac{DM}{DM_\mathrm{0}}\right)^2 + \tau^2_\mathrm{scatt} ,
    \label{eq:effective_width}
\end{equation}
where $\tau^2_\mathrm{samp}$ is the sampling time, $\tau^2_\mathrm{scatt}$ is the ISM scattering time, $DM$ is the dispersion measure in the direction of the pulsar and $DM_\mathrm{0}$ is the diagonal dispersion measure of the survey. PSREvolve uses a fit of $\tau_\mathrm{scatt}$ with respect to $DM$ from \citet{Bhat:2004}. The duty cycle for all pulsars are assumed to be $\frac{W_\mathrm{i}}{P} = 0.05$. Although the duty cycle varies widely across the pulsar population \citep{Lyne:1988}, for simplicity we adopt the stated fixed duty cycle for all pulsars presented in this paper.

\subsubsection{Beaming fraction}
\label{subsubsec:beaming_fraction}

The beam of radio emission from a pulsar has a finite width, and sweeps out a finite area on the sky, so that not all pulsars beam towards the Earth. The fraction of the sky a pulsar sweeps out is known as the beaming fraction $f_\mathrm{beaming}$. We model the beaming fraction as 
\begin{equation}
    f_\mathrm{beaming} = 0.09\left(\log\frac{P}{10}\right)^2 + 0.03 \, , \quad 0 \leq  f_\mathrm{beaming} \leq 1
    \label{fBeaming}
\end{equation}
according to \citet{Tauris:1998}, where $P$ is the spin period of the pulsar in seconds. We calculate $f_\mathrm{beaming}$ for the individual pulsars and use the numerical value as a weight. If $f_\mathrm{beaming} = 1$, it means that the beam is very broad and hence the pulsar is surely detectable. A very narrow beam will have a beaming fraction $f_\mathrm{beaming} < 1$, and thus is less probable to be detected. For our model analysis we use this weighted approach. For visualising the $P\dot{P}$ scatter plots (such as Figure~\ref{PPdotDefault}), we use probabilistic rejection sampling to generate the scatter points from the dataset. 

\subsubsection{Pulsar luminosity}
\label{subsubsec:pulsar_luminosity}

To compute if a pulsar is radio detectable, we also check its radio efficiency $\xi < \xi_\mathrm{max}$ as described in Section~\ref{subsec:pulsar_death}. In order to do this, we must model the luminosity of the pulsar. However, modelling pulsar radio luminosity is one of the most uncertain domains in accounting for radio selection effects. 

\citet{Szary:2014dia} find no correlation between the pulsar parameters ($P$, $\dot{P}$) and the observed radio luminosity distribution. Additionally the observed radio luminosity distribution can be biased by additional radio selection effects. We therefore use the log-normal luminosity distribution 
\begin{equation}
    \log L_{1400} \sim N(0.5, 1.0) \quad -3.0 \leq \log{L_\mathrm{1400}} \leq 4.0
\end{equation}
from \citet{Szary:2014dia} to determine pulsar luminosities, where $L_\mathrm{1400}$ is the radio luminosity at 1400\,MHz. The upper and lower limits of $L_\mathrm{1400}$ are obtained from rounding up the observed maximum and minimum radio luminosity of pulsars.

After calculating the limiting flux $S_\mathrm{min}$ (Equation~\ref{Radiometer}), we compute the pulsar flux from the luminosity ($L$) of the modelled pulsars
\begin{equation}
    F = \frac{L}{4\pi D^2} ,
    \label{FluxLum}
\end{equation}
where $D$ is the distance of the pulsar from the solar-system barycentre. If $F \geq S_\mathrm{min}$, we consider the pulsar to be detected. 

\section{Radio Population}
\label{sec:RadioPopulation} 

\begin{table*}
    \scriptsize
    \centering
    \caption{Description of models used in this paper. Each model varies one parameter from the value assumed in the Initial model.}
    \label{tab:TableModelSpecifications}
    \begin{tabular}{lccccccr}
       \hline
       Model & $B_\mathrm{birth}$  Range (G) & $B_\mathrm{birth}$  Distribution & $P_\mathrm{birth}$  Range (ms)& $P_\mathrm{birth}$  Distribution & $\tau_d$ (Myrs)& $\Delta M_d$ ($\mathrm{M_\odot}$) & CE Accretion \\
       \hline
        Initial & $(10^{10} - 10^{13})$ & Uniform & $(10-100)$&  Uniform  & 1000 & 0.025  & MacLeod \\
        \\
        BMF-R   & $(10^{11} - 10^{13})$ & Uniform & $(10-100)$&  Uniform  & 1000 & 0.025  & MacLeod \\
        BMF-FL  & $(10^{10} - 10^{13})$ & Flat in Log & $(10-100)$& Uniform & 1000 & 0.025 & MacLeod \\
        BMF-FGK06 & - & FGK06 & (10-100) & Uniform & 1000 & 0.025 & MacLeod \\
        \\
        BS-R    & $(10^{10} - 10^{13})$ & Uniform & $(10-1000)$& Uniform & 1000 & 0.025 & MacLeod \\
        \\
        FDT-10 & $(10^{10} - 10^{13})$ & Uniform & $(10-100)$&  Uniform  & 10 & 0.025 & MacLeod \\
        FDT-100 & $(10^{10} - 10^{13})$ & Uniform & $(10-100)$&  Uniform  & 100 & 0.025 & MacLeod \\
        FDT-500 & $(10^{10} - 10^{13})$ & Uniform & $(10-100)$&  Uniform  & 500 & 0.025 & MacLeod \\
        FDT-2000 & $(10^{10} - 10^{13})$ & Uniform & $(10-100)$&  Uniform  & 2000 & 0.025 & MacLeod \\
        \\
        CE-Z & $(10^{10} - 10^{13})$ & Uniform & $(10-100)$&  Uniform  & 1000 & 0.025 & Zero \\
        CE-U & $(10^{10} - 10^{13})$ & Uniform & $(10-100)$&  Uniform  & 1000 & 0.025 & Uniform \\
        \\
        FDM-10 & $(10^{10} - 10^{13})$ & Uniform & $(10-100)$&  Uniform  & 1000 & 0.010 & MacLeod \\
        FDM-15 & $(10^{10} - 10^{13})$ & Uniform & $(10-100)$&  Uniform  & 1000 & 0.015 & MacLeod \\
        FDM-20 & $(10^{10} - 10^{13})$ & Uniform & $(10-100)$&  Uniform  & 1000 & 0.020 & MacLeod \\
        FDM-50 & $(10^{10} - 10^{13})$ & Uniform & $(10-100)$&  Uniform  & 1000 & 0.050 & MacLeod \\
       \hline
	\end{tabular}
\end{table*}

In this section we present a detailed description of the suite of models we have used to explore the radio pulsar parameter space. We simulated 15 models (see Table~\ref{tab:TableModelSpecifications}), each with $10^6$ binaries, and re-used the DNS population $10^3$ times for improved statistics with lowered computational cost. Thus we effectively have a population of $10^9$ binaries for each model. Every model is generated through COMPAS, then evolved in a Galactic potential with NIGO and then analysed by PSREvolve to produce a `survey-observed' population. The resultant population is compared to the catalogued Milky-Way DNS systems. 

Our base model is named ``Initial". For all following models we varied in each only one parameter from the Initial model. This allows us to explore the difference in the resultant population systematically, qualitatively and quantitatively.

For each binary, we draw the zero age main sequence (ZAMS) mass of the initially more massive star from the initial mass function (IMF) of \citet{Kroupa:2000iv}, within the mass range of $4$--$50$\,M$_\odot$. We select this mass range as we focus on binary stars that might form DNS systems. The ZAMS mass of the companion star in the binary is assigned according to a uniform mass ratio distribution \citep{Sana:2012}. The initial separations of the binary systems are assigned from a flat-log distribution \citep{Sana:2012} in the range  $-1.0\leq\log_\mathrm{10}(a/AU)\leq 3.0$. All the binaries are assumed to be born in a circular orbit, thus the initial eccentricities are zero. The metallicity is kept constant across all models to solar metallicity, $ Z = 0.0142 $ \citep{Asplund:2009}, which is a justified assumption since we focus on Milky Way field DNS systems. Additionally \citet{Neijssel2019MNRAS} showed that the DNS formation rate in COMPAS models is unlikely to be strongly affected by the metallicity distribution.

We have standardized a nomenclature for our models. Apart from Initial, all models have a prefix, an acronym for the parameter that we change in the model relative to Initial, and a suffix denoting the actual value/distribution/prescription of the parameter that we change it to. For example, BS-R means that the Birth Spin (BS) distribution of the mentioned model has a different Range (R), with the rest of the model parameters being the same as Initial. More details are given in the subsequent paragraphs where we discuss the models and the inferences we draw from them. A list of the models is given in Table~\ref{tab:TableModelSpecifications}, along with details of the variables for each model.

To statistically compare each model with the observed radio DNSs, we use a one-dimensional Kolmogorov–Smirnov (KS) test. The similarity between two distributions (here, the model population and the catalogued radio data) is estimated by analyzing the maximum vertical distance between the two corresponding cumulative distribution function (CDF) lines, called the $D$-statistic. The KS $p$-value is then the probability of getting a value of $D$ as large or larger than the observed value under the null hypothesis that the two distributions are identical. If the $p$-value is less than a threshold value, we reject the null hypothesis, and state the two distributions completely dissimilar. The maximum $p$-value is 1, obtained for two identical samples. To our precision $p$-values of $5 \times 10^{-3}$ or less will be denoted as 0, indicating that the two distributions are strongly dissimilar, and can be dismissed. We perform the KS test for each pulsar parameter separately (see section~\ref{subsubsec:pulsar_params}). The $p$-values are given in Table~\ref{TableCDFvalues} for each model. The $p$-values obtained by the KS Test may vary due to the limited number of double neutron stars produced by our population synthesis method. We have checked that these variations do not affect our qualitative conclusions.

Table~\ref{tab:rates} shows the predicted number of survey-observed pulsar-NS/double pulsar systems for each model within a simulated Milky-Way. The column `Primary Pulsar' signifies the number of primaries observed and `Secondary Pulsar' identifies the number of secondaries observed. The `Double Pulsar' column estimates the number of observed pulsar-pulsar systems, the primary and secondary of which are already separately included under their individual columns. Hence column-wise, Total Observations = Primary Pulsar + Secondary Pulsar. The number of survey-detected pulsars varies by $\mathcal{O}(2)$ from varying parameters governing pulsar evolution alone. We discuss the total detection rates, and the relative abundance of primaries to secondaries in the following sections. 

\begin{table}\label{TableCDFvalues}
    \caption{Models and $p$-values: The table charts the simulated models and the $p$-values of the six chosen parameters after accounting for radio selection effects. }
    \begin{tabular}{lllllll}
    \hline
    \multicolumn{1}{|l|}{Model} & \multicolumn{1}{l|}{$P$} & \multicolumn{1}{l|}{$\dot{P}$} & \multicolumn{1}{l|}{$B$} & \multicolumn{1}{l|}{$P_\mathrm{orb}$} & \multicolumn{1}{l|}{$e$} & \multicolumn{1}{l|}{$|Z|$} \\ \hline
Initial                   & 0.11                   & 0.03                      & 0.00                   & 0.48                      & 0.00                   & 0.62                   \\
\\
BMF-R                     & 0.02                   & 0.01                      & 0.01                   & 0.47                      & 0.00                   & 0.42                   \\
BMF-FL                    & 0.01                   & 0.05                      & 0.03                   & 0.22                      & 0.01                   & 0.32                   \\
BMF-FGK06                    & 0.02                   & 0.02                      & 0.00                   & 0.34                      & 0.00                   & 0.28                   \\
\\
BS-R                      & 0.13                   & 0.10                      & 0.02                   & 0.24                      & 0.00                   & 0.27                   \\
\\
FDT-10                    & 0.00                   & 0.00                      & 0.00                   & 0.03                      & 0.00                   & 0.60                   \\
FDT-100                   & 0.06                   & 0.02                      & 0.01                   & 0.07                      & 0.00                   & 0.37                   \\
FDT-500                   & 0.00                   & 0.00                      & 0.00                   & 0.27                      & 0.00                   & 0.49                   \\
FDT-2000                  & 0.00                   & 0.01                      & 0.00                   & 0.36                      & 0.00                   & 0.36                   \\
\\
CE-Z                      & 0.00                   & 0.00                      & 0.00                   & 0.27                      & 0.00                   & 0.29                   \\
CE-U                      & 0.00                   & 0.00                      & 0.00                   & 0.37                      & 0.00                   & 0.37                   \\
\\
FDM-10                    & 0.00                   & 0.00                      & 0.00                   & 0.01                      & 0.00                   & 0.26                   \\
FDM-15                    & 0.02                   & 0.04                      & 0.11                   & 0.42                      & 0.00                   & 0.32                   \\
FDM-20                    & 0.83                   & 0.74                      & 0.32                   & 0.16                      & 0.00                   & 0.51                   \\
FDM-50                    & 0.00                   & 0.00                      & 0.00                   & 0.09                      & 0.00                   & 0.35                   \\
\\
    \hline
    \end{tabular}
\end{table}
\begin{table*}
\caption{Predicted number of Galactic double neutron star systems observed in radio for each model}
\begin{tabular}{lrrrr}
\hline
\multicolumn{1}{|l|}{Model} & \multicolumn{1}{l|}{Primary Pulsar} & \multicolumn{1}{l|}{Secondary Pulsar} & \multicolumn{1}{l|}{Double Pulsar} & \multicolumn{1}{l|}{Total Observations} \\ \hline
Initial                   & 64                   & 4                      & 1                   & 68      \\
\\
BMF-R                     & 66                   & 2                      & 0                   & 68      \\
BMF-FL                    & 218                   & 66                      & 15                   & 284      \\
BMF-FGK06                  & 60                   & 3                      & 1                   & 63      \\
\\
BS-R                      & 61                   & 1                      & 0                   & 62      \\
 \\
FDT-10                    & 19                   & 4                      & 1                   & 23      \\
FDT-100                   & 38                   & 2                      & 0                   & 40      \\
FDT-500                   & 43                   & 5                      & 0                  & 48      \\
FDT-2000                  & 48                   & 5                      & 0                   & 53      \\
\\
CE-U                      & 44                   & 4                      & 0                   & 48      \\
CE-Z                      & 7                   & 6                      & 0                   & 13      \\
\\
FDM-10                    & 447                   & 6                      & 1                   & 453      \\
FDM-15                    & 211                   & 7                      & 2                   & 218      \\
FDM-20                    & 41                   & 4                      & 1                   & 45      \\
FDM-50                    & 15                   & 2                      & 0                   & 17      \\
\\
\hline
\end{tabular}
\label{tab:rates}
\end{table*}

\subsection{The ``Initial'' Model}
\label{subsec:initial_model}

In this section, we explain each parameter for our Initial model, and describe the resulting DNS population. The birth magnetic field and the birth spin period of the pulsars are assumed to be drawn from an uniform distribution between $10^{10}$\,G to $ 10^{13}$\,G and $10$--$100$ ms respectively. These ranges match the typically observed ranges for young pulsar populations \citep{2005AJ....129.1993M}. The magnetic field decay time scale $\tau_\mathrm{d}$ is assumed to be $1000$\,Myr and the magnetic field decay mass-scale $\Delta M_\mathrm{d} = 0.025$\,$M_\odot$. We assume a fixed NS radius of $10$\,km, and the \cite{MacLeod:2014yda} prescription for mass accretion during the CE phase as discussed in Section~\ref{subsubsec:common_envelope}. For subsequent models, we change a single variable per model from these Initial model assumptions.

\subsubsection{$P \dot{P}$ for Initial}
\label{subsubec:ppdot_initial}

\begin{figure*}
\includegraphics[width=0.46\textwidth]{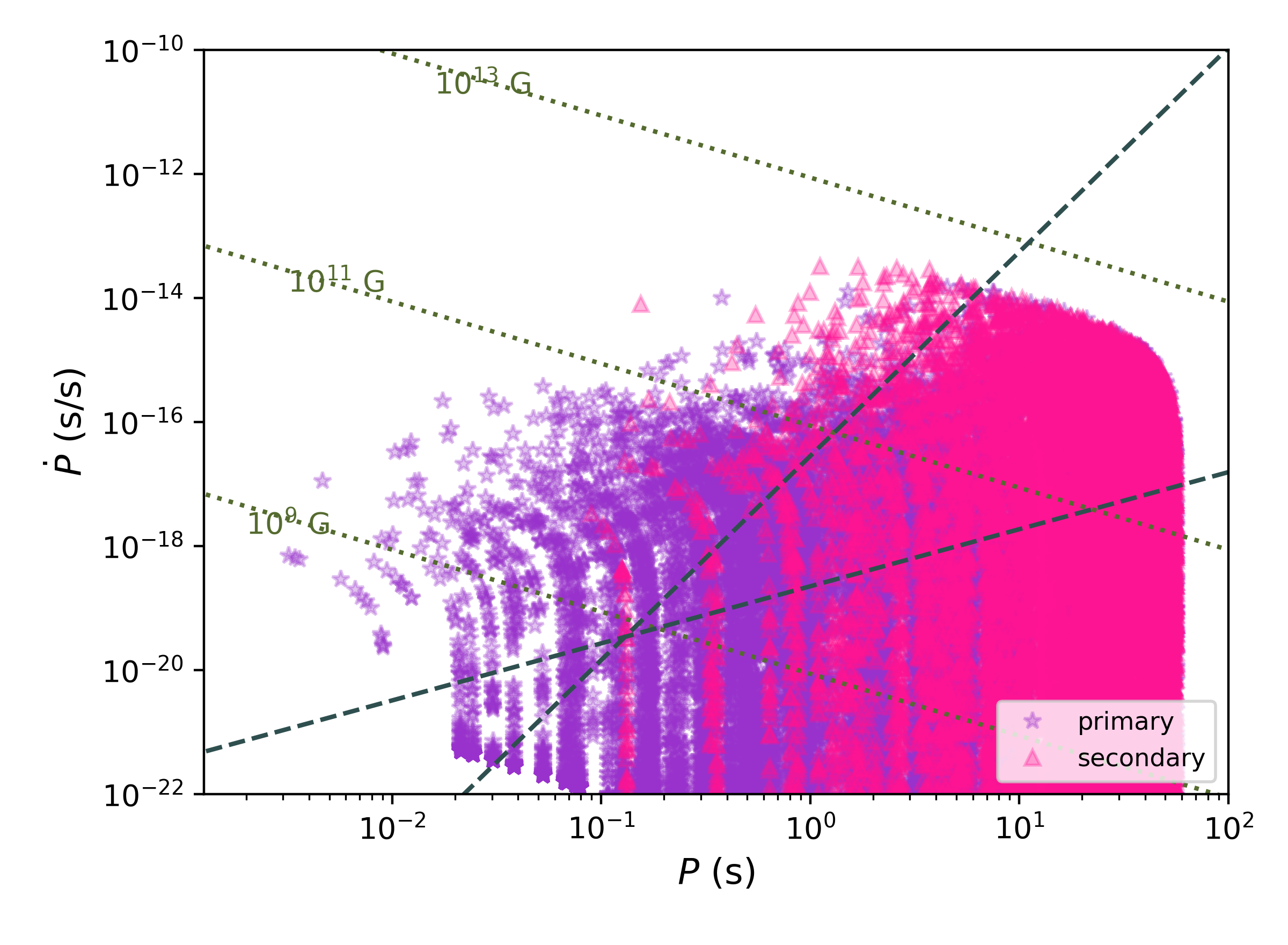}
\includegraphics[width=0.45\textwidth]{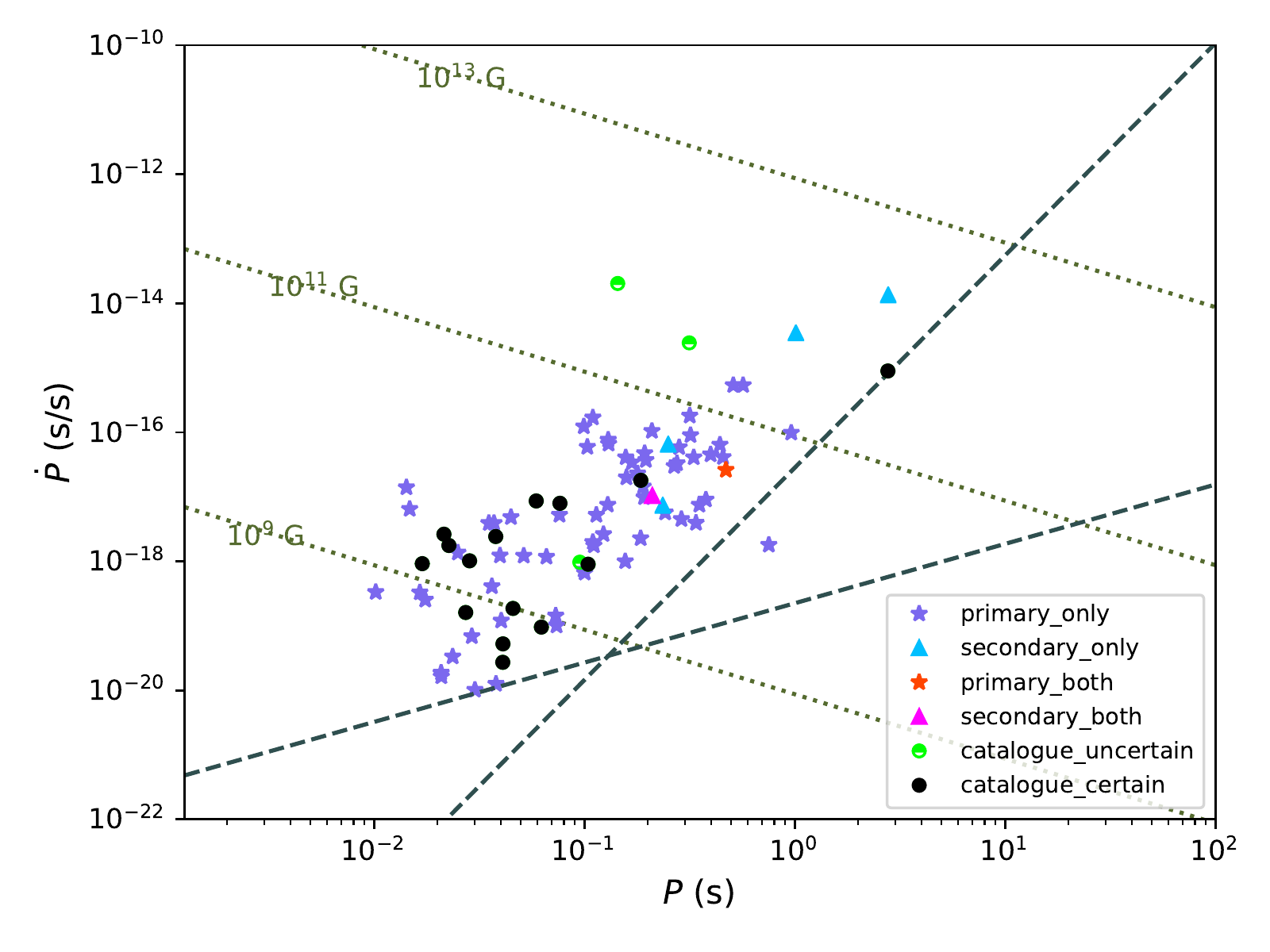}
\caption{The \(P\dot{P}\) diagrams for model Initial, before (left) and after (right) applying radio selection effects. Before applying radio selection effects we can only distinguish between the primary and secondary NSs. Only a fraction of the neutron stars in the left plot are observed by the radio pulsar survey on the right. 
}
\label{PPdotDefault}
\end{figure*}

The $P\dot{P}$ diagram for the model Initial is shown in Figure~\ref{PPdotDefault}. The left panel shows all DNSs in the model whilst the right panel shows the DNS population after accounting for selection effects. The reduction in the sheer number of data points emphasize the importance of accounting for radio selection bias in order to compare DNS models to DNS radio observations. The left panel of Fig.~\ref{PPdotDefault} differentiates only between the primary and the secondary of the DNS. Although the points appear with a track-like feature, they are individual snapshots at the time of observation (13\,Gyr) from the life of the pulsar. Since we re-use each binary, the degeneracy appears in the scatter plot. The right plot of the same figure, shows a particular snapshot in the entire life of the pulsar, if it is predicted to be observed by the pulsar survey. Each pulsar appears as a point. We distinguish the observed pulsars that are `primary-only', where the companion is not observed by the survey and may be an un-detected pulsar or a non-radio NS, and conversely `secondary-only' systems representative of the survey-observed secondary pulsars, whose primary is not detected. We also specify systems that are observed as double pulsars, in which both the primary and secondary of the system are detected by the radio telescope, as `primary-both' and `secondary-both' respectively. We show the radio catalogue DNSs, and distinguish the systems that are definitely DNSs, from those for which the classification is uncertain (Table~\ref{tab:observed_DNSs}). It is noticeable that the primary-only systems are more numerous than other systems, indicating that `primary' pulsars, which contain the recycled pulsar population are more detectable. There are fewer secondary-only points than primary-only. However secondary-only points are more numerous than double pulsar points (primary/secondary-both), signifying that though secondaries are harder to be observed by a pulsar survey, it is more common than observing a double pulsar. This is because firstly, recycled pulsars are greater in number than non-recycled pulsars; since recycling/mass transfer spins a NS up such that even though a pulsar has decayed and may have become `dead', it can be brought back to the radio-emitting regime by the mass transfer. The non-recycled pulsar spins, on the other hand, decay with time and have no possibility of being revived again as a `pulsar' but remain as a NS. Secondly, the radio selection effect is biased towards the detection of recycled pulsars (explained in detail in the following section). For a double pulsar system, both the NSs need to be emitting in the radio regime and detectable by the pulsar survey making them rare, both in the underlying and observed populations.

\begin{figure*}
\includegraphics[width=0.49\textwidth]{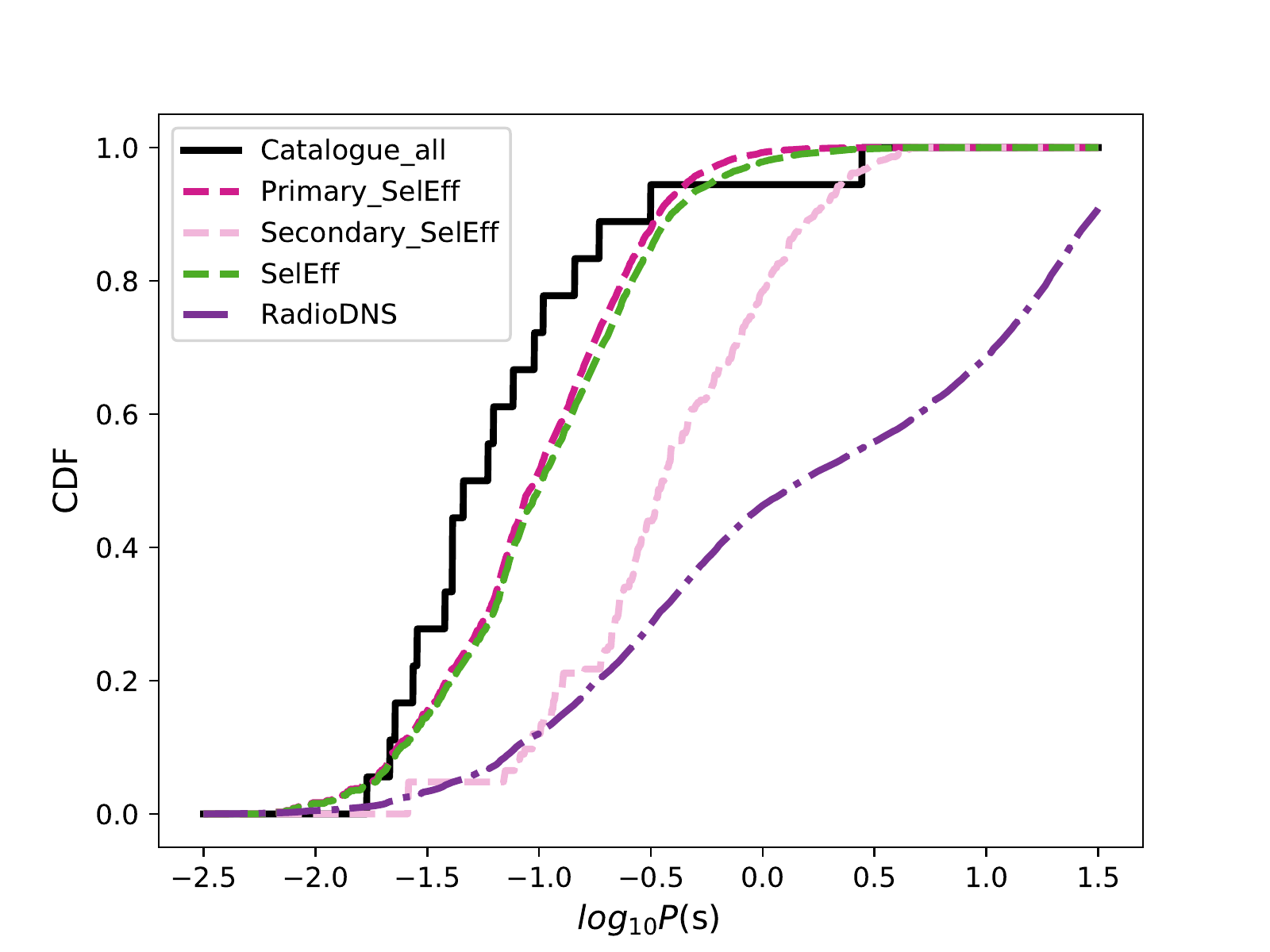}
\includegraphics[width=0.49\textwidth]{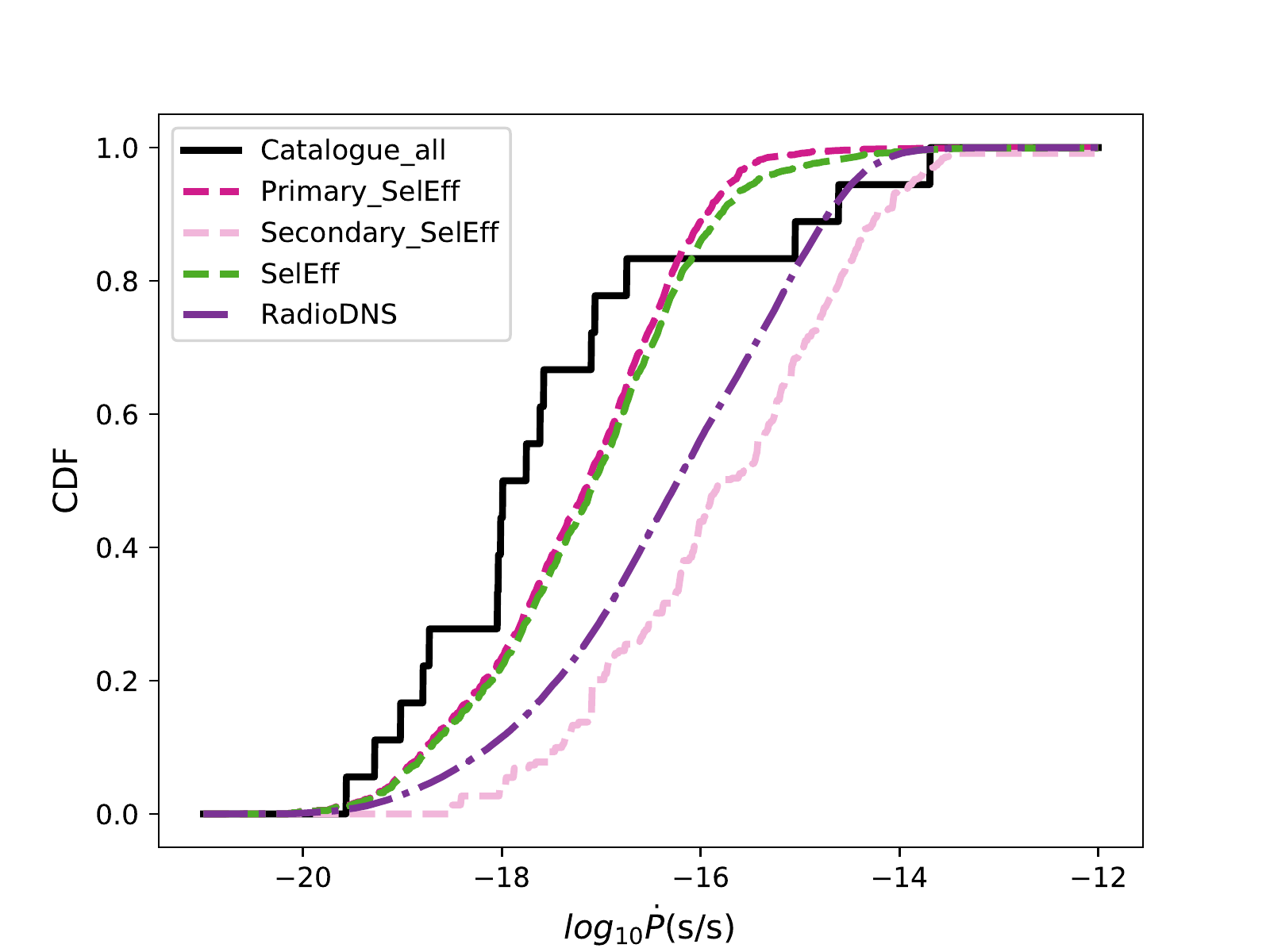}
\includegraphics[width=0.49\textwidth]{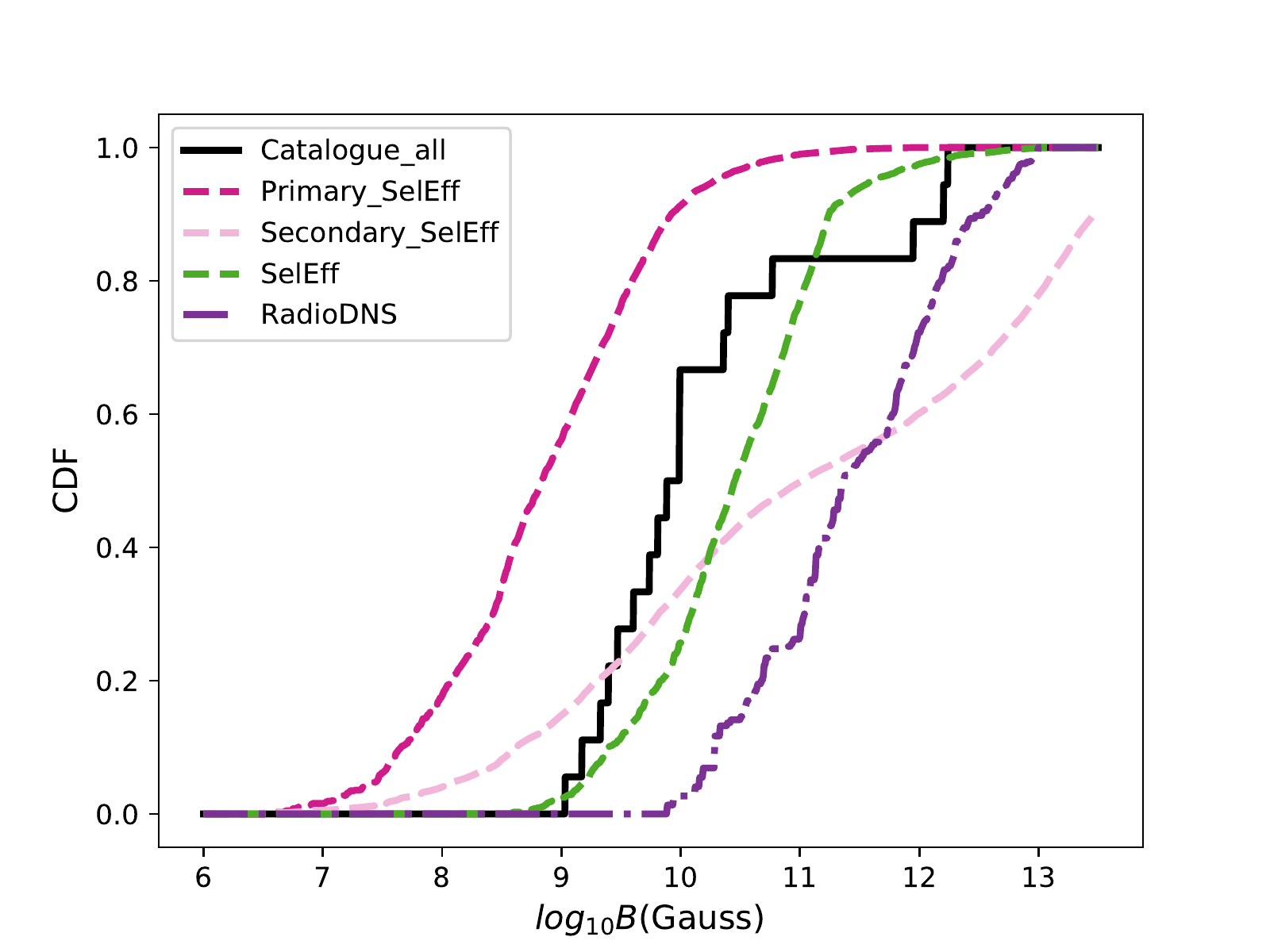}
\includegraphics[width=0.49\textwidth]{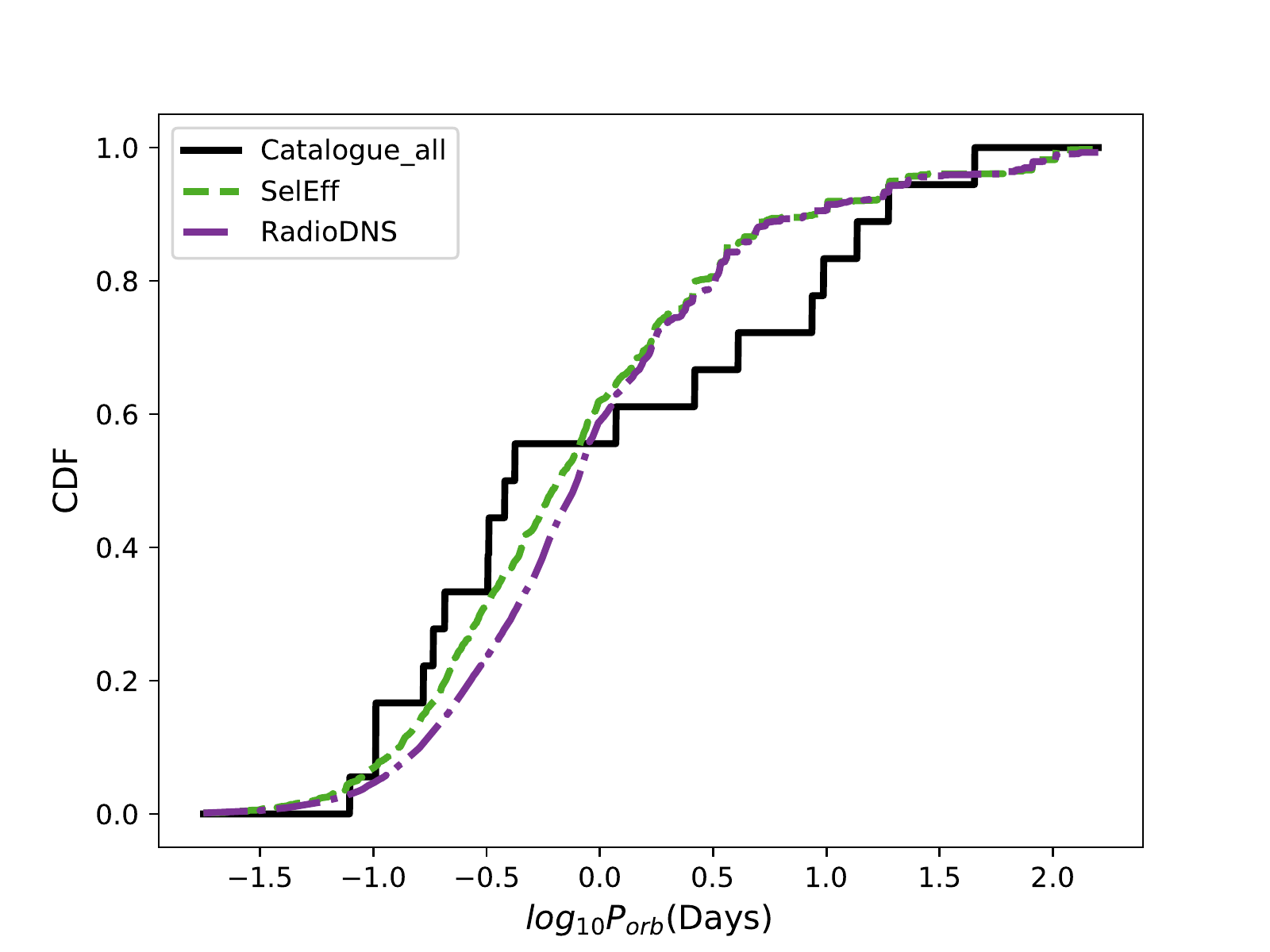}
\includegraphics[width=0.49\textwidth]{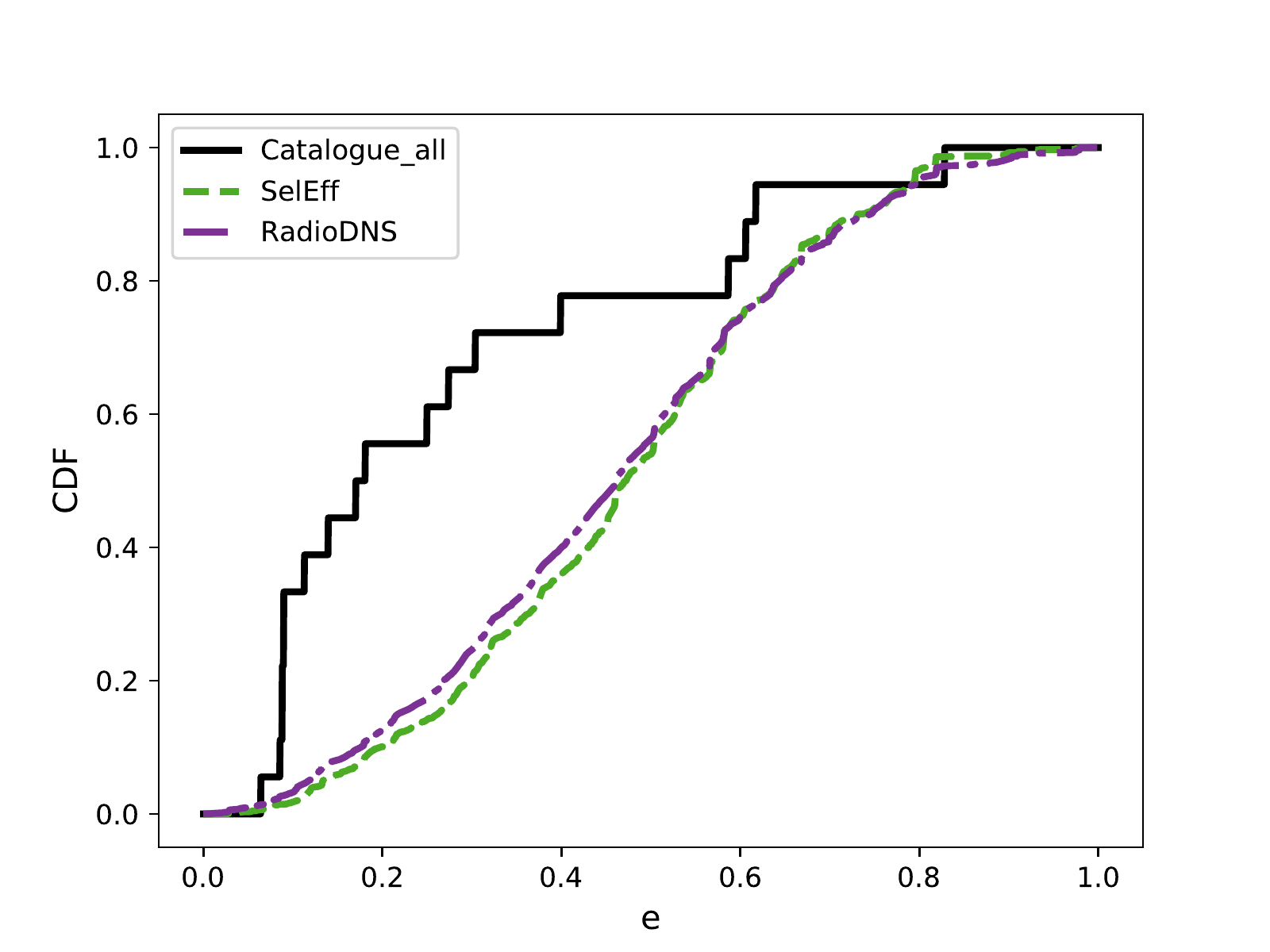}
\includegraphics[width=0.49\textwidth]{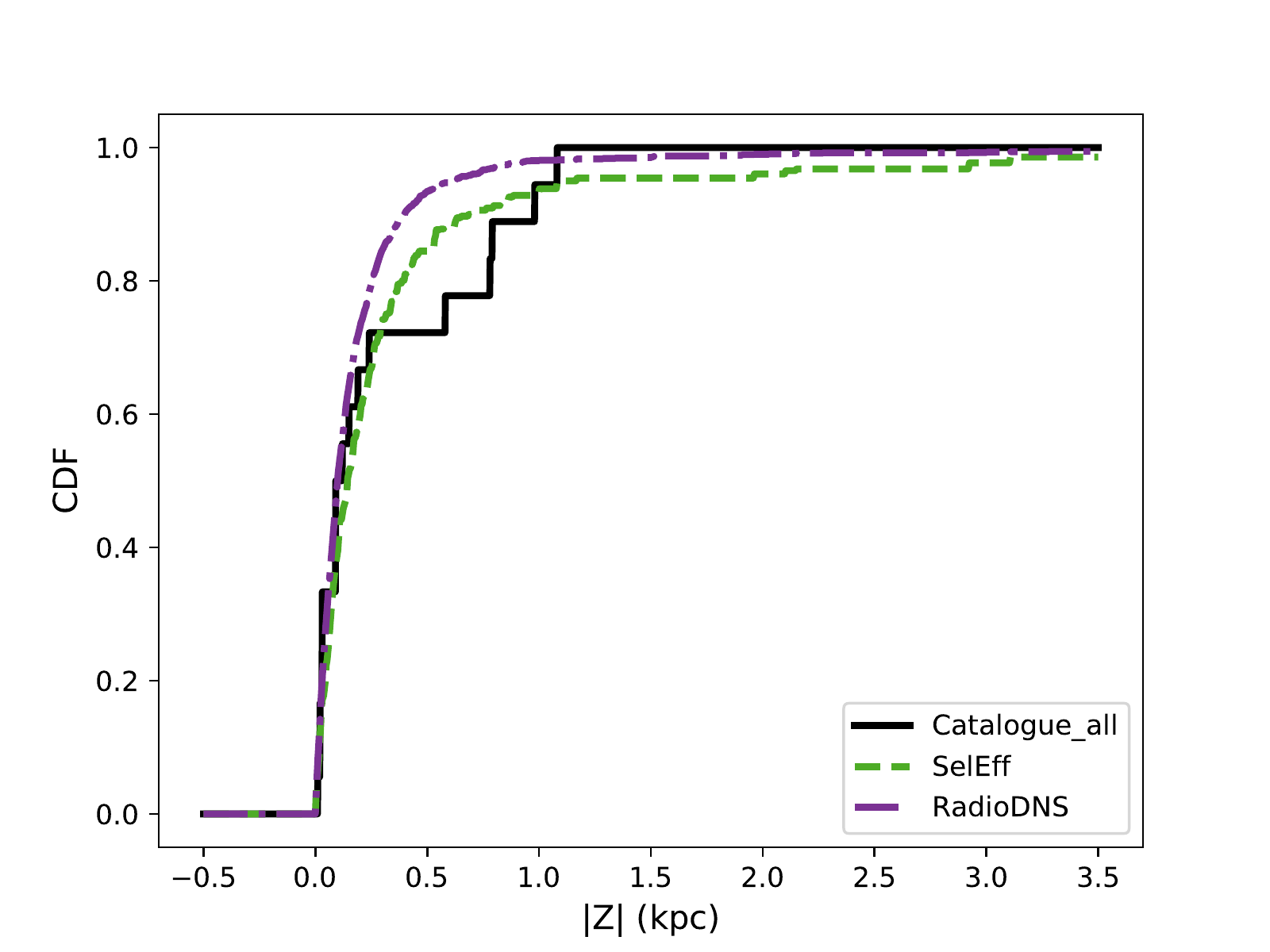}
\caption{The cumulative distribution function (CDF) of the pulsar parameters $P$, $\dot{P}$, $B$, $P_\mathrm{orb}$, $e$, $|Z|$ (from the top, left to right) for model Initial. The black line, `Catalogue-all' denotes the observed Milky Way DNS systems, including those that are uncertain. The purple line `RadioDNS' shows all the radio DNS systems, where at least one NS is a radio pulsar from our Initial simulation, and represents the underlying distribution of observable pulsars. The green dotted line `SelEff' shows the DNS systems after taking into account the radio selection effects, and hence is a subset of RadioDNS. We compare the black `Catalogue-all' and green dotted `SelEff' lines using the KS test to obtain the $p$-values quoted in Table~\ref{TableCDFvalues}. The SelEff population is further subdivided into the deep-pink `Primary-SelEff' and the light pink `Secondary-SelEff' identifying the populations of primaries and secondaries that are observed by the pulsar-survey.}
\label{InitialCDFS}
\end{figure*} 

\subsubsection{Pulsar Parameters}
\label{subsubsec:pulsar_params}

In this section, we compare the predictions of our Initial model to the observed DNSs. For each DNS parameter---the pulsar spin period ($P$) and spin-down rate ($\dot{P}$), surface magnetic field ($B$), orbital period ($P_\mathrm{orb}$), orbital eccentricity ($e$) and scale height ($|Z|$)---we plot cumulative distribution functions (CDFs) in Fig.~\ref{InitialCDFS}. We calculate corresponding $p$-values using the KS test. The individual $p$-values for each parameter are shown in Table~\ref{TableCDFvalues}. 

In each CDF, we show the observed radio catalogue data-set, and compare to the `RadioDNS' model population---meaning those systems are emitting in radio but may or may not be detected by a pulsar survey---as well as the `SelEff' population which is a subset of the same after accounting for the radio selection effects (as described in Section~\ref{subsec:radio_selection_effects}) and hence detectable. The latter population can further be decomposed into two categories, the primaries (`PrimarySelEff') and the secondaries (`SecondarySelEff').
\\
\\
$P$ : The CDF for the pulsar spin is shown in the top left panel of Fig.~\ref{InitialCDFS}. The RadioDNS population is the original DNSs where at least one is a (radio loud) pulsar. Hence RadioDNS signifies the population which would have been detectable as pulsar-NS/double pulsar systems, if no selection bias existed. It is dominated by old, slow pulsars with long spin periods (median spin period $\approx 1$\,s). The SelEff population biases towards faster spinning pulsars (median spin period of $\approx 0.1$\,s). This is because pulsars with shorter spin periods (lower $P$) have higher values of $f_\mathrm{beaming}$ (Equation~\ref{fBeaming}), and hence broader beams that are more likely to be detected. Thus, though there are more slow pulsars, radio selection effects bias the detection towards faster spins (lower values of $P$). This fact is even more apparent when we identify the PrimarySelEff and SecondarySelEff sub-populations. The primaries have faster spins due to the presence of recycled pulsars amongst them. Also, for the same reason, more primaries are detected (see Table~\ref{tab:rates}, this is true for all models). Thus the SelEff CDF (Primary-SelEff and Secondary-SelEff combined) is very similar to the Primary-SelEff CDF. The plot also shows that although the underlying pulsar population $P$ values are very different from the observed distribution (the CDFs for RadioDNS and the Catalogue-all have a relatively large vertical separation), taking radio selection effects into account reduces the dissimilarity between the  corresponding CDFs (SelEff and Catalogue-all). The model Initial produces a $p$-value of 0.11 for the spin parameter, when the SelEff population is compared with respect to the Catalogue-all data.\\
\\
$\dot{P}$ : The top right plot of Fig.~\ref{InitialCDFS} shows the CDF for the spin down rate of the Initial model. We see similar features of the post-radio selection effect population (SelEff) being closely aligned with the distribution of the primaries rather than the secondaries, as observed for the CDF of $P$. The reason is again due to increased detection of the primaries relative to secondaries as explained for $P$. Since $\dot{P}$ is correlated to $P$ through Equation~\ref{SpinDownIsolated}, it is not a surprise that the effect is propagated into the spin-down rate distribution. The $p$-value of $\dot{P}$ for this model is 0.03.\\
\\
$B$ : The CDF for the surface magnetic field strength is shown in the middle left plot of Fig.~\ref{InitialCDFS}. The sub-population of primaries typically have a lower value of $B$, than the secondaries. This is due to the recycled pulsars present in the PrimarySelEff population. Mass transfer buries the surface magnetic field of the pulsar (Equation~\ref{MagneticFieldAccretion}). This is more apparent in Fig.~\ref{PPdotEvolution}, where the primary is a recycled primary pulsar. Once again, as for $P$ and $\dot{P}$, the radio detection of more recycled pulsars shifts the underlying RadioDNS population towards lower values of $B$. Though accounting for the radio selection effect shifts the population towards the catalogue population---the CDF of SelEff is further left of RadioDNS---the $p$-value of $B$ for this model remains lower than $5\times10^{-3}$ (our threshold), and hence we conclude the distribution not to be similar to the observations.\\
\\
\begin{figure}
\includegraphics[width=0.99\columnwidth]{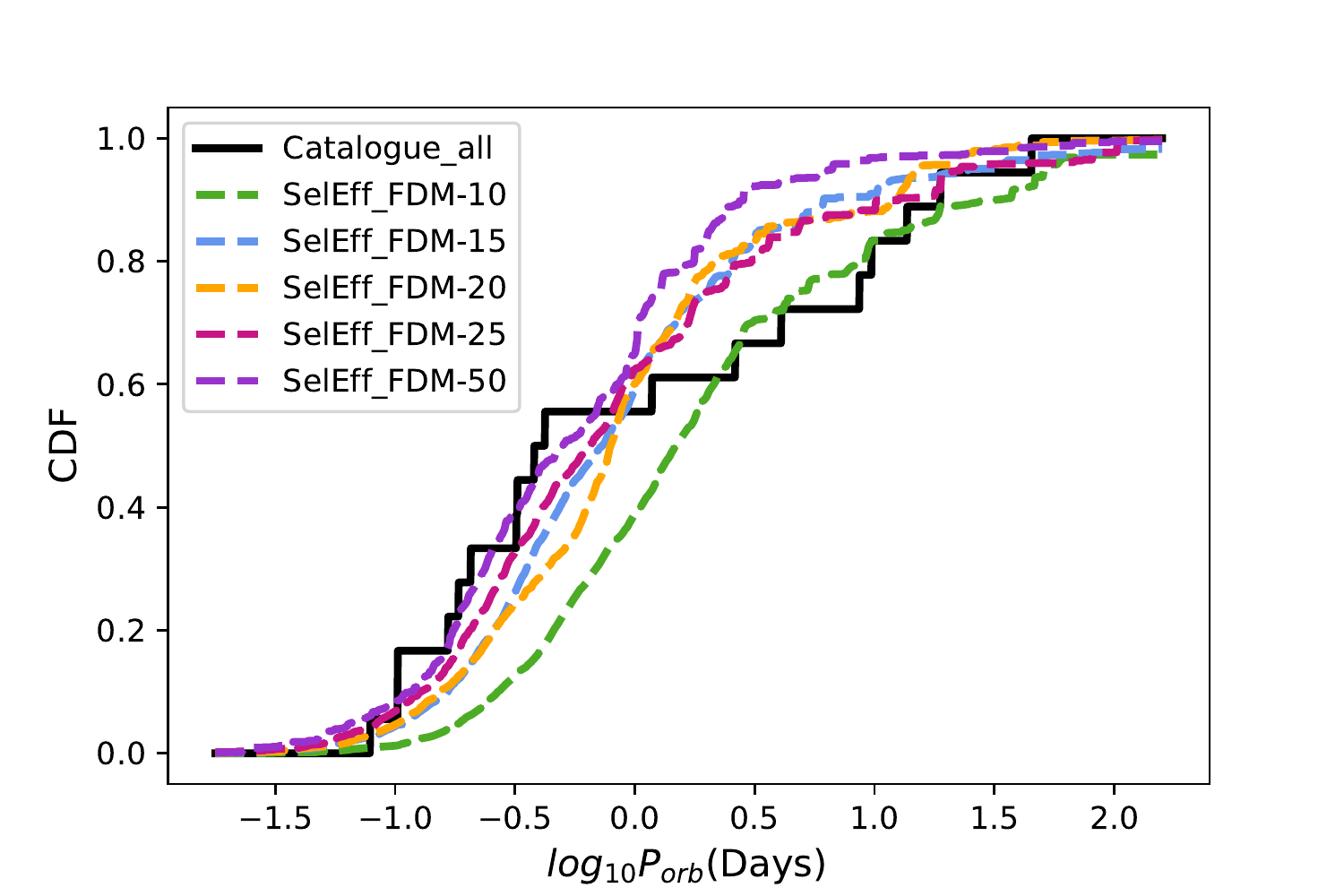}
\caption{The orbital period CDFs of all the FDM models, including Initial (FDM-25).}
\label{fig:Porb}
\end{figure}
$P_\mathrm{orb}$ : The middle right plot of Fig.~\ref{InitialCDFS}, shows the CDF for the orbital period of Initial. There are no distinctions between the sub-populations of primary and secondary pulsars because $P_\mathrm{orb}$ is a composite parameter of the entire binary system. Comparing the SelEff and RadioDNS CDFs, though very similar, shows a small shift towards lower $P_\mathrm{orb}$ values for the SelEff population. This is because binaries with recycled pulsars have lower value of $P_\mathrm{orb}$. The mass transfer, and especially the CE phase, if present, reduces the orbital separation and period of the binary system. Since recycled pulsars are more likely to be detected, such systems have higher detection probability. This effect, however is not very strong, because unlike the previous three pulsar parameters $P$, $\dot{P}$ and $B$, $P_\mathrm{orb}$ is not pulsar specific but binary specific. NS natal kicks also play an essential role in determining the orbital period of the binary. The $p$-value for $P_\mathrm{orb}$ for Initial is 0.48. The CDFs of $P_\mathrm{orb}$ of all FDM models, including Initial (which is FDM-25, since $\Delta M_\mathrm{d} = 0.025$\,$M_\odot$) are shown in Fig.~\ref{fig:Porb}. Model FDM-10 preferentially matches the long orbital period systems, whilst FDM-50 preferentially matches the short orbital period systems. Models FDM-15, FDM-20 and FDM-25 (Initial) all produce similar $P_\mathrm{orb}$ distributions, resulting in similar $p$-values, with all three models providing an adequate match across the full range of observed $P_\mathrm{orb}$ distribution. The differences in $p$-values for these models can be attributed primarily to fluctuations due to the number of DNSs produced by our population synthesis method ($\sim$500--1000), amplified by the fact that radio observable DNSs represent a small fraction of the total population (see section~\ref{subsec:merger_time}).
\\
\\
$e$ : The CDF for the eccentricity distribution is shown in the bottom left panel of Fig.~\ref{InitialCDFS}. As for the $P_\mathrm{orb}$, $e$ is also binary system specific and hence has no sub-population of primaries and secondaries. The population including observational selection effects (SelEff) closely represents the underlying distribution (RadioDNS). This shows that the selection effects we have modelled are largely decoupled from the observed eccentricity distribution. It is clear by eye (and confirmed by the $p$-value) that our model does not provide a good match to the eccentricity distribution of Galactic DNSs \citep[as shown before by][]{Kiel:2010MNRAS,2017AcA....67...37C,Vigna-Gomez:2018dza}. We believe there are two reasons for this discrepancy.

Firstly, helium stars which are ultra-stripped in our model leave behind CO cores which are more massive compared to detailed simulations \citep{Tauris:2015,Vigna-Gomez:2018dza}. This causes them to lose too much mass during the ultra-stripped SN, resulting in a large Blaauw kick\footnote{SN natal kick solely due to mass loss \citep{Blaauw:1961}} that increases the orbital eccentricity. Population synthesis models including updated prescriptions for the core masses of ultra-stripped helium stars do not show this discrepancy \citep[e.g.][]{2018MNRAS.481.1908K,2019arXiv190611299Z}.

Secondly, tight binary pulsars with large orbital eccentricities produce higher orbital acceleration and `jerk' (time derivative of acceleration) rendering the pulsar more difficult to be observed by pulsar searches \citep{Bagchi:2013wga}. This biases pulsar searches against binaries with the highest eccentricities. We have not modelled these selection effects \citep{Bagchi:2013wga} in our radio selection bias for $e$. Therefore, all of our suite of models show disagreement in $e$ distribution from the radio population. \\
\\
$|Z|$ : The bottom right plot of Fig.~\ref{InitialCDFS} shows the CDF for the vertical heights $|Z|$, of the pulsar-NS/pulsar-pulsar binaries, in a Galactocentric Cartesian co-ordinate system. The value of $|Z|$ signifies how far away the pulsar is located from the Galactic plane. In the CDF, we also observe that the scale height for the SelEff is larger than for RadioDNS. This is because the sky temperature $T_\mathrm{sky}$ (section ~\ref{subsec:radio_selection_effects}, equation~\ref{Radiometer}) is smaller in regions above the Galactic disc with higher scale height, rendering such radio pulsars easier to observe. We see that our model agrees well with the observed heights of DNSs ($p=0.4$), lending support to our model for NS natal kicks.\\ 

We next explore how each parameter that we have varied affects the resultant population.

\subsection{Birth Magnetic Field (BMF)}
\label{subsec:birth_magnetic_field}

We use a uniform distribution for the pulsar birth surface magnetic field strength with a range between  $10^{10}$\,G to $10^{13}$\,G for model Initial. This range is based on the observed surface magnetic field range of young pulsars \citep{2005AJ....129.1993M}. We have varied the assumption of the BMF range (R) for model BMF-R, where we use a uniform distribution but within the range $10^{11}$\,G to $10^{13}$\,G. As we see from Table~\ref{TableCDFvalues}, we do not observe any noticeable improvement in the subsequent $p$-values of the pulsar parameters for model BMF-R. This is because pulsars in our model experience exponential decay and spend very little time $\mathcal{O}$(Myr) in the region of the parameter space where they are born; irrespective of whether the magnetic braking is governed by the the magnetic field decay time-scale (see Equation~\ref{MagneticFieldIsolated}) or magnetic field decay mass-scale too (see Equation~\ref{MagneticFieldAccretion}) for recycled pulsars. Therefore, it is not surprising that model BMF-R does not produce significantly different results from Initial. The spin period $P$($=\frac{2\pi}{\Omega}$) however, is correlated to the initial value of $B$ through Equation~\ref{SpinIsolatedIntegrated}. A different initial $B$ value shifts the evolved $P$ distribution and we see a decrease in the corresponding $p$-value.

\begin{figure}
\includegraphics[width=0.99\columnwidth]{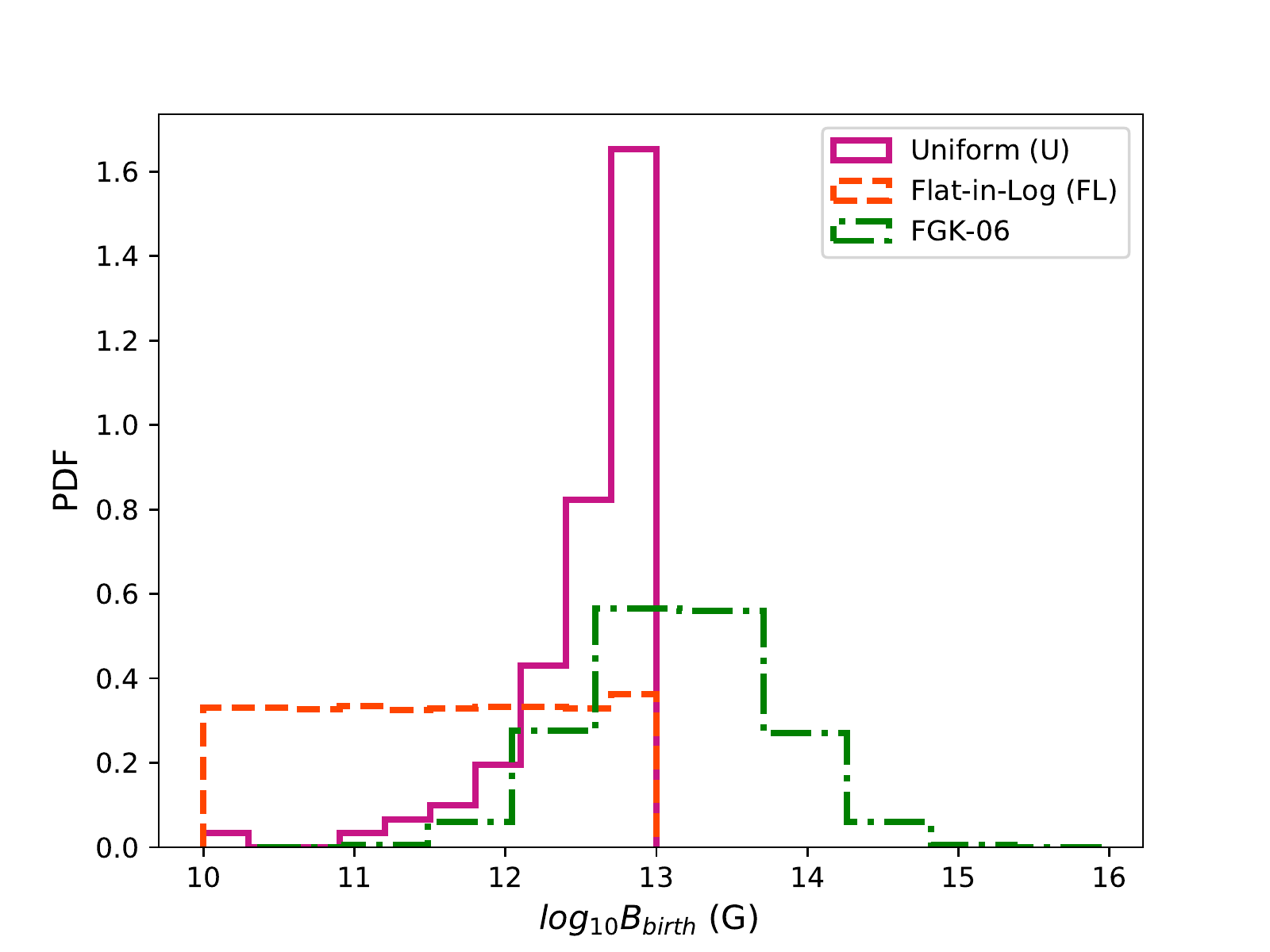}
\caption{Birth magnetic field distribution of modelled pulsars. Model Initial uses a uniform distribution (magenta line), model BMF-FL uses a flat in log distribution (orange, dashed line). Both distributions have  $10^{10} < \log_{10} B/\mathrm{G} < 10^{13}$. Model BMF-FGK06 \citep{FaucherGiguere:2005ny} uses a log-normal distribution with a mean of 12.65 and standard deviation of 0.55 (green dash-dotted, line).}
\label{fig:BirthMagnetic}
\end{figure}
 
In addition to our Initial model, we also modelled BMF-FL and BMF-FGK06. BMF-FL has the same range of birth magnetic field magnitudes as Initial ($10^{10}$\,G to $10^{13}$\,G) but uses a flat-in-the-log (FL) distribution rather than a uniform distribution. BMF-FGK06 uses the prescription given by \citet{FaucherGiguere:2005ny}, referred as FGK06 here, who model the birth magnetic field distribution as a log normal distribution with a mean of 12.65 and standard deviation of 0.55.

\begin{figure*}
\includegraphics[width=0.49\textwidth]{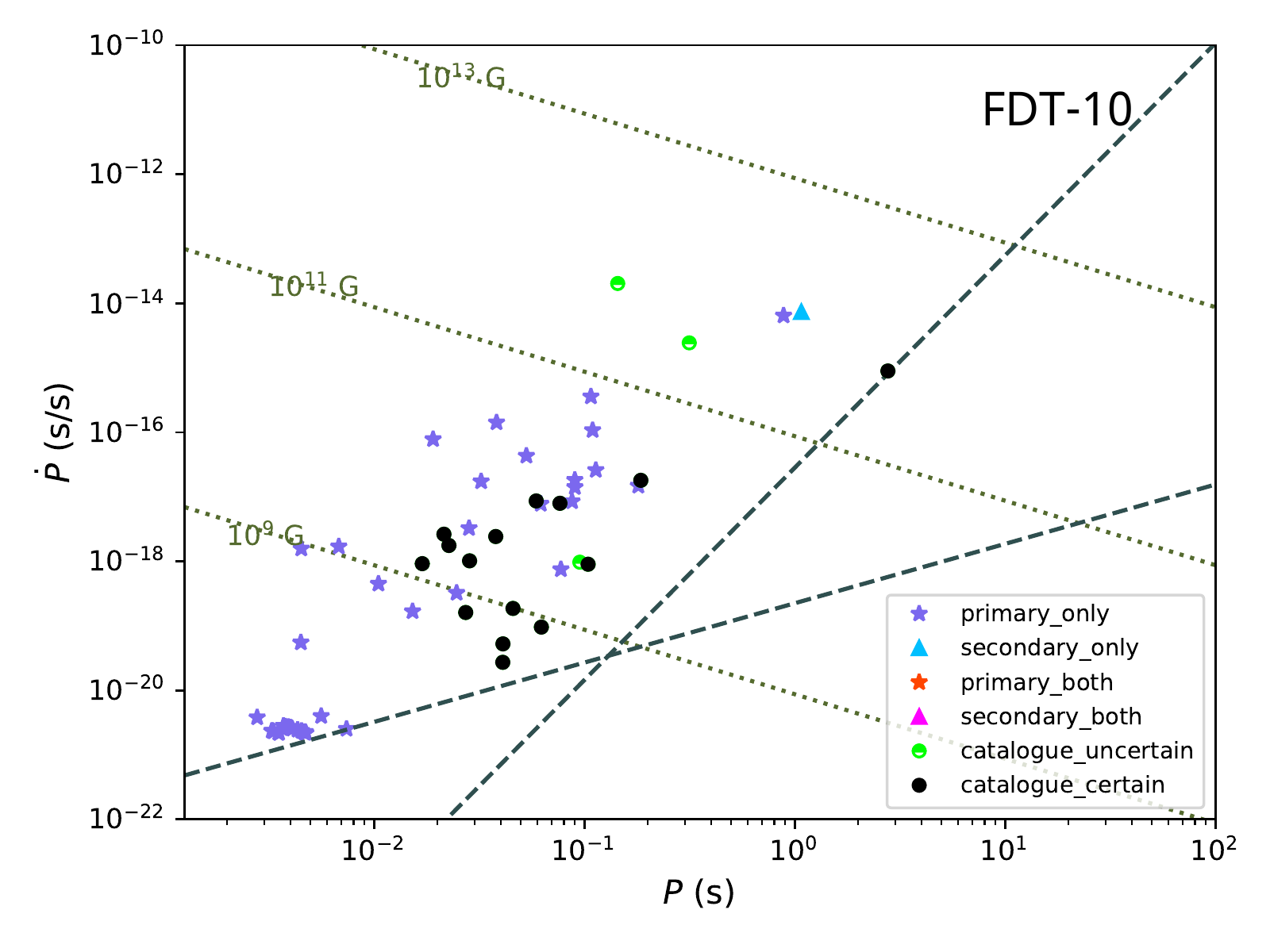}
\includegraphics[width=0.49\textwidth]{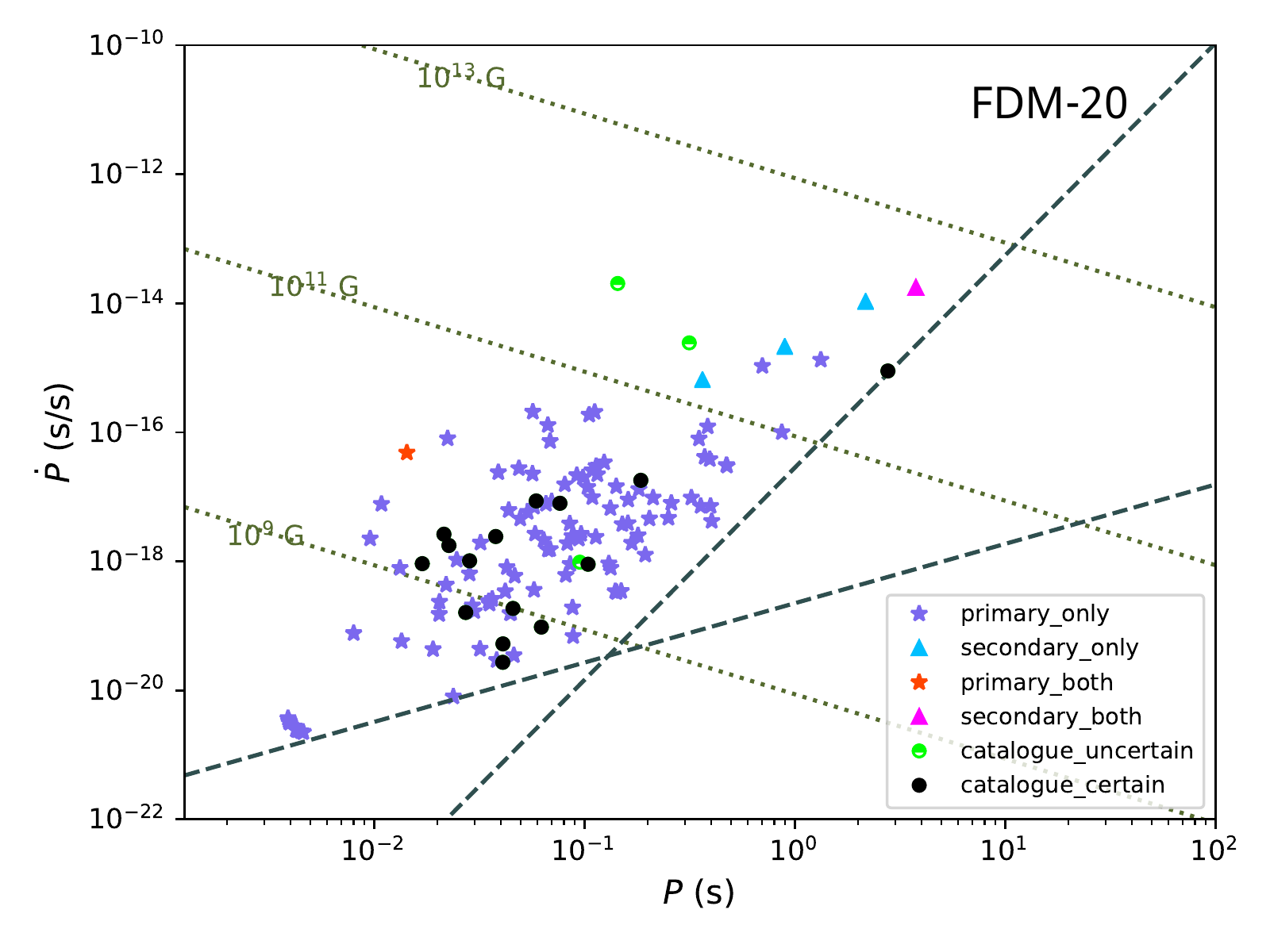}
\includegraphics[width=0.49\textwidth]{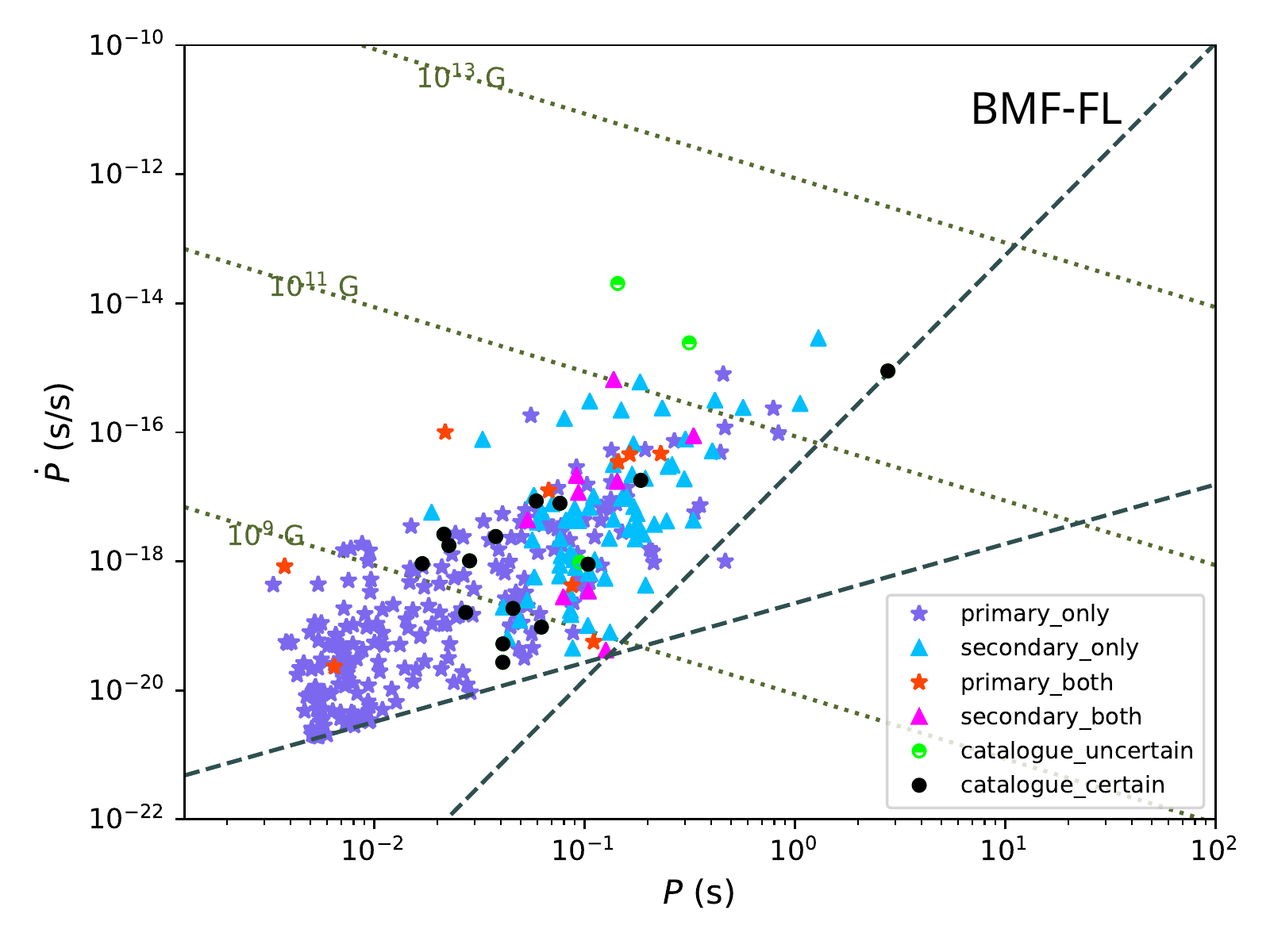}
\includegraphics[width=0.49\textwidth]{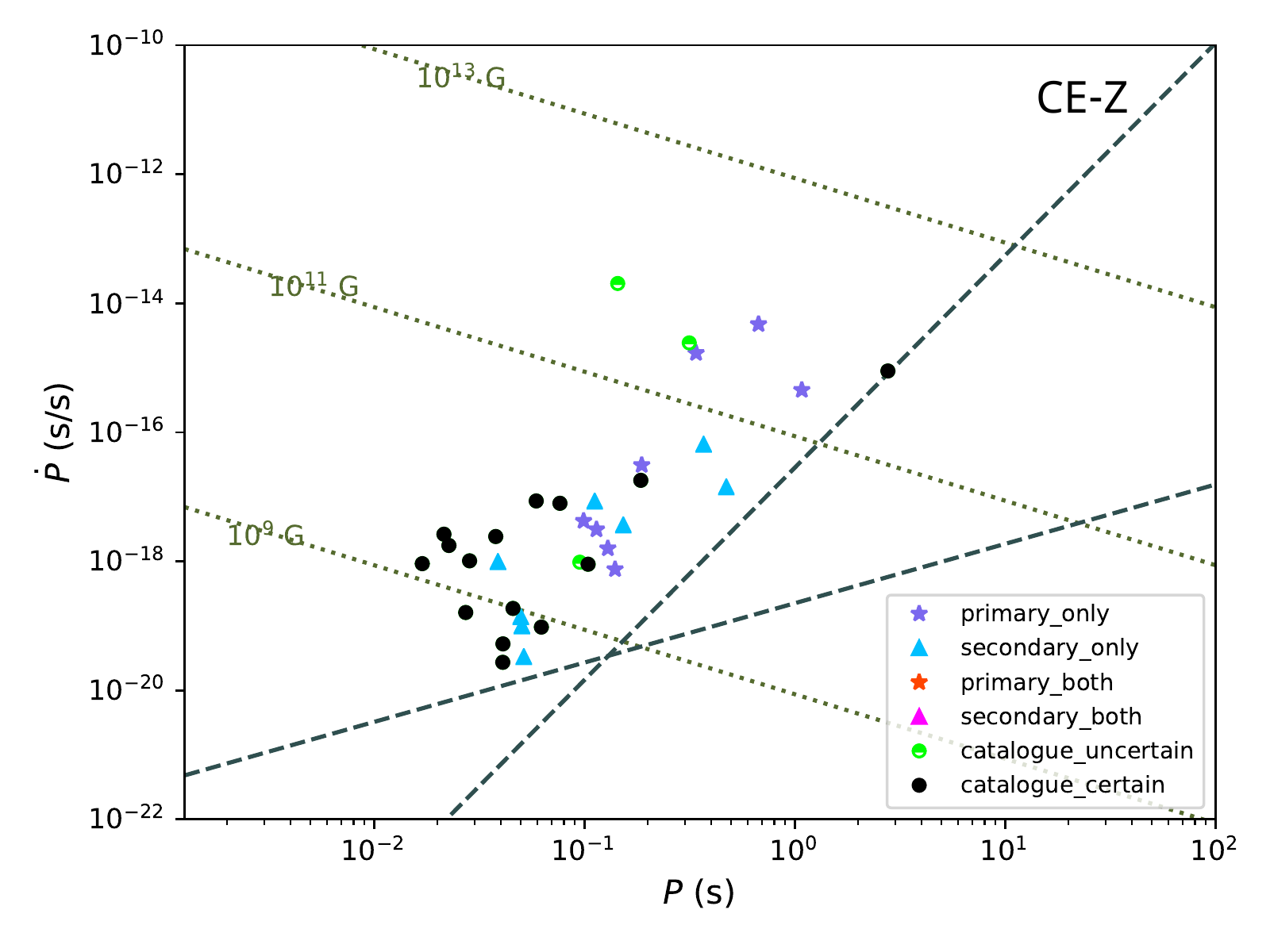}
\caption{The \(P\dot{P}\) diagrams for models FDT-10 (top left), FDM-20 (top right), BMF-FL (bottom left) and CE-Z (bottom right). All the points shown in the plots are after accounting for the radio selection effects, thus, essentially showing the 'observed' data-points. The legend 'primary/secondary-only' denotes points where either the primary or the secondary of the binary is observable after the radio selection effects have been applied, while 'primary/secondary-both' is where both pulsars in the pair are observable by the survey. Only a small fraction of pulsar-neutron star binaries are double pulsar-like binaries with two observable pulsars. The primary candidates from the simulations are marked with a star symbol, while the secondaries are marked by an upright triangle. Data points crossing the black dashed death lines are discarded from further analysis. The terms 'catalogue-certain' and 'catalogue-uncertain' indicate the Australia Telescope National Facility (ATNF) catalogued Galactic pulsar systems \citep{2005AJ....129.1993M}. The uncertainty arises from the fact that the minimum companion mass measured from observations can either make the companion a white dwarf or a NS.}
\label{PPdotObs}
\end{figure*}

The three distributions are visualized in Fig.~\ref{fig:BirthMagnetic} as probability density functions (PDFs). The average birth magnetic field strength in FL is lower than for the uniform distribution; this shifts the observed population to a lower magnetic field range which can be seen by comparing the $P\dot{P}$ diagram for BMF-FL which we show in Fig.~\ref{PPdotObs} (bottom left panel) to that for the Initial model in Fig.~\ref{PPdotDefault} (right panel). We also find that BMF-FL produces more observable pulsars in total, of which a comparatively larger fraction are secondaries (Table~\ref{tab:rates}). 

The fractional increase in the number of secondaries is due to the fact that the secondaries are always non-recycled pulsars and thus are affected more by the birth distributions than the primaries (which may/may not be recycled). Though this improves the $p$-value for the magnetic field, this model produces pulsars with spin periods which are shorter than the observed population, and thus the $p$-value becomes too low. BMF-FGK06 does not show a remarkable change in the $p$-values of the pulsar parameters compared with the Initial model (see Table~\ref{TableCDFvalues}). 

\subsection{Birth Spin (BS) Period} 
\label{subsec:BS_period}

For the Initial model, we assume an uniform birth  spin period of pulsars between 10--100\,ms. We have altered the range (R) of birth spins in model BS-R, where it is between 10--1000\,ms, but still with an uniform distribution. Apart from the $p$-value of the surface magnetic field $B$, BS-R does not show a significant change in the $p$-values from Initial (see Table~\ref{TableCDFvalues}) because the initial spin evolves according to Equation~\ref{SpinDownIsolated}. Pulsars very rapidly spin down over timescales of $0.1$--$1$\,Myr, depending on their birth magnetic field strength (see Section~\ref{subsec:birth_magnetic_field}). This is much shorter than the age of typical DNS systems (see Figure~\ref{fig:TBirthAndTmerger}). The $p$-value of $B$ shows an order-of-magnitude improvement. This is because as discussed in Section~\ref{subsec:birth_magnetic_field}, the surface magnetic field and the spin of the pulsar are related through Equation~\ref{SpinIsolatedIntegrated}, and hence changing the initial range of spin period results in a shift in the final evolved range of surface magnetic field. The birth spin range, shows as an important parameter through the effect on the $p$-value of $B$.

\subsection{Magnetic Field Decay Time (FDT) Scale}
\label{subsec:FDTscale}

The pulsar magnetic field decays over time on a characteristic timescale $\tau_d$ in our model (see Equation~\ref{MagneticFieldIsolated}). The spin down rate $\dot{\Omega}$ depends explicitly on the magnetic field (Equation~\ref{SpinDownIsolated}), and thus there is implicit dependence on $\tau_d$. Hence, the magnitude of $\tau_d$ not only governs the magnetic field of the pulsar but also influences its spin and spin down rate, affecting both the non-recycled and the recycled pulsars (i.e. both the primary and secondary population).

We selected $\tau_d=1000$\,Myr for our Initial model. We varied it to $10$\,Myr, $100$\,Myr, $500$\,Myr and $2000$\,Myr across models FDT-10, FDT-100, FDT-500 and FDT-2000 respectively. The time-scale $\tau_d$ describes the exponential magnetic field decay that determines the path of the pulsar in the $P\dot{P}$ diagram. Shorter decay timescales (smaller $\tau_d$) lead to a sharper decay curve of the pulsars. Hence, too low $\tau_d$ pushes most systems to have higher radio efficiency $\xi$, and also makes them cross the death-line sooner - ending in the graveyard region of non-radio dead pulsars. This explains the low number of observable systems for FDT-10 shown in Fig.~\ref{PPdotObs} and Table~\ref{tab:rates}. Conversely, long magnetic field decay timescales (large values of $\tau_d$) lead to pulsars spinning down along lines of constant magnetic field strength, and pushes most systems to a magnetic field range not observed in pulsar--NS systems.

From Table~\ref{TableCDFvalues} we note that FDT-10 is a poor fit to the radio observations with the exception of the $p$-value of $Z$. Since $Z$ solely depends on the positional distribution of the DNSs in the Galactic potential, which is model independent, the $p$-value of $Z$ remains consistent for all models. Models FDT-100, FDT-500, FDT-2000, show lower $p$-values for both $P$ and $\dot{P}$ than model Initial, showing the latter is a better match to the observations. The $p$-values for $B$ and $e$ remain negligible for all the FDT models. The $p$-value of $P_\mathrm{orb}$ improves as we move towards higher values of $\tau_\mathrm{d}$. Though the change is very slight, it emphasizes the importance of the pulsar parameters, even when comparing models to the orbital properties of Galactic DNSs \citep[e.g.][]{2015ApJ...801...32A,Vigna-Gomez:2018dza}.

\subsection{Common Envelope (CE) Mass Accretion}
\label{subsec:CE_accretion}

As discussed in Section~\ref{subsubsec:common_envelope}, mass accretion during the CE phase may play a role in the surface magnetic field burial and spinning up process of recycled pulsars. Since the amount of mass that can be accreted through CE may be significantly larger than through RLOF, its effect can be non-negligible on the pulsar parameters. To understand the effect of CE mass accretion on the resultant pulsar population, we have assumed a prescription based on \citet{MacLeod:2014yda} for CE mass accretion in the model Initial (Section~\ref{subsubsec:common_envelope}) and varied that assumption to zero (Z) CE mass accretion for model CE-Z and, to a uniform (U) distribution for model CE-U. We notice that for CE-Z, the total number of `observed' systems is significantly lower (Table~\ref{tab:rates}), and this can be explained by the fact that no accretion during CE takes away the possibility of the pulsar to be spun up through CE. With the reduction of one possible source of pulsar spin-up and hence recycling, there are a lower number of radio systems that do not cross the death line and have the optimum value of $\xi$ (see Section~\ref{subsec:pulsar_death}). Due to low recycling, the number of observed primaries and secondaries are nearly equal for CE-Z, which is uncharacteristic of all other models where typically more primaries are observed (see Table~\ref{tab:rates}).

Apart from $P_\mathrm{orb}$ and $|Z|$, the negligible $p$-values of all pulsar parameters show that both CE-Z and CE-U match poorly with observed Galactic DNS systems. We hence conclude that CE mass accretion plays a non-trivial role in the resultant pulsar population, and allowing mass accretion during CE, using the \cite{MacLeod:2014yda} prescription gives a better match with the observations. 

\subsection{Magnetic Field Decay Mass Scale (FDM)}
\label{subsec:magn}

The magnetic field decay mass scale $\Delta M_\mathrm{d}$ (c.f. Equation~\ref{MagneticFieldAccretion}) determines how much mass a pulsar needs to accrete to bury the magnetic field. There is a huge uncertainty associated with $\Delta M_d$ due to the lack of a comprehensive understanding of the process of mass transfer in binaries and the exact process of quenching the pulsar's magnetic field. 

The model Initial has $\Delta M_\mathrm{d}= 0.025$\,M$_\odot$, and we have varied it to 0.010\,M$_\odot$, 0.015\,M$_\odot$, 0.020\,M$_\odot$ and 0.050\,M$_\odot$ for models FDM-10, FDM-15, FDM-20 and FDM-50 respectively. Since $\Delta M_\mathrm{d}$ governs the field decay equation for the case of mass-transfer, changing it affects only the recycled pulsar population. Lower $\Delta M_\mathrm{d}$ results in a steeper decay, since even a small amount of accreted mass will bury the magnetic field; thus more recycled pulsars are pushed towards a lower magnetic field for lower values of $\Delta M_\mathrm{d}$. If the magnetic field becomes too low, the pulsars cross the death lines and no longer emit in radio. Both FDM-15 and FDM-10 show enhanced recycled populations in the lower magnetic field region than the Initial model. Comparing quantitatively, the mean magnetic field of the observed pulsar population for model Initial is $8.8\times10^{10}$\,G, while for models FDM-10 and FDM-15 are $6.0\times10^{9}$\,G and $1.9\times10^{10}$\,G respectively.  FDM-15 and FDM-20 show considerable order-of-magnitude improvement in the $p$-value of $B$, FDM-50 however shows a decrease in the same, showing $\Delta M_\mathrm{d} \approx 0.015$--$0.020$\,M$_\odot$ gives the best fit for $B$ (when all other parameters are constant). FDM-50 also shows a decline in the $p$-values of $P$, $\dot{P}$ and $P_\mathrm{orb}$ relative to the Initial model. Model FDM-20, with $\Delta M_\mathrm{d} = 0.020$\,M$_\odot$  shows a significant improvement in the $p$-values of all parameters, aside from $P_\mathrm{orb}$, which remains in the same order of magnitude as Initial. 

Since FDM-20 shows an overall agreement to the catalogued population of DNSs, it is our current `best-fit' model. We are aware that a more detailed investigation on modulating the initial variables may result in a more precise best-fit model. However, we focus more on exploring the effects of the pulsar parameter space on the resultant population and hence running more simulations to determine a best-fit model across all parameters will be left for future work.

\subsection{Estimating the number of DNSs in the Milky Way} 
\label{subsec:total_number}

We have evolved 10$^{6}$ binaries in a mass range of 4--50\,M$_\odot$ using the IMF of \citet{Kroupa:2000iv} with COMPAS. As discussed in Section~\ref{sec:RadioPopulation}, we re-use every binary that forms a DNS 1000 times, by assigning each a birth-time drawn from an uniform distribution between 0--13 Gyr. This is assuming an uniform star-formation history of the Milky Way \citep{Vigna-Gomez:2018dza}. By re-cycling the binaries, and postulating that the same binary with different birth times are unique, we are essentially creating $10^{9}$ binaries and increasing the computational efficiency. The fraction of stars in the Milky Way that are in the mass range of $4$--$50$\,M$_\odot$ is about 0.01--0.02. Thus, for a galaxy of $\approx10^{11}$ stars \citep{Flynn:2006tm, Irrgang:2013}, the number of stars in the said mass range is $\approx 10^{9}$. However, we assume a 100\% gravitationally bound binary fraction, whereas the Milky Way binary fraction of the said stellar range is $\approx 30$--$50$\%\footnote{Our chosen orbital separation distribution means that around half of our binaries do not interact through mass transfer and evolve effectively as two single stars, though gravitationally bound.} \citep{Raghavan:2010, Sana:2012}. 

The assumption of isolated binary evolution may not hold true for the Galactic bulge. The Milky Way bulge accounts for $\approx 20$\% of the total number of stars of the Galaxy \citep{Flynn:2006tm}. Putting the numbers in, we obtain $\approx 10^{8}$ stars of the Milky Way that are in isolated binaries within the said mass range. Hence, we roughly evolve $\approx 1$ Milky Way set of stars when calculating the detection rates in Table~\ref{tab:rates}.

We observe that even after using PSREvolve to account for the radio selection effects, we overestimate the observable pulsars per model survey by an order of magnitude for some of our models (see Table~\ref{tab:rates}). This may indicate that the parameters of these models are disfavoured by the observed DNS population. Some of the discrepancy may be due to additional un-modelled radio selection effects, and our simplification of only modelling a single radio survey. Our best fit model FDM-20 produces a factor of $\approx 2$ more observed DNSs than the radio catalogue. 

Uncertainties in the initial distributions of binary properties lead to a factor of two uncertainty in predictions for the rates of double neutron star formation \citep[e.g.][]{deMink:2015yea, Klencki:2018zrz}. This factor is negligible compared to binary evolution uncertainties  (e.g. mass transfer, common envelope) which  typically lead to a factor of 10--100 uncertainty \citep{2012ApJ...759...52D,deMink:2015yea,Vigna-Gomez:2018dza}. As such, we do not use the total number of observable double neutron stars as a factor in determining our best-fit model.

\section{Gravitational Wave Population}
\label{sec:grav_waves}

Radio surveys of the DNS population are limited to studying only those DNSs which are in our Galaxy, and have at least one pulsar beaming towards the Earth. Current gravitational-wave observatories are observing merging DNSs in other galaxies, with very different selection effects to radio surveys. In this section, we make predictions for the population of DNSs observable in gravitational-waves. We choose to present this analysis for our best fit model FDM-20. Using this model calibrated to the Galactic DNS population, we analyse the entire evolved population that exists as a DNS system at the present time, i.e. that has not already merged, without applying radio selection effects. In this section the term `DNS' (Double Neutron Star) is to signify the entire population of neutron star-neutron star binaries, `non-radio' for the systems where both NSs in the binary have crossed the death-lines to the `graveyard' region and have thus ceased emitting in radio wavelength, and `radio' for the systems where at least one of the NSs emit in radio (hence, a pulsar), irrespective of whether it is observed by the survey telescope. In places we do use the post-radio-selection effect population to compare with the radio catalogue data-set, in which case we name this the `RadioSelection' population.

\subsection{Merger Time}
\label{subsec:merger_time}

\begin{figure*}
\includegraphics[width=0.49\textwidth]{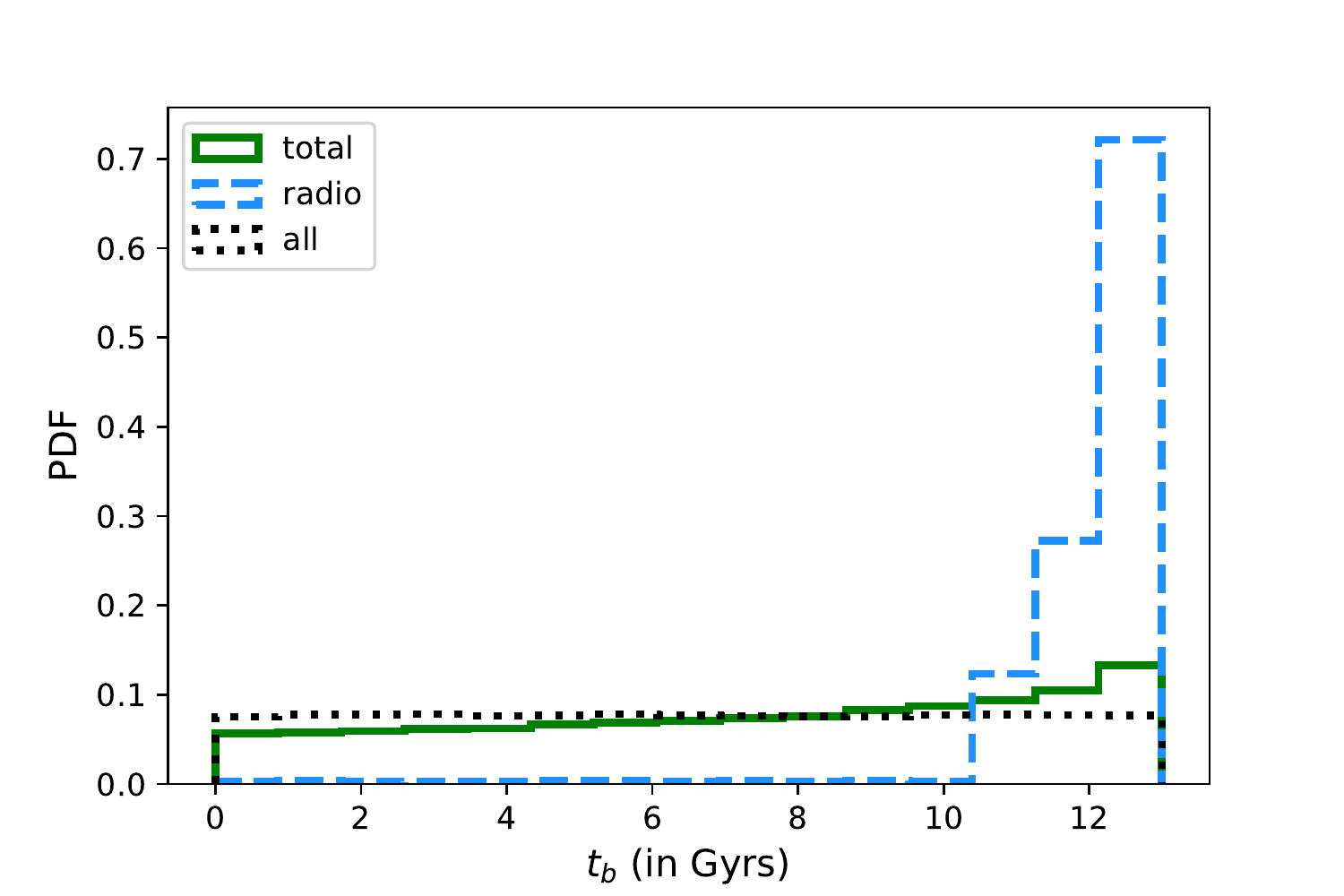}
\includegraphics[width=0.49\textwidth]{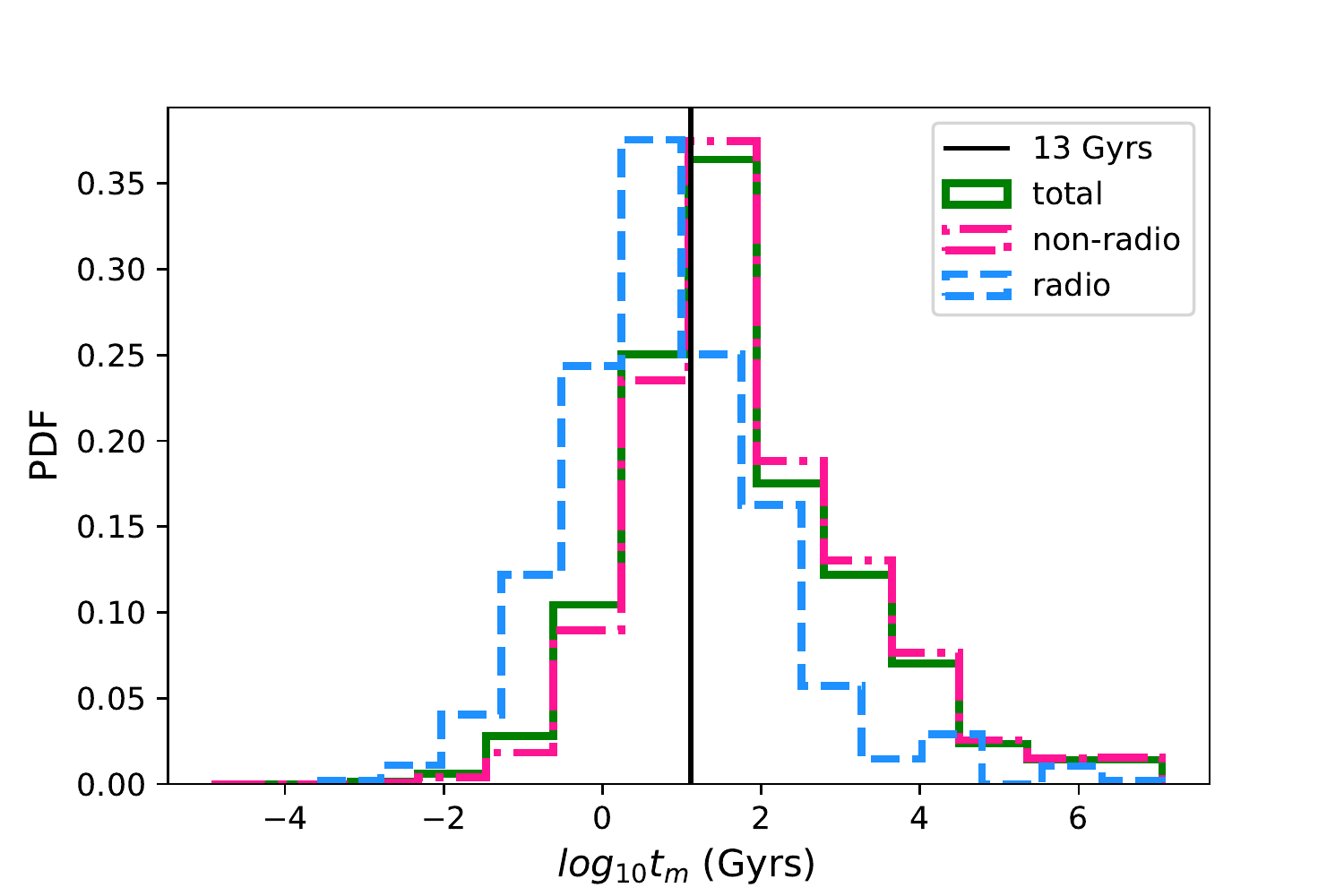}
\includegraphics[width=0.49\textwidth]{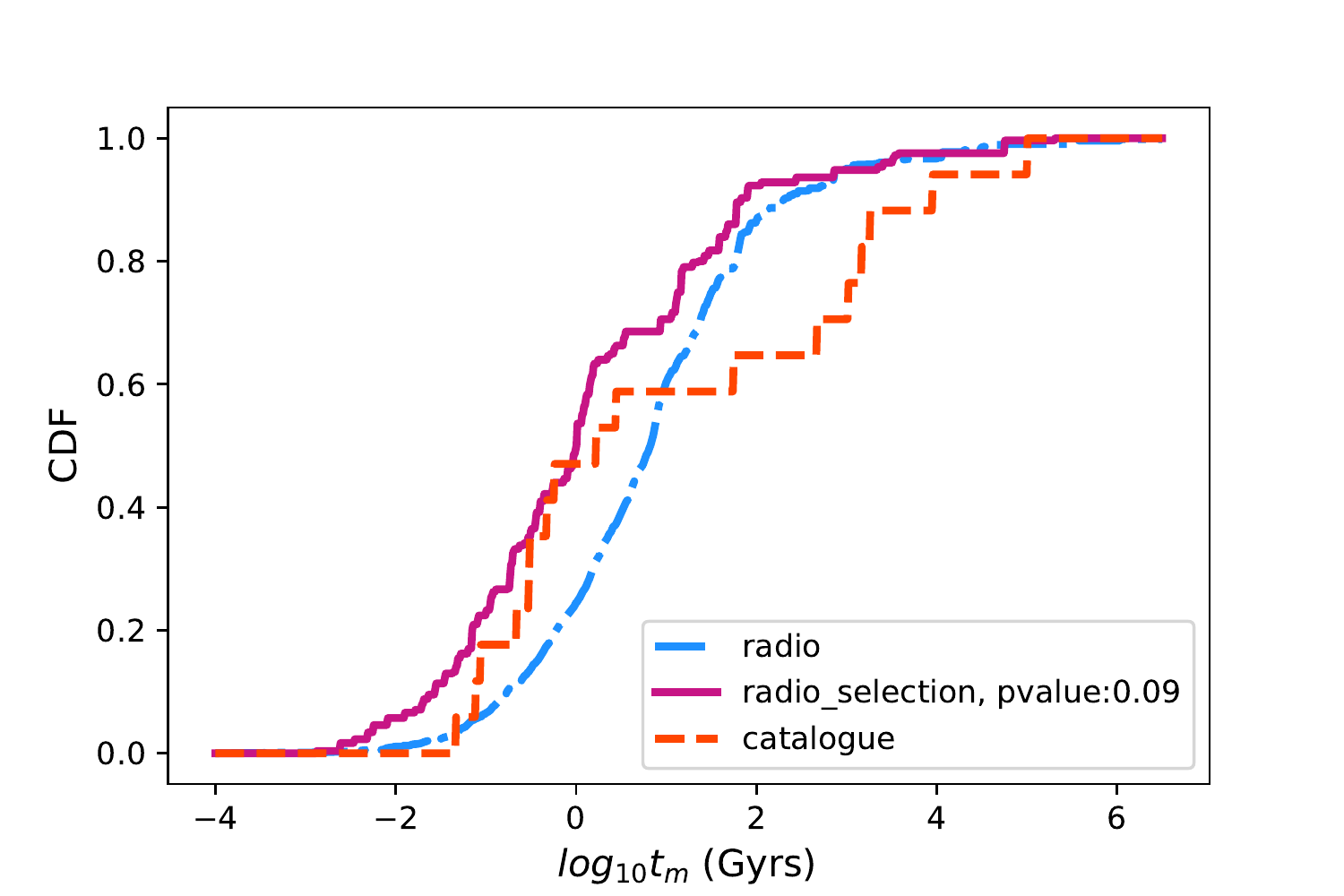}
\includegraphics[width=0.49\textwidth]{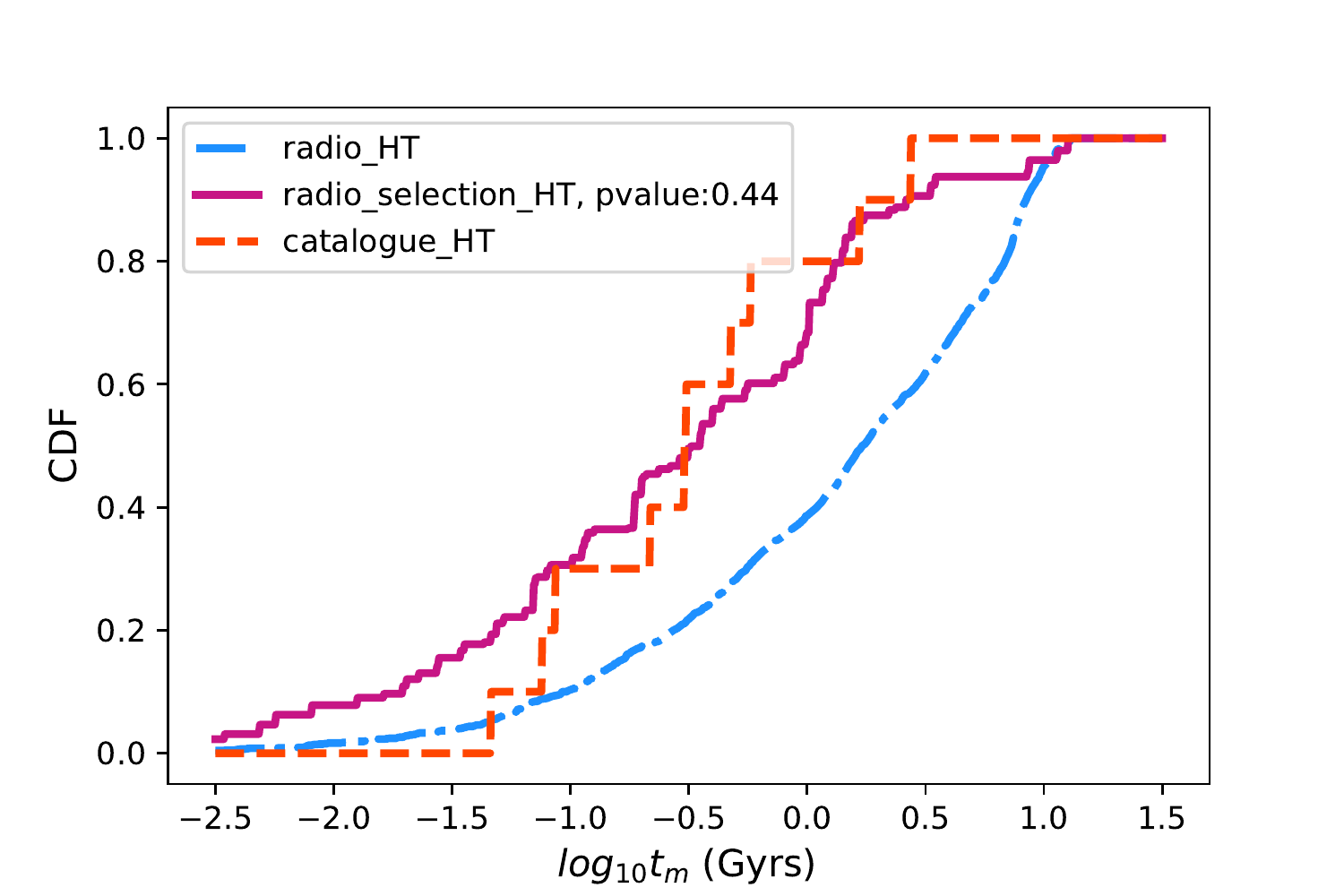}
\caption{The top left panel shows the normalized birth time distribution of the binaries. The black dotted line labeled `all' signifies the normalized distribution for all systems. The green line for `total' denotes the total number of DNS systems that exists as a DNS at the present time, while the blue broken line indicates the radio systems amongst the former. The top right panel shows the normalized distribution of $\log_\mathrm{10}t_\mathrm{m}$ for the total DNS systems, radio and non-radio. Though the median of the total and non-radio distributions are quite similar ($\approx 30$\,Gyr), the radio distribution median is a significantly lower value ($\lesssim 10$\,Gyr). The bottom left panel shows CDFs for $\log_\mathrm{10}t_\mathrm{m}$ of the radio and radio-selection distributions and compares the latter with the catalogue data-set. The bottom right panel is similar to the bottom left, but only for the sub-population that merges within a HT (i.e. have $t_\mathrm{m} < t_\mathrm{HT}$). 
}
\label{fig:TBirthAndTmerger}
\end{figure*}

We define birth time $t_\mathrm{b}$ as the time of DNS formation, i.e. the time of the second SN and the Hubble Time (HT) $t_\mathrm{HT}=13$\,Gyr. Approximating the age of the Milky Way to be a HT, we take $t_\mathrm{HT}$ as the present time. Although we assume a uniform star formation history for the Milky Way and thus a uniform $t_\mathrm{b}$, there is a bias for higher values of $t_\mathrm{b}$ for the systems selected for analysis. This feature of the $t_\mathrm{b}$ distribution is apparent in the top left normalized histogram of Fig.~\ref{fig:TBirthAndTmerger} - the `all' distribution denoting the uniform distribution of all DNSs that exist/ever existed in the galaxy, and the `total' distribution denoting the net DNSs that exist as a binary at $t_\mathrm{HT}$ (the current time) --- the latter showing a bias towards higher values of $t_\mathrm{b}$ and hence are younger systems. As apparent again from the top left plot of Fig.~\ref{fig:TBirthAndTmerger}, the bias of higher $t_\mathrm{b}$ is even stronger for the `radio' population, since at least one of the NSs in the radio systems is required to be on the `alive' side of the death-line. Thus the probability of radio emission is higher for younger non-recycled pulsars.

We define merger time $t_\mathrm{m}$ as the time required by the DNS to merge, calculated from the present time, due to the loss of energy as gravitational radiation \citep{Peters:1964}. If the binaries have $t_\mathrm{m}\leq t_\mathrm{HT}$, we state that it merges in a HT, if $t_\mathrm{m}> t_\mathrm{HT}$, we state that the binaries that take longer than a HT to merge, will not merge during the age of the universe. 

The total number of DNSs formed in the Milky Way up to the present day for model FDM-20 is $\approx 71,000$ (${\approx} 70,000$--$80,000$ for all models). Out of these, we find that ${\approx} 64$\% of DNS systems have already merged in the history of the Milky Way. Out of the $\approx36$\% ($=25638$, for Model FDM-20) that exist as DNSs at the current time, $\approx 34$\% will merge in a HT (i.e. $t_\mathrm{m}\leq t_\mathrm{HT}$). 

These fractions and the order of magnitude of the total numbers do not vary sufficiently within our suite of models. This is because the DNS formation and merger rates are independent of the pulsar parameters that we alter from model to model. Instead, the DNS formation rate depends on the initial properties of the binaries (e.g. their initial orbital separations and masses) and the treatment of uncertain binary evolution physics such as mass transfer stability and SN natal kicks \citep[see][for details]{Vigna-Gomez:2018dza}.

Gravitational-wave observables other than the DNS merger rate (such as the chirp mass and effective spin distributions, discussed in the following Sections~\ref{subsec:chirp_mass} and ~\ref{subsec:spin}) do vary between our models. We present results for model FDM-20 since it provides the best match to the Galactic DNS population.

We exclusively analyse the binaries that exist as DNSs at the present observation time, i.e. have undergone the second SN but have not merged. Only such existing DNS systems that form and merge within the age of the universe can be sources for gravitational wave detectors. Very old DNSs that have already merged at a time before the present time, are excluded from the data-set before further analysis. 

The $\log t_\mathrm{m}$ distribution for existing DNS systems is shown in the top right plot of Fig.~\ref{fig:TBirthAndTmerger}. While the total and non-radio populations show similar median values of $t_\mathrm{m}$ (33.1 Gyr and 39.8 Gyr respectively) the radio population has a significantly lower median of 6.6 Gyr. Since 90\% of the total population is non-radio, the net population and non-radio sub-population behaviours tend to be quite similar. The shorter median merger time of the radio population can be explained by the fact that radio systems also constitute recycled pulsar systems that went through CE and/or RLOF processes, that not only spin up the slow/dead pulsars, but also reduce orbital separation. Since the time taken for a system to merge just by the emission of gravitational waves is a function of the fourth power of the separation, reduction in the separation results in faster mergers. 

Amongst the Milky Way DNSs at the present time that merge in a HT, 19\% are radio systems. Out of those radio systems that merge in a HT, only 2\% are radio-selection systems. Out of all the radio-selection systems (that may or may not merge within a HT), we note that 0.64 of the fraction of the net population merge within a HT. The fraction of radio catalogued systems that merge within the age of the universe is 0.59. We note that 0.64 is within less than $ 10\% $ of 0.59, and thus our model is in agreement with the radio catalogue population.

The bottom left plot of Fig.~\ref{fig:TBirthAndTmerger} shows the CDF of $\log t_\mathrm{m}$ for the radio, radio-selection and catalogued radio DNSs. The bottom right plot of Fig.~\ref{fig:TBirthAndTmerger} shows the CDFs for the sub-populations of the radio, radio-selection and catalogued radio DNSs that merge in a HT. The $p$-value of $\log t_\mathrm{m}$ of the net radio-selection population compared to the catalogued radio DNSs is 0.09 and 0.44 for the systems that merge within a HT.

\subsection{Chirp Mass}
\label{subsec:chirp_mass}

\begin{figure}
\includegraphics[width=0.95\columnwidth]{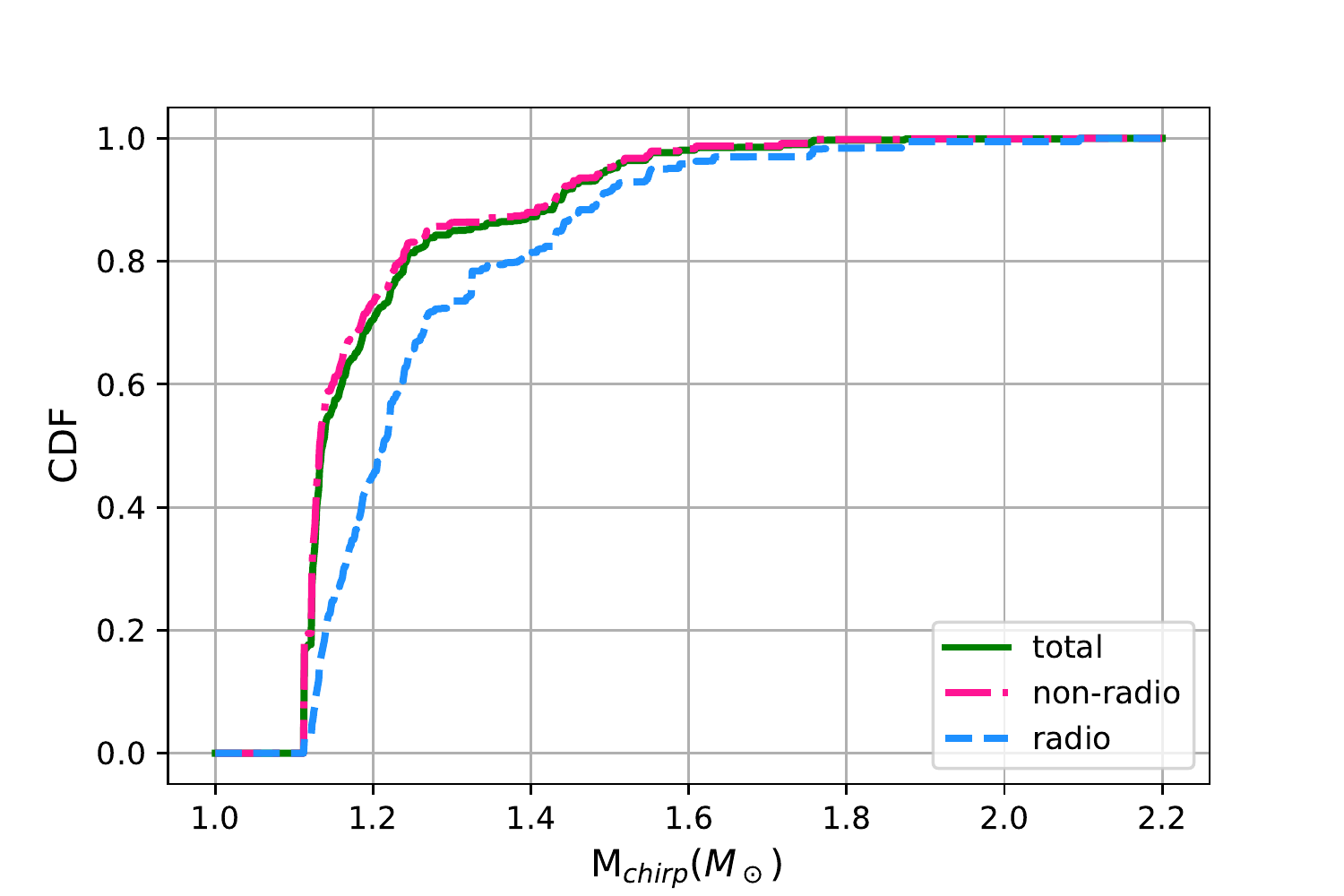}
\includegraphics[width=0.95\columnwidth]{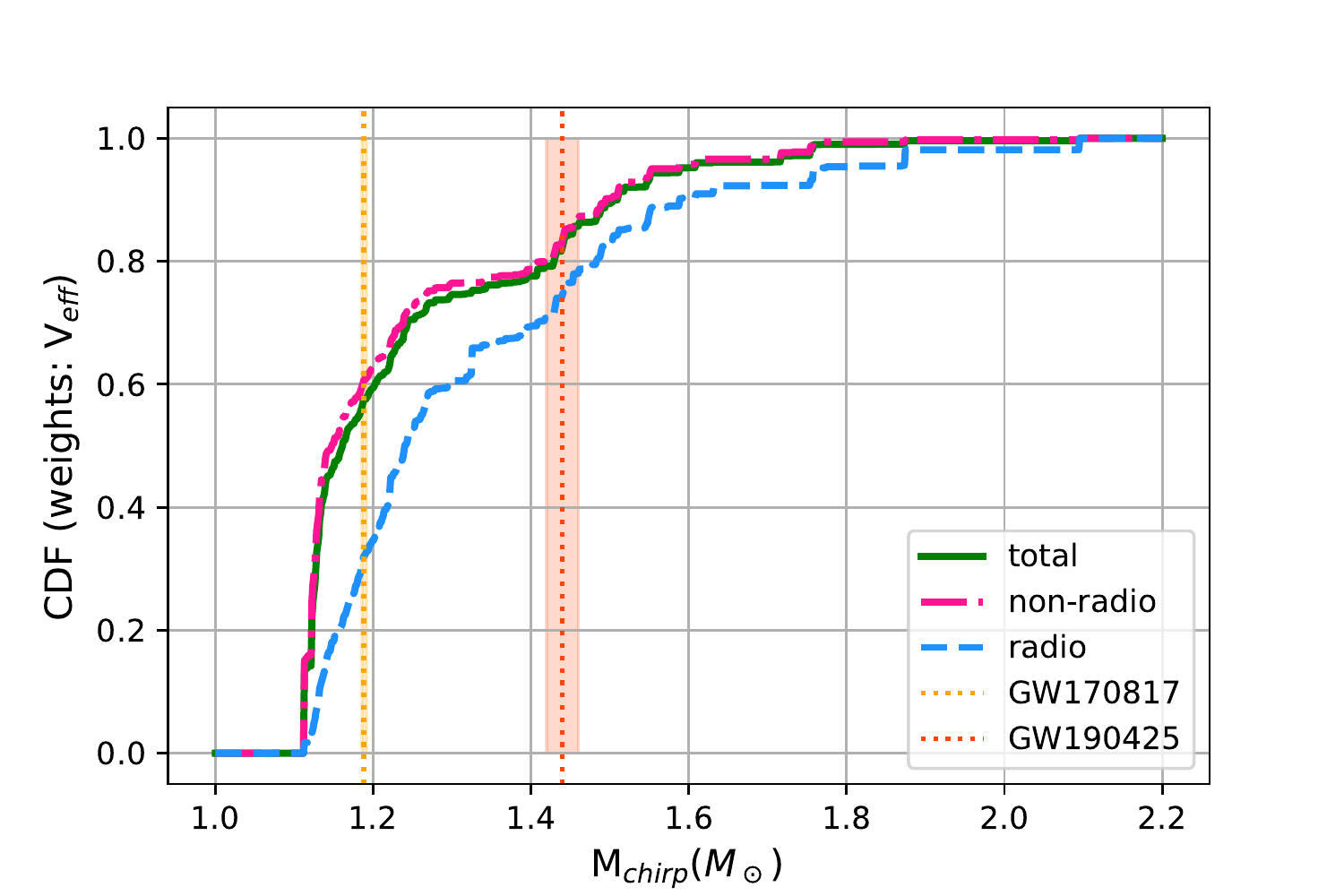}
\includegraphics[width=0.95\columnwidth]{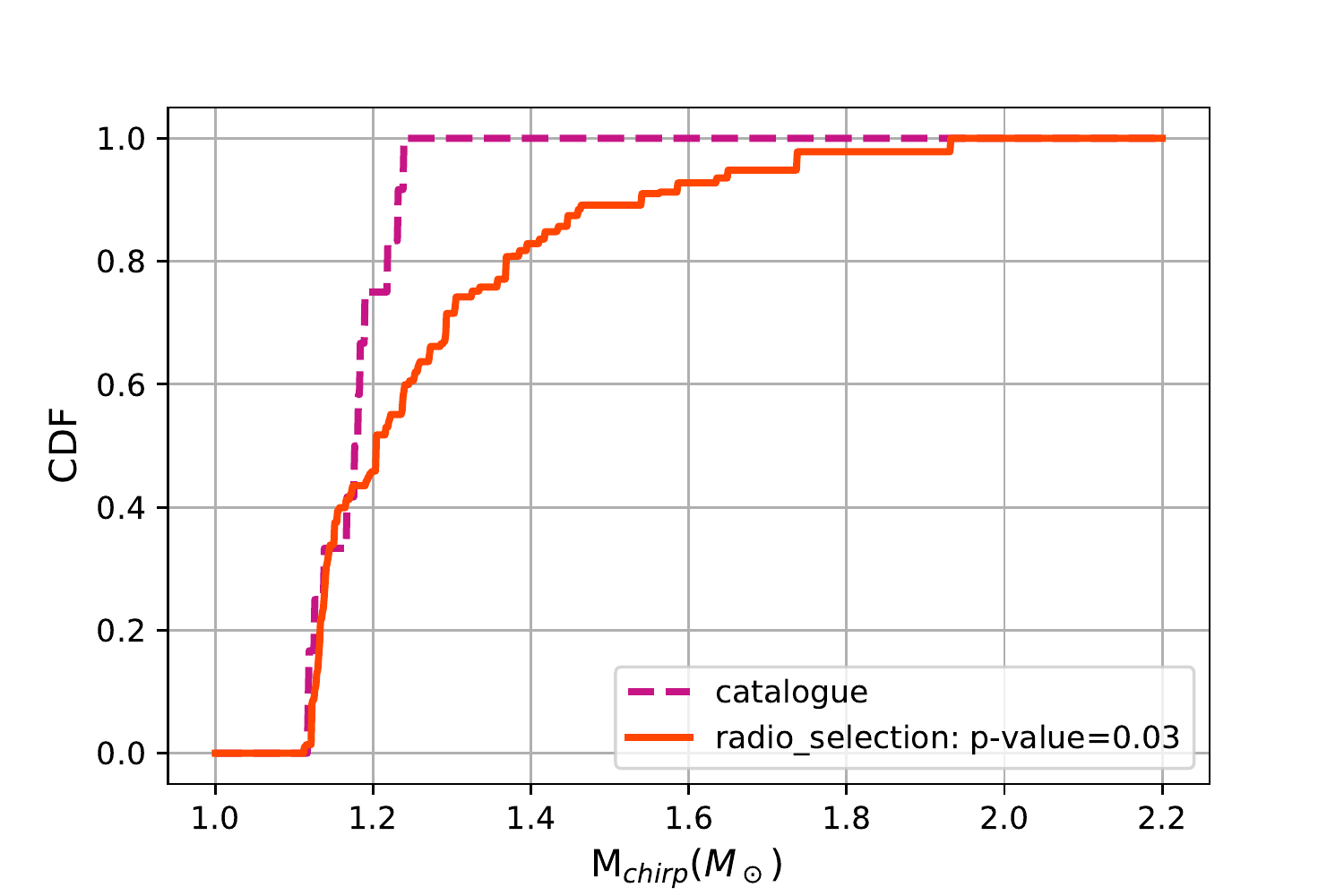}
\caption{The chirp mass distributions for model FDM-20. Shown are the normalized CDFs for the non-radio, radio and the entire population of DNSs (top) and the normalized CDFs for the same populations weighted by effective volume (middle). The chirp masses of GW170817 (yellow) and GW190425 (orange) with their respective confidence intervals are also shown in the latter. Comparison between the chirp mass distributions of the RadioSelection sub-population (after accounting for the radio selection effects) and the radio catalogue data-set are also shown (bottom).}
\label{fig:ChirpMass}
\end{figure}

Energy loss in a compact binary system due to gravitational quadrupole radiation gives a mass dependent term, called the `Chirp Mass' in the expression for the largest order-of-magnitude contribution to the mathematical expression for orbital decay. Chirp mass determines the frequency of the gravitational wave detected, and is one of the well-measured parameters by the detectors. If the masses of the two compact objects in the binary are given by $M_\mathrm{1}$ and $M_\mathrm{2}$, the chirp mass $M_\mathrm{chirp}$ is given by
\begin{equation}
    M_\mathrm{chirp} = \frac{(M_\mathrm{1}M_\mathrm{2})^{3/5}}{(M_\mathrm{1}+M_\mathrm{2})^{1/5}} \,  .
    \label{eq:chirp_mass}
\end{equation}

There remains an observational bias in the chirp mass distribution of the detected compact binary mergers. More massive system mergers can be detected further away. If $d$ is the distance from the location of a compact object merger to the detector, the signal-to-noise ratio of aLIGO and aVirgo can be expressed in terms of chirp mass as
\begin{equation}
    \frac{S}{N} \propto M_\mathrm{chirp}^{5/6}\times\frac{1}{d} .
\end{equation}
Since the 3-D volume \(V_\mathrm{d}\) enclosed by a radius $ d $, is \(d^3\), the chirp mass distribution histograms are weighted by
\begin{equation}
    V_\mathrm{eff} \propto M_\mathrm{chirp}^{5/2} .
    \label{Veff}
\end{equation}

Fig.~\ref{fig:ChirpMass} shows the CDF of chirp mass distributions for model FDM-20. The top plot of Fig.~\ref{fig:ChirpMass} shows the intrinsic CDF, whilst the middle plot shows the distribution weighted by gravitational-wave selection effects (according to Equation~\ref{Veff}). Both panels distinguish between the radio and non-radio sub-populations. The bottom panel of Fig.~\ref{fig:ChirpMass} compares the chirp mass distribution of the observed Galactic DNS population to our model FDM-20 after applying radio selection effects.

The $M_\mathrm{chirp}$ distribution of the non-radio DNSs is similar to the total population, since only $10$\% of DNSs are in principle radio systems. Both the total and non-radio populations have about $60$\% of the population with $M_\mathrm{chirp}\leq1.2$. Only $\approx 0.1$\% of the total population are observed after selection effects for model FDM-20. 

More than about 80$\%$ of NSs in our model are in the mass range 1.3--1.4\,$M_\odot$. Hence for systems with equal mass NSs, and the said mass range, a majority (about 70\% for the net DNS population, see Fig.~\ref{fig:ChirpMass} top plot) of the DNSs have 1.1$<M_\mathrm{chirp}/M_\odot\leq$1.2. For reference, GW170817 and GW190425 were detected to have chirp masses of 1.188\,M$_\odot$ \citep{TheLIGOScientific:2017qsa} and 1.44\,M$_\odot$ \citep{Abbott:2020uma} respectively. 

On average, we find that radio DNSs are around $0.1$\,M$_\odot$ more massive than non-radio DNSs. This is because the first born neutron star in these systems has accreted more mass than in the non-radio DNSs, recycling them to shorter spin periods and lower magnetic fields, increasing the probability of them remaining radio loud pulsars and increasing their detection probability.

Our model predicts a range of DNS chirp masses from $\approx$1.1M$_\odot$ to $\approx$2.1M$_\odot$ (see also the discussion in \citealp{Vigna-Gomez:2018dza}). The upper limit is larger than most massive DNSs observed in radio surveys \citep{Farrow:2019}. Fig.~\ref{fig:ChirpMass} can be used to infer the proportional abundance of different chirp mass ranges in the model.


The bottom plot of Fig.~\ref{fig:ChirpMass}  shows the chirp mass CDF of radio-selection sub-population and compares that to the radio catalogued population. The radio catalogue data-set shows about 70\% of its population to have $M_\mathrm{chirp} \leq 1.2$, while the modelled radio selection population shows about 50\% of it to similar $M_\mathrm{chirp}$ range. The $p$-value obtained is 0.03. 


\subsection{Spin}
\label{subsec:spin}

The magnitude of the spin of a compact object can be expressed as a `dimensionless spin' parameter
\begin{equation}
    \chi = \frac{cJ}{GM^2} = 2 \pi \frac{c I}{G P M^2}
    \label{chi}
\end{equation}
where, $J$ is the angular momentum of the object, $I$ is the moment of inertia, $P$ is the spin period and $M$ is the mass of the object. For a black hole, $0 \leq \chi \leq 1$.

Since pulsars spin down over time (equation~\ref{SpinDownIsolated}), $\chi$ decreases as a function of time. We calculate the distribution of $\chi$ of the DNS population at the present time, which allows us to compare the same with the radio catalogued data-set.

\begin{figure*}
\includegraphics[width=0.49\textwidth]{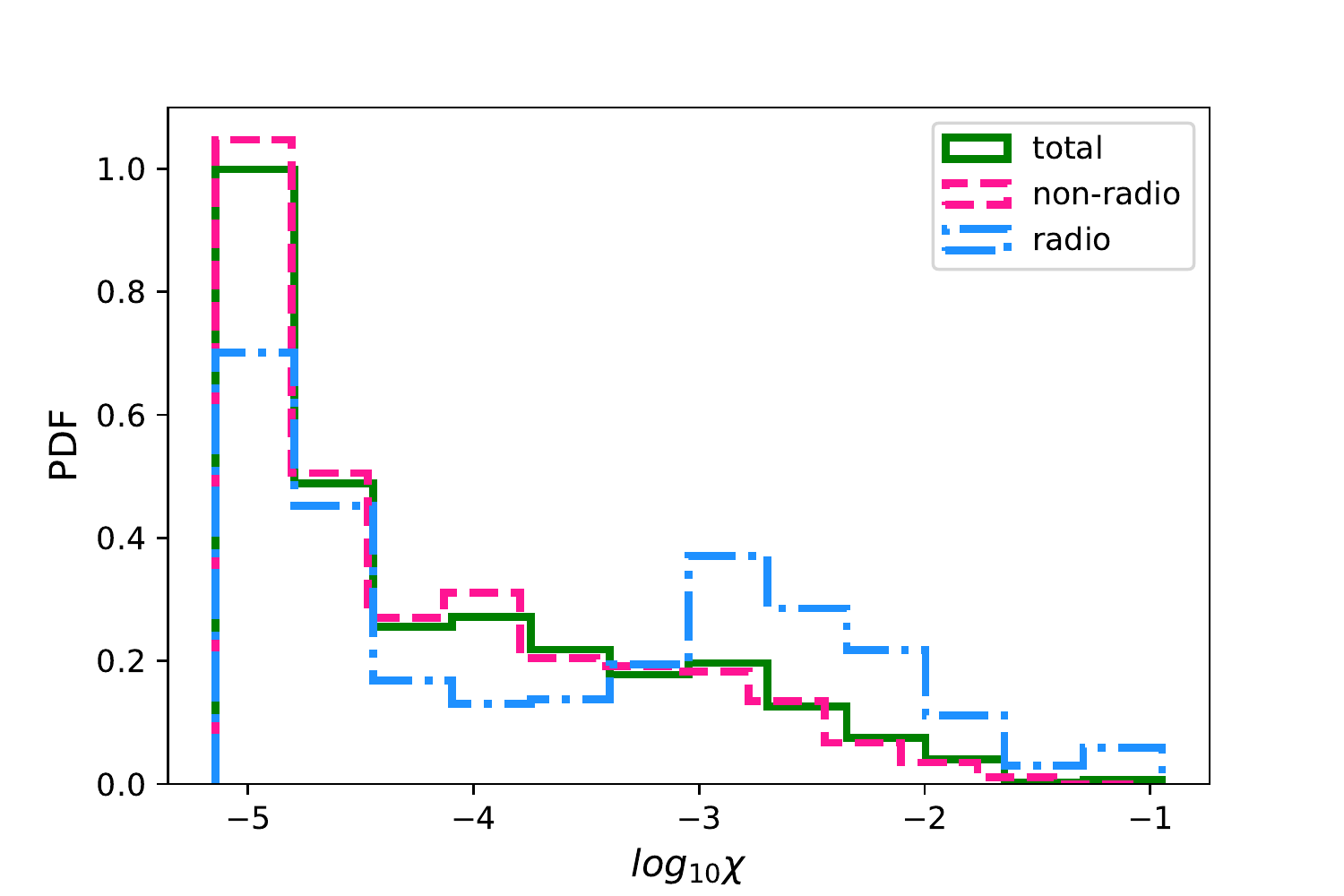}
\includegraphics[width=0.49\textwidth]{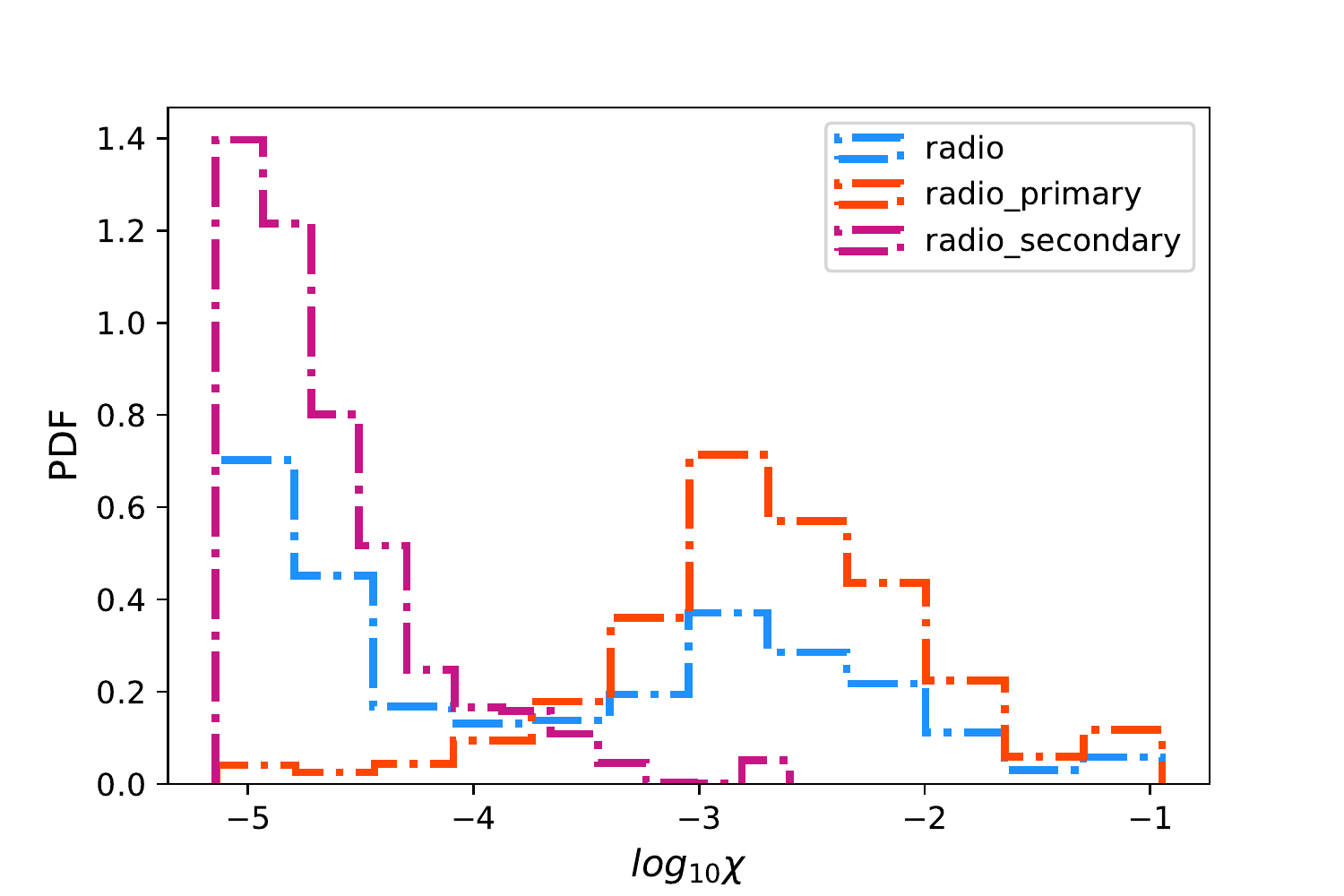}
\includegraphics[width=0.49\textwidth]{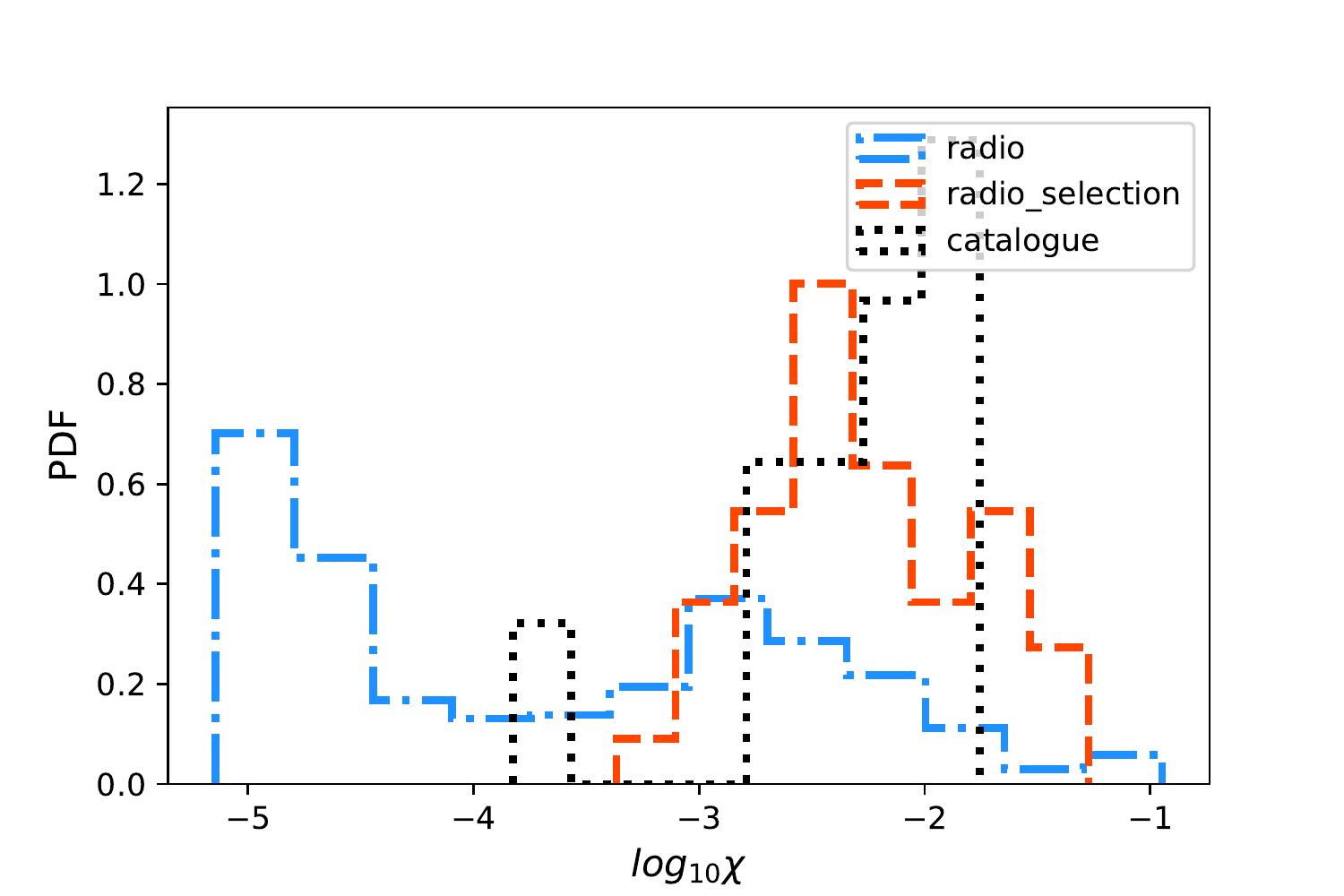}
\includegraphics[width=0.49\textwidth]{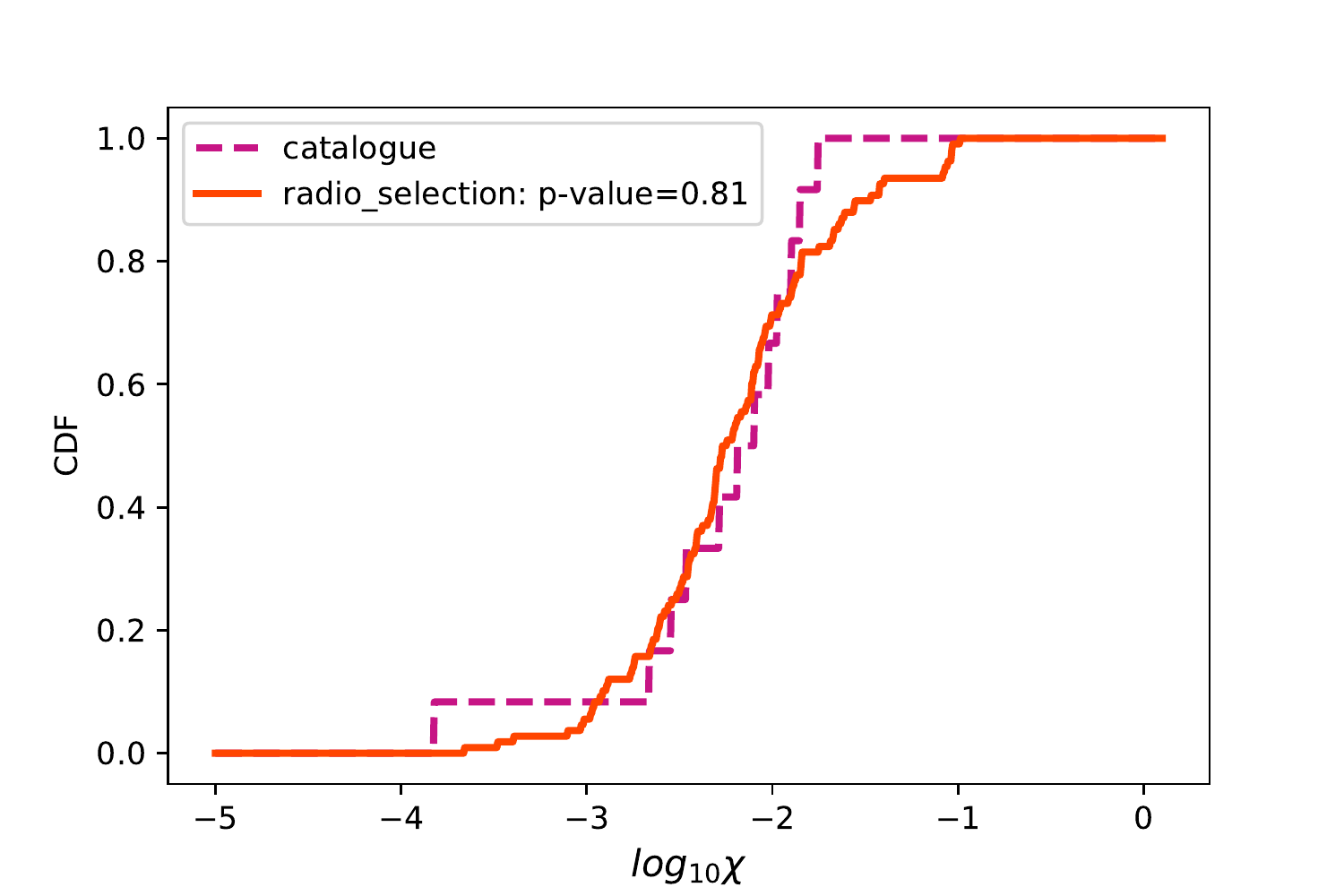}
\caption{The distributions of pulsar dimensionless spins $\chi$ for model FDM-20 measured at the current time. The top left panel shows the normalized histograms for the non-radio, radio and the entire populations. There is a bump in the blue line of the radio population around $\log_{10} \chi \approx -3$ due to recycled pulsars. The top right panel shows the radio population segregated into individual primary and secondary populations. The primaries, being constituted of the spun up recycled pulsars show higher values of $\log_{10}\chi$, which is mimicked in the entire radio population. In the lower left panel we show the spins for the same radio population, before and after accounting for the radio selection effects, as well as the comparison to the radio catalogue systems. The lower right panel shows the CDF of $\log_{10}\chi$ that gives a $p$-value of 0.81.}
\label{fig:Spins}
\end{figure*}

\begin{figure}
\includegraphics[width=0.99\columnwidth]{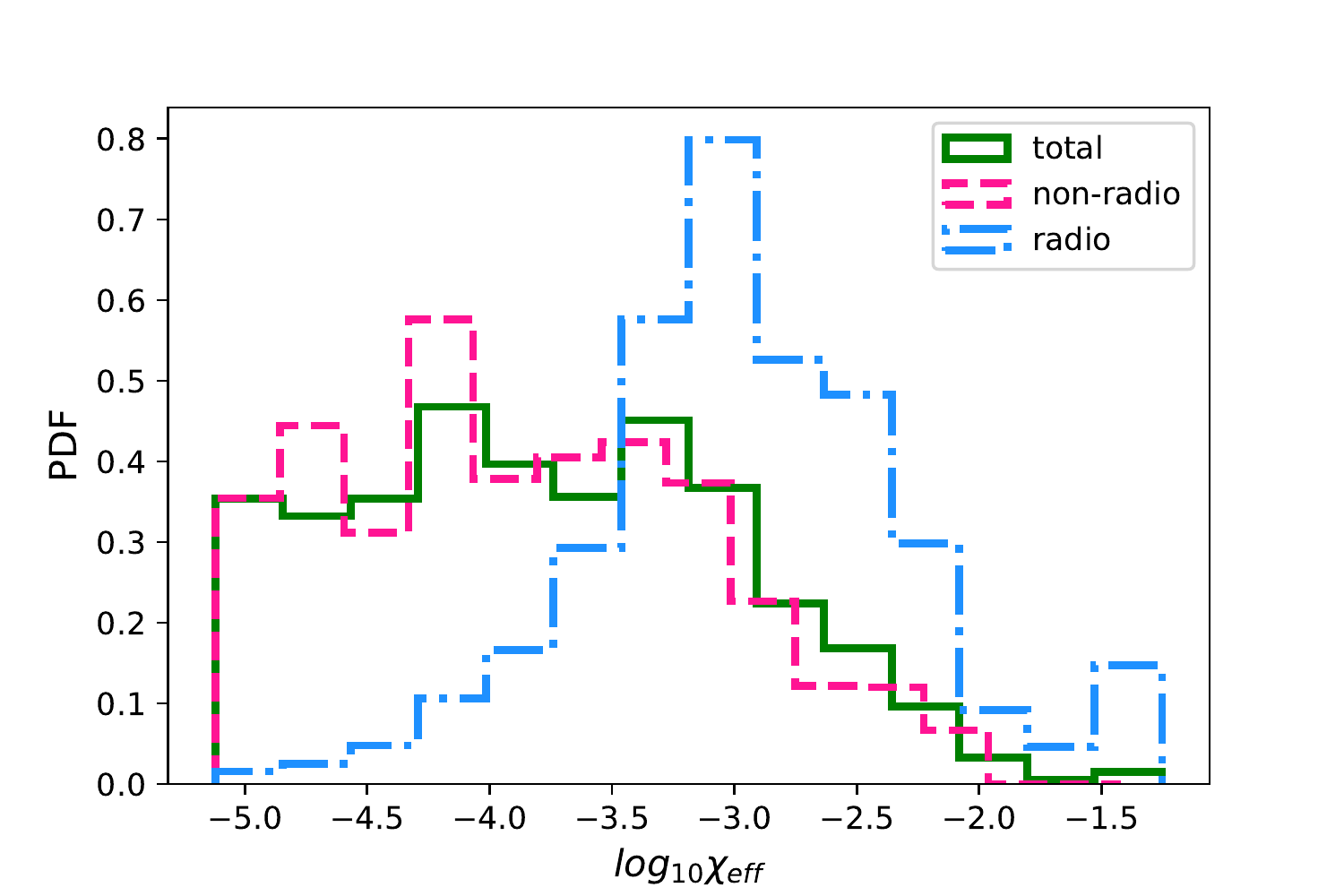}
\caption{Distribution of the effective spin parameter $\chi_\mathrm{eff}$ for DNSs measured at the time of merger. The trend of higher individual spin values within the radio population seen in Fig.~\ref{fig:Spins} is propagated to the effective spins as well.}
\label{fig:chieff}
\end{figure}

In the top left panel of Fig.~\ref{fig:Spins} we show the total distribution of $\log_{10} \chi$, separating the sub-populations of radio and non-radio DNSs as described at the beginning of Section~\ref{sec:grav_waves}. The bump for the radio population at $\log_{10} \chi \approx -3$ is caused by the spinning up of recycled pulsars which accrete matter from their companion and gain angular momentum $J$, resulting in a higher value of $\chi$ (c.f. Equation~\ref{chi}). This is evident in the top right panel of Fig.~\ref{fig:Spins} where we plot the distribution of $\log_{10} \chi$ for primaries and secondaries that constitute the radio systems. Since all recycled pulsars are primaries (though not all primaries are recycled), their influence causes the primaries to show a stronger bias towards higher values of $\log_{10} \chi \approx -3$. 

Since the `radio' population is a composite of the `radio-primary' and `radio-secondary' sub-populations, it inherits the same bump in the distribution. We also show the $\log_{10}\chi$ values after radio selection effects and compare it to the radio catalogue data set (Fig.~\ref{fig:Spins}, bottom left). We observe a similar build up of recycled pulsars with high values of $\log_{10} \chi \approx -3$ for the radio catalogue data set. 

All the $\log_{10}\chi$ plots are computed with the spins of the DNSs as at the current observation time. This allows us to compare our model predictions to the radio catalogue of DNSs. Comparing the $\log_{10}\chi$ distribution of our model FDM-20 after accounting for the radio selection effects to the radio catalogue data-set, we obtain a $p$-value of 0.81; the corresponding CDFs are shown in the bottom right plot of Fig.~\ref{fig:Spins}. 

We also observe from the bottom plots of Fig.~\ref{fig:Spins} the radio-selection population not predicting very low spins ($\log_\mathrm{10}\chi\approx$-3.8, the data-point in `catalogue' is from pulsar-B J0737--3039B of the double pulsar system). Although from the same Fig.~\ref{fig:Spins} the radio population is showing such smaller spin systems, radio-selection effects are removing them. This is because the radio efficiency $\xi$ cut-off we selected in Section~\ref{subsec:pulsar_death} makes it less-likely to find populations at the first death line (equation~\ref{eq:deathline1}), though not impossible. A higher value of $\xi_\mathrm{max}$ solves this. It is therefore the uncertainty in when the radio emission mechanism of pulsars shuts down that creates the issue.

Similar to $M_\mathrm{chirp}$, another well-measured quantity from the observed gravitational wave data is the `effective spin' parameter of the merging compact objects, $\chi_\mathrm{eff}$, the projection of the mass weighted individual spins of the merging binary. If the individual spins of the binary stars are expressed as $\chi_\mathrm{1}$ and $\chi_\mathrm{2}$, $\chi_\mathrm{eff}$ is given by \citep{Ajith:2009bn, Cutler:1992tc}
\begin{equation}
    \chi_\mathrm{eff}=\frac{m_\mathrm{1}\chi_\mathrm{1}\cos\theta_\mathrm{1}+m_\mathrm{2}\chi_\mathrm{2}\cos\theta_\mathrm{2}}{m_\mathrm{1}+m_\mathrm{2}} ,
    \label{chiEff}
\end{equation}
where $m_\mathrm{1,2}$ are the masses of the two objects in the binary and $\cos\theta_\mathrm{1,2}$ are the angles subtended by their respective spins on the orbital angular momentum unit vector. We assume aligned NS spins $\cos\theta_{1,2} = 1$.  The effective spin $\chi_\mathrm{eff}$ is a constant of motion at least to the second order of the post-Newtonian terms \citep{Blanchet:2013haa}. Given that $\chi_{1,2}$ both decrease with time for NSs, so does $\chi_\mathrm{eff}$.

Both aLIGO and aVIRGO can only measure the $\chi_\mathrm{eff}$ when the compact binary is merging, thus we evolve our DNS systems past the current time and plot the $\log_{10}\chi_\mathrm{eff}$ distribution of the DNSs at their respective merger times. This gives more insight to the distribution that the gravitational wave detectors will observe. 

We show the PDF of $\log_{10}\chi_\mathrm{eff}$ for our DNS population in Fig.~\ref{fig:chieff}. The bump of the radio distribution around about -3.5 to -2.0 is due to recycled pulsars as explained for individual $\chi$ values (see Fig.~\ref{fig:Spins}). We find that DNSs are expected to have $\chi_\mathrm{eff} < 0.03$ at merger. Our models are in good agreement with both the Galactic radio population, the $\chi_\mathrm{eff}$ value inferred for GW170817 whose 90 percent confidence interval lies between $-0.01$ and $0.02$ \citep{TheLIGOScientific:2017qsa} and for GW190425 the effective spin is deduced to be between $-0.008$ and $0.012$ with 90 percent credibility \citep{Abbott:2020uma}. 

Binaries with $\chi_\mathrm{eff} > 0$ produce a gravitational wave merger signal that is long-lived and of higher frequency, while those with  $\chi_\mathrm{eff} < 0$ result in a comparatively shorter lived, lower-frequency gravitational wave \citep{Zhu:2017znf}. For two systems with the exact same masses and magnitude of spins but different spin alignment angles, the one with spin orientation $\chi_\mathrm{eff} > 0$ is more likely to be observed than the one with $\chi_\mathrm{eff} < 0$ \citep{Ng:2018neg}, leading to a selection bias. However this selection bias is expected to be small ($<10$\%, \citealp{Ng:2018neg}) and we neglect it here. 

More observations of gravitational-waves from DNS mergers will make it possible to obtain a distribution of $\chi_\mathrm{eff}$. This will allow constraints on the magnetic field decay timescale, spin misalignment angle distribution and neutron star equation of state to be drawn \citep{2017Natur.548..426F, Zhu:2017znf}. 

\subsection{Post-merger remnant}
\label{subsec:post_merger_remnant}

\begin{figure} 
\includegraphics[width=\columnwidth]{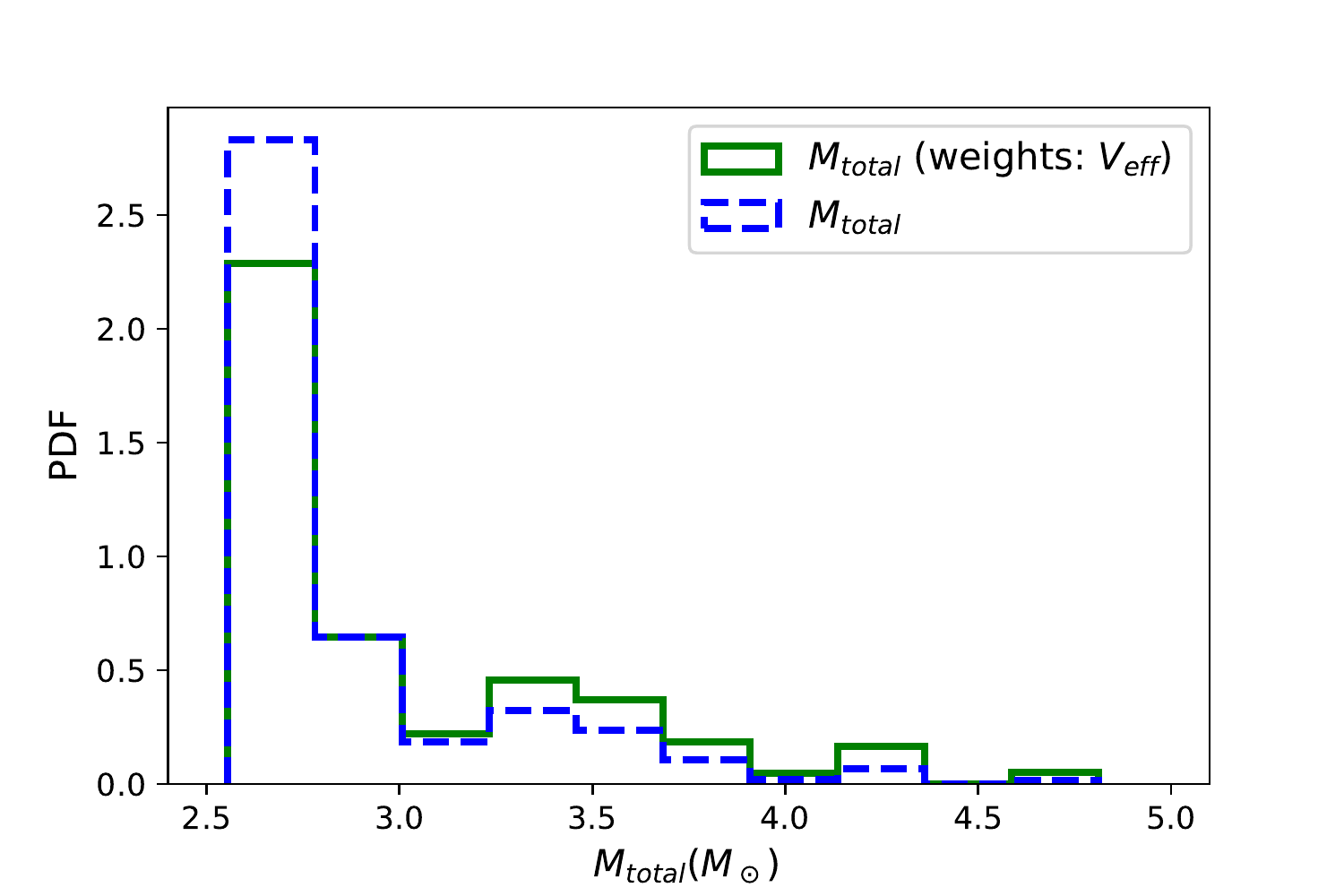}
\includegraphics[width=\columnwidth]{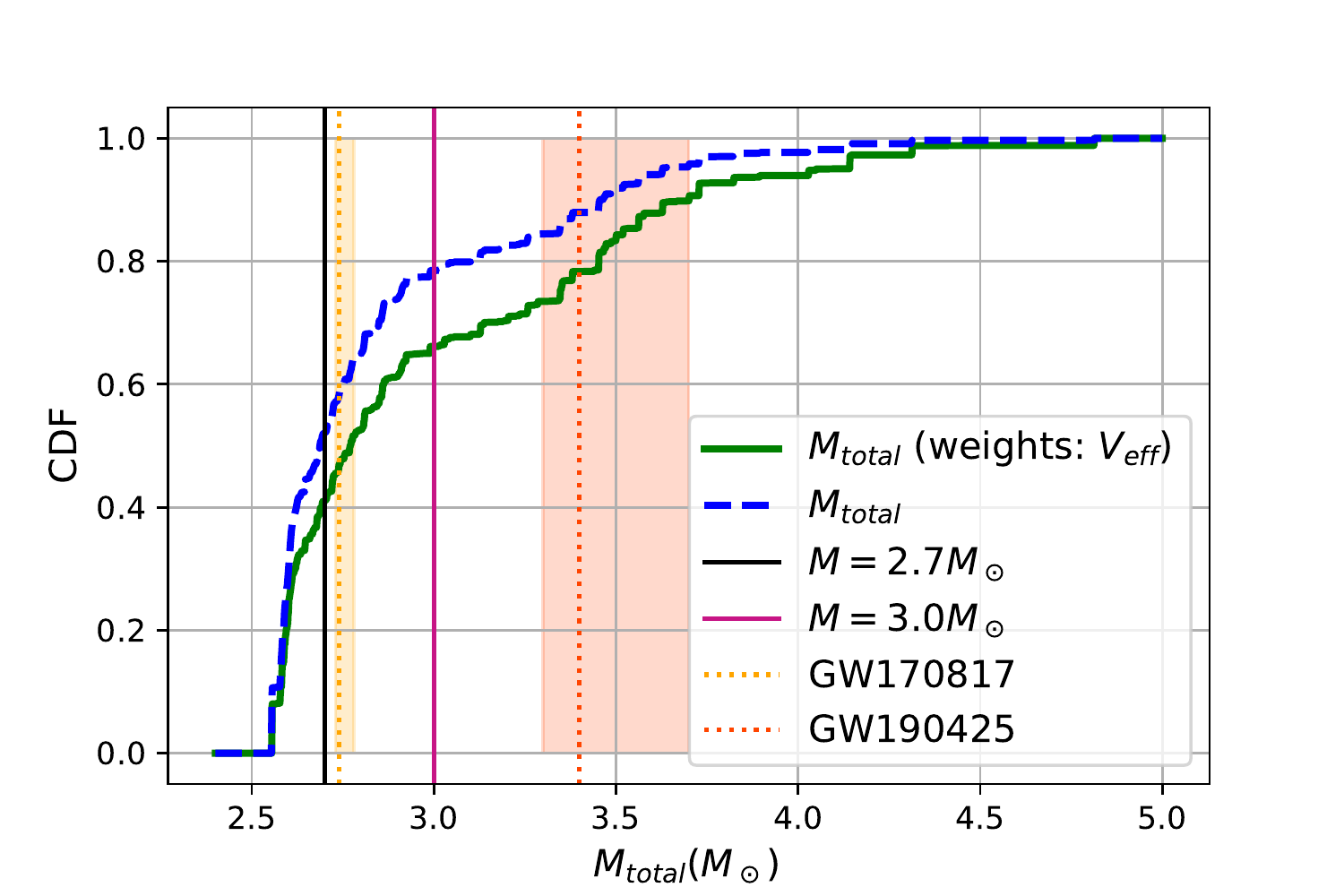}
\caption{Distribution of the total mass $M_\mathrm{tot}$ of double neutron stars (that merge in the next 13 Gyr) - unweighted (blue broken line) and weighted by effective volume (green line) in the PDF (top) and the CDF (bottom). The black and the magenta lines in the bottom panel mark the masses 2.7$M_\mathrm{\odot}$ and 3.0$M_\mathrm{\odot}$ respectively. The total masses for gravitational wave DNS observations GW170817 (yellow) and GW190425 (orange) are also shown with their upper and lower limits.}
\label{fig:totalMass}
\end{figure}

The remnant of a DNS merger can either be a (stable, supermassive or hypermassive) NS or a black hole. If the post merger remnant is a NS, then it may emit gravitational-waves observable by aLIGO and Virgo. There have been searches for gravitational waves from the remnant of GW170817, though no further GWs were detected \citep{Abbott:2017dke}. The post-merger remnant may be a rapidly rotating NS with a high magnetic field (a magnetar), visible in X-ray and gamma rays \citep[e.g.][]{Xue:2019nlf}. In addition, DNS post merger remnants may also be a source of Fast Radio Bursts \citep[FRBs, e.g.][]{Wang:2016dgs, Yamasaki:2018, 2019arXiv190700016M}. 

The fate of the DNS merger remnant depends on the currently unknown maximum NS mass \citep[see e.g.][]{2019ApJ...880L..15M}. Remnants more massive than this collapse to a black hole. The most massive known NS detected in radio is $2.14$\,M$_\odot$ \citep{Cromartie:2019kug}. Constraints from causality place an upper limit on the maximum NS mass of $\lesssim 3$\,M$_\odot$ \citep[][]{Kalogera:1996}. Assuming the merger remnant of GW170817 formed a short lived hypermassive neutron star that then collapsed to a black hole 
\footnote{Though some late time observations point to a long lived remnant \citep{Yu:2018, Piro:2019}.}, the maximum NS mass can be constrained to $\lesssim 2.3$\,M$_\odot$ \citep[e.g.][]{Margalit:2017dij,Ruiz:2017due,Rezzolla:2017aly,Shibata:2019ctb}.

From our model FDM-20, we plot the CDF of the total mass of the DNS systems that merge within a HT. The total binary mass for GW170817 was $\approx 2.7$\,M$_\odot$ \citep{TheLIGOScientific:2017qsa,Abbott:2018wiz} , while for GW190425 was $\approx 3.4$\,M$_\odot$ \citep{Abbott:2020uma}. Although the post-merger remnant mass is lower than the total binary mass, the mass loss is small compared to the total ($\approx 0.05$\,M$_\mathrm{\odot}$ for GW170817 \citealp[e.g.][]{Smartt:2017fuw}). We hence approximate the total binary system mass as the remnant mass. Fig.~\ref{fig:totalMass} thus represents the unweighted and weighted CDF distribution of total/remnant mass. We obtain that about 40\% of the merging DNSs observable by ground based GW detectors will have a remnant mass $\leq 2.7$\,M$_\mathrm{\odot}$ (the same for the original un-weighted population is about 50\%). The weighted population has $<$70\% DNSs with total mass $\leq 3$\,M$_\mathrm{\odot}$, while the unweighted population shows about 80\% for the same. 

\section{Summary}
\label{sec:discussion_conclusion}

In this paper we have modelled the evolution of double neutron star (DNS) systems, focusing on isolated pulsar-NS/pulsar-pulsar binaries, using the population synthesis code COMPAS. DNSs are one of the most interesting astrophysical systems in both electromagnetic and gravitational-wave astronomy. 

We have implemented the canonical magnetic dipole model for pulsar evolution within COMPAS. Our model contains several free parameters. By varying one parameter in our model at a time, we explored the parameter space to understand the effect of different pulsar evolution prescription parameters on the resultant DNS population. We compare our models to the catalogue DNSs observed in the radio (Table~\ref{tab:observed_DNSs}) accounting for radio observation selection effects. This analysis identified a `best-fit' model which can simultaneously reproduce the distributions of $P$, $\dot{P}$, $P_\mathrm{orb}$ and $Z$ for radio DNSs in the Galactic field. However, preliminary investigations show that there are strong degeneracies between many model parameters (e.g. $\Delta M_d$, $\tau_d$) that our simple exploration is unable to map out. A more detailed study could employ Bayesian parameter estimation (e.g. using stochastic sampling) to compare the observed DNS population to our models, fully exploring the parameter space (see e.g. \citealp{2019MNRAS.486.1086M}). In this study we have not varied the uncertain binary evolution parameters relating to e.g. the stability of mass transfer and the efficiency of common envelope evolution (see \citealp{Vigna-Gomez:2018dza} for an investigation using COMPAS). Future work should explore possible degeneracies between pulsar parameters and binary evolution parameters. 

We have compared our models to the population of field Galactic DNSs, excluding those which are found in globular clusters, since dynamics may play a non-trivial role in their formation. Our working hypothesis is that all field Galactic DNSs formed through isolated binary evolution. However, it is still possible that some of the DNSs in the observed sample were formed through an alternative evolutionary channel \citep[e.g.][]{2019ApJ...883...23H,Andrews:2019vou}.

Assuming a uniform star formation history and the present age of the Milky-Way to be 13 Gyr, our model predicts ${\approx} 25000$ DNSs exist in the Milky Way at present, of which ${\approx}$10\% are either pulsar-NS/pulsar-pulsar systems. 

There are large uncertainties in the exact physical process of magnetic field burial during mass accretion onto a pulsar. Here, we have assumed a phenomenological model in which the magnetic field decays exponentially with accreted mass. We find that in our models, mass accretion during common envelope evolution plays an important role in recycling a pulsar and producing the resultant radio DNS population. The importance of pulsar parameters is more apparent from Table~\ref{tab:rates}, which shows that there can be an order-of-magnitude variation in the total number of pulsar-survey predicted detections by only altering one parameter.

We have shown that after accounting for radio selection effects, the observed distributions of pulsar parameters (spin period and period derivative) are strongly biased with respect to the underlying population. In contrast to this, we find that selection effects are unimportant for the binary properties such as the orbital period and eccentricity. This means that the observed distributions of orbital periods and eccentricities of DNSs are representative of the intrinsic population. This insight is useful when comparing to models of DNS formation.

We find that our models produce typical DNS eccentricities which are too high compared to the observed Galactic DNS population (as noted previously by \citealp{Kiel:2010MNRAS,2017AcA....67...37C,Vigna-Gomez:2018dza}). We argued that this is due to our treatment of ultra-stripped helium stars, which leave behind more massive CO cores than detailed simulations suggest  \citep[e.g.][]{Tauris:2015,Vigna-Gomez:2018dza}. This causes these stars to lose too much mass during the ultra-stripped SN, resulting in a large Blaauw kick that leads to a high orbital eccentricity. Future work should incorporate better models of ultra-stripped SN progenitors \citep[e.g.][]{Tauris:2015,2018MNRAS.481.1908K,2019arXiv190611299Z}. One caveat to this conclusion is that we did not model radio selection effects for highly accelerated binaries with large eccentricities \citep[see][]{Bagchi:2013wga}.

Our analysis uses a one-dimensional KS test to compare our models to the observations. This is a simple test commonly used in the pulsar literature \citep[e.g.][]{FaucherGiguere:2005ny,Szary:2014dia}. However, this test does not allow us to account for correlations between multiple DNS parameters (e.g. $P_\mathrm{orb}$--$e$). Some authors have performed two-dimensional analyses \citep[e.g.][]{2015ApJ...801...32A,Vigna-Gomez:2018dza}, but model comparison in the full eight-dimensional parameter space is currently computationally challenging.

DNSs are rare (we have $\sim$500--1000 unique DNSs per model), resulting in minor statistical fluctuations. Evolving a larger ensemble of binaries in order to produce a greater number of modelled DNSs would reduce these statistical fluctuations but becomes computationally expensive. However, employing sampling algorithms such as STROOPWAFEL \citep{Broekgaarden:2019MNRAS} can assist in producing a more massive data-set using similar computation time.  

Our best-fit model FDM-20 that showed the closest match to the observed catalogued radio DNSs had the following pulsar parameters: \\
a) Uniform birth magnetic field distribution ($10^{10}$--$10^{13}$ G);\\
b) Uniform birth spin period distribution ($10$--$100$ ms);\\
c) Magnetic field decay time-scale ($1000$ Myrs);\\
d) Magnetic field decay mass-scale ($0.02$\,M$_\mathrm{\odot}$);\\
e) CE mass accretion following \citet{MacLeod:2014yda};\\
f) A lognormal pulsar luminosity function following \citet{Szary:2014dia}.

Using our best fit model calibrated to the Galactic radio DNS population, we analyse the entire DNS population from the perspective of gravitational waves. The median merger-time for the entire DNS population is 33.1 Gyr, while that for radio-DNSs (${\approx}$10\% of total DNSs) is 6.6 Gyr. We note that ${\approx}$34\% of the present DNSs will merge within the next 13 Gyr. For the net DNS population ${\approx}$40\% have $M_\mathrm{chirp}\geq$1.2\,M$_\odot$. The median of the chirp mass distribution for the net DNS population is $\approx$1.14\,M$_\odot$ The median of the chirp mass distribution for radio-DNSs is greater than that for the net DNS chirp mass by ${\approx}$0.07\,M$_\odot$. The DNSs show the maximum value of effective spin; $\chi_\mathrm{eff}{\approx}0.03$, with the radio sub-population showing a biased peak around $\chi_\mathrm{eff}{\approx}0.01$. 

Of the DNSs that merge in a Hubble Time (HT), ${\approx}$40\% have a total (remnant) mass $\leq 2.7$\,M$_\odot$, the mass of the remnant of GW170817 \citep{TheLIGOScientific:2017qsa,Abbott:2018wiz}. Neutron star merger remnants are of interest as they may be sources of gravitational-waves and progenitors of magnetars or Fast Radio Bursts. The uncertainty in the maximum neutron star mass induces ambiguity in the fraction of merger remnants which will collapse to a black hole.  

\section*{Acknowledgements}

We thank Matthew Bailes, Ilya Mandel, Alejandro Vigna-G\'{o}mez, Floor Broekgaarden, Coenraad Neijssel and Serena Vinciguerra for useful comments and suggestions. We thank Stefan Os{\l}owski, Andrew Cameron and Rahul Sengar for instructive discussions on pulsar selection effects. The authors are supported by the Australian Research Council Centre of Excellence for Gravitational Wave Discovery (OzGrav), through project number CE170100004. This work made  use  of  the  OzSTAR  high  performance  computer at  Swinburne  University  of  Technology. OzSTAR  is funded by Swinburne University of Technology and the National Collaborative Research Infrastructure Strategy(NCRIS). We thank the referee for their helpful insights and constructive suggestions.




\bibliographystyle{mnras}
\bibliography{bib_new}







\bsp	
\label{lastpage}
\end{document}